\documentclass[aps,pra,twocolumn,superscriptaddress]{revtex4-1}
\usepackage{amsmath}
\usepackage[pdftex]{graphicx}
\usepackage[pdftex,colorlinks=true,bookmarks=true,citecolor=blue,linkcolor=blue,urlcolor=blue,breaklinks=true]{hyperref}

\begin{document}
\title{High-harmonic generation in quantum spin systems}
\author{Shintaro Takayoshi}
\affiliation{Max Planck Institute for the Physics of Complex Systems,
Dresden 01187, Germany}
\affiliation{Department of Quantum Matter Physics, University of Geneva,
Geneva 1211, Switzerland}
\author{Yuta Murakami}
\affiliation{Department of Physics, University of Fribourg,
Fribourg 1700, Switzerland}
\author{Philipp Werner}
\affiliation{Department of Physics, University of Fribourg,
Fribourg 1700, Switzerland}

\date{\today}

\begin{abstract}
We theoretically study the high-harmonic generation (HHG)
in one-dimensional spin systems.
While in electronic systems the driving by AC electric fields produces radiation 
from the dynamics of excited charges, 
we consider here the situation where spin systems excited 
by a magnetic field pulse 
generate radiation via a time-dependent magnetization. 
Specifically, we study the magnetic dipole radiation 
in two types of ferromagnetic spin chain models, 
the Ising model with static longitudinal field and the XXZ model, 
and reveal the structure of the spin HHG and its relation to spin excitations.
For weak laser amplitude, a peak structure appears which can be explained by 
time-dependent perturbation theory.
With increasing amplitude, plateaus with well-defined cutoff energies emerge. 
In the Ising model with longitudinal field, 
the thresholds of the multiple plateaus in the radiation spectra 
can be explained by the annihilation of multiple magnons. 
In the XXZ model, which retains the $\mathbf{Z}_{2}$ symmetry, 
the laser magnetic field can induce a phase transition of the ground state 
when it exceeds a critical value,
which results in a drastic change of the spin excitation character.
As a consequence,
the first cutoff energy in the HHG spectrum changes from a single-magnon 
to a two-magnon energy at this transition.
Our results demonstrate the possibility of 
generating high-harmonic radiation from 
magnetically ordered materials and 
the usefulness of high-harmonic signals for extracting information on
the spin excitation spectrum.
\end{abstract}

\maketitle

\section{Introduction}

%General introduction of HHG
The dynamics induced by light-matter coupling 
is an important problem in optical physics 
as well as nonequilibrium condensed matter and statistical physics. 
The application of strong laser pulses to 
a broad range of materials, including metals, semiconductors, 
and superconductors, results in rich physics and new phenomena, such as  
collective excitations~\cite{Matsunaga2014Science,Lu2017PRL}, 
the control of order parameters~\cite{Kirilyuk2010RMP,Kampfrath2013NatPhoton}, 
and fundamental changes in material properties~\cite{Nasu2004Book,Oka2009PRB,Lindner2011NatPhys}.
In particular, the high-harmonic generation (HHG), 
which is a nonlinear optical phenomenon observed in periodically driven systems, 
is attracting interest because of the underlying nontrivial charge dynamics 
and its technological relevance for attosecond laser science 
and the spectroscopy of charge dynamics~\cite{Krausz2009RMP,Cavalieri2007}.

%HHG from solids
HHG has originally been observed and studied in atoms and molecular gases \cite{McPherson1987,Ferray1988}.
Its mechanism can be understood by the so-called three step model, 
where tunnel-ionization occurs in the presence of a strong electric field, 
the released electrons are accelerated by the periodic field 
and eventually recombine with the ionized atoms 
by emitting the high-harmonic light~\cite{Corkum1993PRL,Lewenstein1994}.
Recently the interest in this field has been renewed by the observation of HHG in various solids, 
in particular band insulators~\cite{Ghimire2011NatPhys,Schubert2014,Luu2015,Vampa2015Nature,Langer2016Nature,Hohenleutner2015Nature,Ndabashimiye2016,Liu2017,You2016,Kaneshima2018,Luu2018}.
Although the HHG in this case also originates from the dynamics of excited charges, 
the spatially periodic arrangement of atoms in solids leads to 
qualitative differences compared to atomic gases.  
Theoretical studies assuming weak correlations or employing 
an effective single particle picture have been performed to discuss 
the origin of the HHG in these band 
insulators~\cite{Golde2008,Ghimire2011NatPhys,Kemper2013NJP,Higuchi2014,Vampa2014PRL,Wu2015,Tamaya2016,Vampa2015PRB,Luu2016,Otobe2016,Ikemachi2017,Osika2017,Hansen2017,Tancogne-Dejean2017b,Tancogne-Dejean2017,Ikemachi2018,Ikeda2018PRA}. 
(For recent reviews, see Refs.~\cite{Huttner2017,Kruchinin2018,Ghimire2019}.)
It has been revealed that HHG originates from the intraband charge dynamics 
reflecting the non-parabolic shape of the  bands~\cite{Luu2015,Golde2008,Ghimire2011NatPhys} 
and the interband dynamics corresponding to the recombination 
of excited charges~\cite{Vampa2014PRL,Vampa2015PRB,Otobe2016,Ikemachi2017,Osika2017}. 
Furthermore, the existence of multiple bands and the interference 
between different excitation paths can play 
an important role~\cite{Wu2015,Luu2016,Hansen2017,Ikemachi2017}.
Even though the details of its mechanism are still actively discussed, 
HHG in solids can be used to obtain important information about these solids, such as band 
and lattice structures~\cite{Ghimire2011NatPhys,Ndabashimiye2016,Liu2017,Hohenleutner2015Nature,You2016}.
In addition, potential applications in new high-frequency laser sources are expected 
due to the high concentration of atoms compared to atomic gases~\cite{Ndabashimiye2016}.
Stimulated by these developments and prospects, both experimentalists and theorists 
are making intensive efforts to understand the mechanism of HHG in greater detail
and to explore new classes of materials, e.g., 
liquids~\cite{Heissler2014NJP,Luu2018}, 
graphene~\cite{Yoshikawa2017Science,Hafez2018}, 
topological systems~\cite{Chacon2018arXiv}, 
strongly correlated systems~\cite{Silva2018NatPhoton,Murakami2018PRL,Murakami2018PRB,Tancogne-Dejean2018,Zhu2018}, 
impurity-doped systems~\cite{Yu2019PRA} 
and magnetic metals~\cite{Zhang2018NatComm}. 

%%%%%%%%%% Fig : Concept of this paper %%%%%%%%%%
\begin{figure*}[t]
\includegraphics[width=0.9\textwidth]{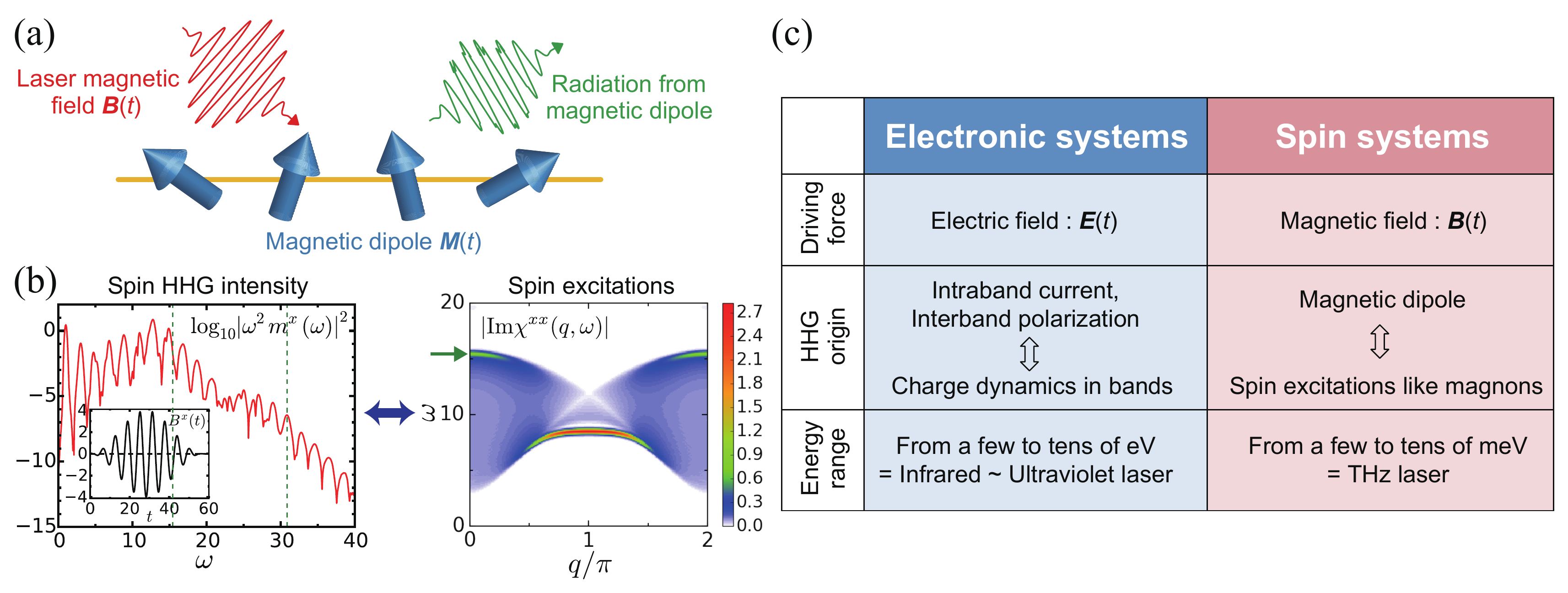}
\caption{(a) Schematic picture of the HHG from quantum magnets 
discussed in this paper.  The spins are excited by a magnetic field pulse, 
and the induced magnetization dynamics results 
in electromagnetic radiation with high frequency components.  
(b) Example of the correspondence between the spin HHG intensity 
and the spin excitation spectrum 
(XXZ model with $J_{xy}=2$ and $J_{z}=10$, pulse with $\Omega=1$ and $B=4$). 
The thresholds (vertical dashed lines) of the multiple plateaus in the HHG signal 
correspond to multiples of the magnon excitation energy at $q=0$, see horizontal arrow. 
(Inset: The shape of the magnetic field 
for the linearly polarized pulse laser %~\eqref{eq:laser_mag_field} 
with $B=4$ and $N_{\mathrm{cyc}}=9$.) 
(c) Comparison between the HHG in electronic systems 
and that in spin systems. 
}
\label{fig:Concept}
\end{figure*}
%%%%%%%%%%%%%%%%%%%%

%Aim and general message of this paper
In this paper, we explore a new avenue for HHG in solids, 
by considering the dynamics of the {\it spin degrees of freedom} in magnetic insulators, 
i.e. quantum spin systems. 
We theoretically study the excitation of these systems 
by time-periodic external {\it magnetic} fields, 
and evaluate the HHG signal resulting 
from the change of the {\it magnetic} moments [Fig.~\ref{fig:Concept}(a)]. 
This setup is relevant for materials with a large charge excitation gap 
whose low energy excitations are governed by the spin degrees of freedom. 
Recent developments in the field of metamaterials~\cite{Mukai2014APL} 
and plasmonics~\cite{Ciappina2017RepProgPhys} 
enable the generation of strong magnetic field pulses with small electric field, 
which can be used to realize the setup considered in our study.
The nonequilibrium dynamics of quantum spin systems, 
especially the dynamical control of the magnetization by laser fields, 
has been intensively studied 
both in the experimental~\cite{Hohlfeld1997PRL,Kimel2005Nature,Kirilyuk2010RMP} 
and theoretical communities~\cite{Takayoshi2014PRBa,Takayoshi2014PRBb}. 
In Refs.~\cite{Takayoshi2014PRBa,Takayoshi2014PRBb}, 
the magnetization dynamics in antiferromagnets has been calculated 
for laser fields with a frequency comparable to the exchange coupling. 
On the other hand, for the study of HHG, a lower photon energy is advantageous 
since it results in spectra with higher energy resolution and thus 
allows to elucidate the excitation structure. 
In this paper, 
we reveal that the HHG signal from spin systems can be associated 
with elementary spin excitations like magnons, 
just as the HHG in electronic systems reflects 
the dynamics of excited charges [see Fig.~\ref{fig:Concept}(b)].
These results suggest that the spin HHG can be potentially 
used as a probe of spin dynamics as well as 
for new laser sources in the THz regime. 
In Fig.~\ref{fig:Concept}(c), we summarize the similarities and differences 
to HHG in electronic systems, which are useful to keep in mind 
in the following discussion.

%Concrete contents
The present study focuses on one-dimensional ferromagnetic quantum spin systems 
described by the Ising model with longitudinal field and the XXZ model. 
These models are simple but fundamental, 
and can be realized in materials such as  $\mathrm{Dy(C_{2}H_{5}SO_{4})_{3}\cdot 9H_{2}O}$, 
$\mathrm{LiTbF_{4}}$, $\mathrm{LiHoF_{4}}$ ~\cite{Wolf2000BJP} 
and $\mathrm{CoNb_{2}O_{6}}$~\cite{Coldea2010Science}.
We numerically investigate the nonequilibrium dynamics 
and the radiation spectrum resulting from the time-dependent magnetization by means of 
the infinite time-evolving block decimation (iTEBD)~\cite{Vidal2007PRL}, 
exact diagonalization (ED) calculations, 
and time-dependent mean-field theory (tdMF).
To understand the relation between the HHG signal and elementary spin excitations,
we also calculate the low-energy excitation structure of these systems 
by combining the density matrix renormalization group (DMRG)~\cite{White1992PRL} 
and the time-evolving block decimation (TEBD)~\cite{Vidal2003PRL}.
When the laser field is weak, a peak appears 
around the energy of the single-magnon excitation in both models, 
which can be explained by time-dependent perturbation theory. 
With increasing strength of the laser field, 
this peak structure changes to a plateau.
We also find indications for multiple plateaus, 
whose thresholds are associated with the annihilation of
(multiple) elementary spin excitations (magnons). 

%structure of this paper
This paper is organized as follows. 
In Sec.~\ref{sec:HHGinQSS}, 
we discuss general properties of the HHG 
in quantum spin systems. 
Section~\ref{sec:Ising} presents the HHG signals 
resulting from the application of a linearly polarized laser 
to Ising models with longitudinal static field. 
Section~\ref{sec:XXZweakJxy} is devoted to an analysis of the HHG signal 
from the laser application to the XXZ models.
We summarize our results and 
discuss future extensions in Sec.~\ref{sec:Summary}. 

\section{HHG in quantum spin systems}
\label{sec:HHGinQSS}

In this section, we present the theory of HHG 
in quantum spin systems. 
In usual HHG, the electric field of a laser pulse  
induces a change of the electric polarization, 
which in turn produces electromagnetic waves. 
The total instantaneous radiated power is proportional to  
$|d\boldsymbol{j}(t)/dt|^{2}$, 
where $\boldsymbol{j}(t)$ is the electric current. 
If $\boldsymbol{j}(t)$ is a polarization current 
$d\boldsymbol{P}(t)/dt$ 
(with $\boldsymbol{P}$ the electric interband polarization), 
the power is proportional to 
$|d^{2}\boldsymbol{P}(t)/dt^{2}|^{2}$. 
In a similar way, we can consider the radiation of 
electromagnetic waves from a time-dependent magnetic dipole. 
The total instantaneous radiated power 
from the change of a localized magnetic dipole 
$\boldsymbol{M}(t)$ is proportional to 
$|d^{2}\boldsymbol{M}(t)/dt^{2}|^{2}$~\cite{Jackson1998Book}. 

To study quantum spin systems 
in the presence of a time-dependent magnetic field $\boldsymbol{B}(t)$
we consider the Hamiltonian
\begin{align}
 \mathcal{H}(t)=\mathcal{H}_{\mathrm{spin}}
   -\boldsymbol{B}(t)\cdot
   \boldsymbol{S}_{\mathrm{tot}},
\label{eq:Hamil_tdep}
\end{align}
where $\mathcal{H}_{\mathrm{spin}}$ is the spin Hamiltonian 
and the last term represents the Zeeman coupling of the spins in the material 
with the magnetic field produced by the laser.  
We calculate the time evolution of the magnetization 
$\boldsymbol{M}(t)\equiv\langle\boldsymbol{S}_{\mathrm{tot}}(t)\rangle$, 
where $\boldsymbol{S}_{\mathrm{tot}}=\sum_{j}\boldsymbol{S}_{j}$ 
represents the summation over all spins and 
$\boldsymbol{S}_{\mathrm{tot}}(t)\equiv U^{-1}(t)\boldsymbol{S}_{\mathrm{tot}}U(t)$ 
($U(t)=\mathcal{T}\int_{0}^{t}dt'\exp[-i\mathcal{H}(t')t']$ 
is the time evolution operator with $\mathcal{T}$ the time ordering). 
From this we obtain the Fourier transform of the magnetization 
$\boldsymbol{M}(\omega)=\int dt e^{i\omega t} \boldsymbol{M}(t)$ 
and the radiation power  
\begin{align}
 I\propto|\omega^{2}\boldsymbol{M}(\omega)|^{2}.
\nonumber
\end{align}

The symmetry of the system may impose 
constraints on the structure of the HHG signal.
For example, the inversion symmetry limits the HHG signal 
in electronic systems to odd harmonics. 
Now, let us consider the case when the time dependent Hamiltonian has a symmetry
which can be represented as the combination of 
time translation and $\pi$ rotation around the $S^{z}$ axis
\begin{align}
\begin{cases}
 \mathcal{H}(t)\to\mathcal{H}(t+T_{\mathrm{per}}/2),\\
 (S^{x},S^{y},S^{z})\to(-S^{x},-S^{y},S^{z}),
\end{cases}
\label{eq:symmetry}
\end{align}
where $T_{\mathrm{per}}\equiv 2\pi/\Omega$ 
is the period of the laser.
Then, the magnetization satisfies 
\begin{align}
 &M^{x}(t+T_{\mathrm{per}}/2)=-M^{x}(t),\nonumber\\
 &M^{y}(t+T_{\mathrm{per}}/2)=-M^{y}(t),\nonumber\\
 &M^{z}(t+T_{\mathrm{per}}/2)= M^{z}(t),\nonumber
\end{align}
if we assume a time-periodic steady state having the same symmetry as the Hamiltonian. 
In this case, the temporal Fourier transform of $M^{x}$ 
becomes 0 for $\omega=2n\Omega$ ($n$ is an integer) since 
\begin{align}
 &M^{x}(2n\Omega)\propto
   \int_{0}^{T_{\mathrm{per}}}dt\hspace{0.5mm}
     e^{i2n\Omega t}M^{x}(t)\nonumber\\
   &=\int_{0}^{\frac{T_{\mathrm{per}}}{2}}dt \hspace{0.5mm}e^{i2n\Omega t}
     (M^{x}(t)+M^{x}(t+T_{\mathrm{per}}/2))=0.
\label{eq:SteadyCondMx}
\end{align}
In the same way, 
the temporal Fourier transform of $M^{z}$ becomes 0 
for $\omega=(2n+1)\Omega$ ($n$ is an integer) since 
\begin{align}
 &M^{z}((2n+1)\Omega)\propto
   \int_{0}^{T_{\mathrm{per}}}dt\hspace{0.5mm}
     e^{i(2n+1)\Omega t}M^{z}(t)\nonumber\\
   &=\int_{0}^{\frac{T_{\mathrm{per}}}{2}}dt \hspace{0.5mm}e^{i(2n+1)\Omega t}
     (M^{z}(t)-M^{z}(t+T_{\mathrm{per}}/2))=0.
\label{eq:SteadyCondMz}
\end{align}
In the case of a finite pulse width, 
these arguments are strictly speaking not valid. 
Still we will see that in practice, 
these rules are satisfied except around the HHG peak 
in the case of weak laser fields. 

In the following two sections, 
we will use numerical calculations to study the HHG 
in specific one-dimensional quantum spin systems.

\section{HHG in Ising models}
\label{sec:Ising}

Let us start by investigating the HHG in Ising models,
which are among the simplest and most important models of magnets. 
In this case, the spin Hamiltonian $\mathcal{H}_{\mathrm{spin}}$ 
in Eq.~\eqref{eq:Hamil_tdep} explicitly reads  
\begin{align}
 \mathcal{H}_{\mathrm{Ising}}
   =-J\sum_{j}S_{j}^{z}S_{j+1}^{z}
   -HS_{\mathrm{tot}}^{z},
\label{eq:IsingModel}
\end{align}
where $J>0$ is the ferromagnetic exchange coupling 
and $H>0$ is a static external magnetic field. 
$S^{x}$, $S^{y}$, and $S^{z}$ are spin-1/2 operators. 
The ground state of $\mathcal{H}_{\mathrm{Ising}}$ is 
a ferromagnetic state 
($\langle S_{j}^{z}(0)\rangle= 1/2$ for all $j$ ) 
and this state is perturbed by the application of a linearly polarized pulse laser 
$\boldsymbol{B}(t)=(B^{x}(t),0,0)$ in the $x$ direction.
Because of the longitudinal field $H>0$, 
the $\mathbf{Z}_{2}$ symmetry of the system is broken. 
Hence, there is no quantum phase transition 
as a function of the transverse magnetic field, 
i.e., the ground state of the snapshot Hamiltonian 
$\mathcal{H}(t)=\mathcal{H}_{\mathrm{Ising}}-B^{x}(t)S_{\mathrm{tot}}^{x}$ 
remains gapped at any time. 
Though the main objective of this section is the theoretical analysis  
of the magnetization dynamics and HHG mechanism, 
the obtained results are relevant for materials having 
ferromagnetic dipole order such as 
$\mathrm{Dy(C_{2}H_{5}SO_{4})_{3}\cdot 9H_{2}O}$, 
$\mathrm{LiTbF_{4}}$ and $\mathrm{LiHoF_{4}}$~\cite{Wolf2000BJP}. 

We consider a magnetic field pulse of the form  
\begin{align}
 B^{x}(t)=
\left\{
\begin{array}{ll}
\displaystyle
 B\sin^{2}\Big(\frac{\Omega t}{2N_{\mathrm{cyc}}}\Big)
   \cos(\Omega t)\;&
   (0<t<T_{\mathrm{f}})\\
 0&(\mathrm{otherwise})
\end{array}
\right.,
\label{eq:laser_mag_field}
\end{align}
where $T_{\mathrm{f}}=2\pi N_{\mathrm{cyc}}/\Omega$, 
$\Omega$ is the laser frequency, 
$N_{\mathrm{cyc}}$ the number of laser cycles, 
and $B\sin^{2}(\frac{\Omega t}{2N_{\mathrm{cyc}}})$  
the envelope of the pulse. 
In this paper, the parameters are fixed as 
$N_{\mathrm{cyc}}=9$ and $\Omega=1$ ($\Omega$ is also used as the energy scale 
by employing the units $\hbar=c=1$). 
The magnetic field pulse with $B=4$ 
is shown in the inset of  Fig.~\ref{fig:Concept}(b). 

In this section, the other parameters are set to $J=2$ and $H=6$, 
so that the gap is much larger than $\Omega=1$ 
and heating effects are suppressed. 
In addition, since we anticipate that the width of the plateau 
in the HHG signal is of the order of the characteristic energy scales of the spin system, 
we expect to observe several harmonics if $J$ and $H$ are chosen large compared to $\Omega$. 
The Ising model with smaller longitudinal field $H$, 
where the lifting of the two-fold degeneracy and hence the gap is smaller, 
is discussed in Appendix~\ref{sec:IsingWeakField}. 
We numerically calculate the magnetization dynamics and 
report hereafter the normalized magnetizations  
$m^{x,y,z}\equiv M^{x,y,z}/N$, where $N$ is the number of spins.
As explained in Sec.~\ref{sec:HHGinQSS}, 
the radiation power of a magnetic dipole is proportional to 
$|\omega^{2}m^{\alpha}(\omega)|^{2}$. 

%%%%%%%%%% Fig : susceptibility with high Hz %%%%%%%%%%
\begin{figure}[t]
\includegraphics[width=0.48\textwidth]{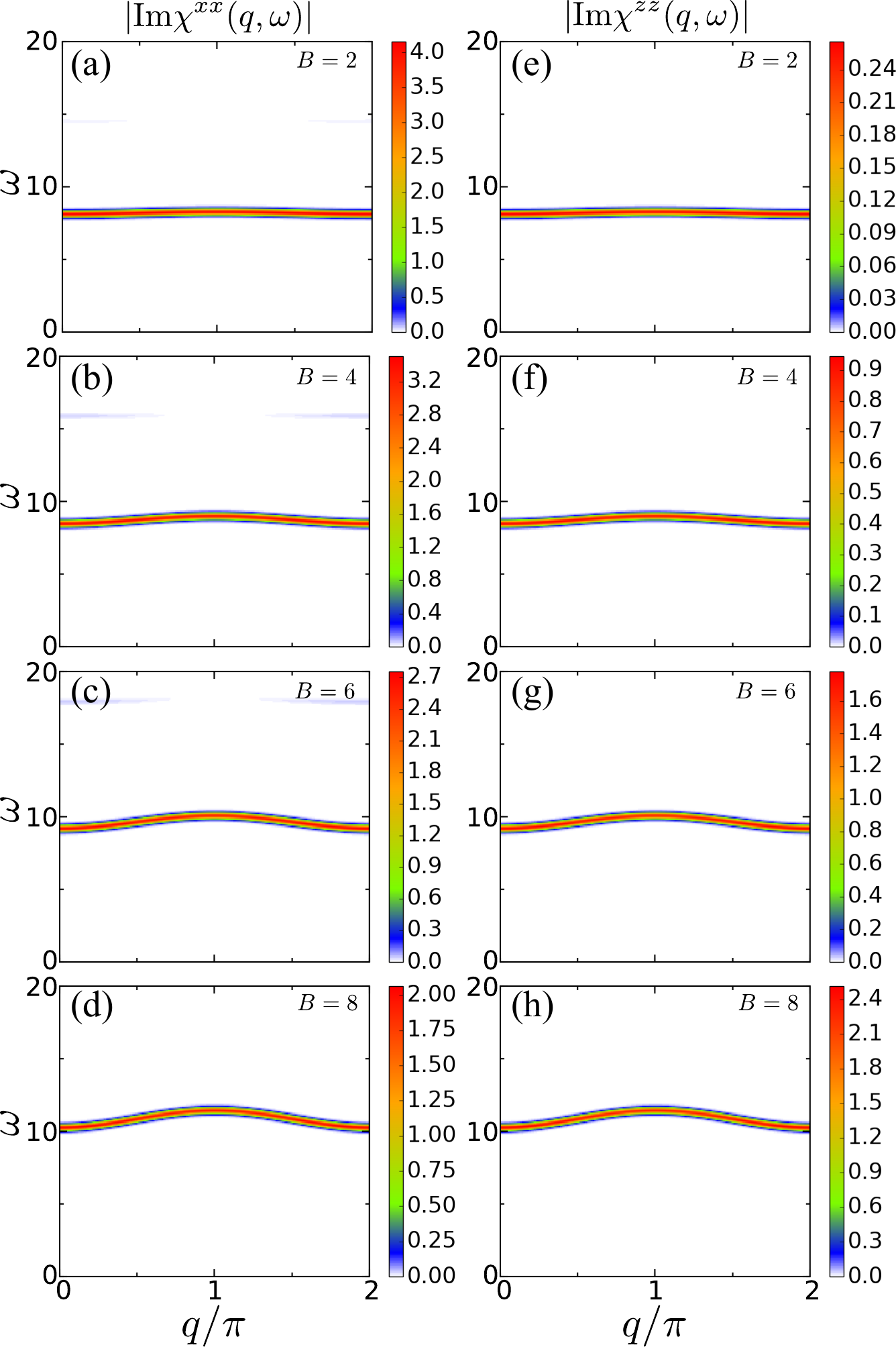}
\caption{DSFs
(a)-(d) $|\mathrm{Im}\chi^{xx}(q,\omega)|$ and 
(e)-(h) $|\mathrm{Im}\chi^{zz}(q,\omega)|$ 
for the Ising model with $J=2$ and $H=6$. 
}
\label{fig:Ising_high_suscep}
\end{figure}
%%%%%%%%%%%%%%%%%%%%

Before studying the dynamics induced by the laser field, 
we investigate the excitation structure of the equilibrium system. 
To study excitations, 
we numerically calculate the dynamical structure factor (DSF), 
which is the imaginary part of the dynamical susceptibility. 
The method is as follows. 
We first obtain the ground state of the system 
by the DMRG~\cite{White1992PRL}, 
and then calculate the retarded correlation function 
\begin{align}
 \chi^{\alpha\beta}(r,t)
   =-i\vartheta(t)
     \langle[S_{r}^{\alpha}(t),S_{0}^{\beta}(0)]\rangle,
\label{eq:RetCorrel}
\end{align}
where $\vartheta(t)$ is the step function, 
by the TEBD method~\cite{Vidal2003PRL} for finite size systems. 
The dynamical susceptibility is the Fourier transform
of the retarded correlation function,
\begin{align}
 \chi^{\alpha\beta}(q,\omega)
   =\int_{-\infty}^{\infty}dt\sum_{r}
     e^{i(\omega t-qr)}
     \chi^{\alpha\beta}(r,t).
\nonumber
\end{align}
In this paper, we consider systems with size $N=120$, 
which are large enough that finite size effects can be neglected. 

The DSFs $|\mathrm{Im}\chi^{xx}(q,\omega)|$ and 
$|\mathrm{Im}\chi^{zz}(q,\omega)|$ for 
the ground state of the Ising model 
in both longitudinal and transverse fields 
\begin{align}
 \mathcal{H}=\mathcal{H}_{\mathrm{Ising}}
   -BS_{\mathrm{tot}}^{x}
\label{eq:Hamil_Ising_LandT}
\end{align}
with $H=6$ are shown in Fig.~\ref{fig:Ising_high_suscep}. 
Equation~\eqref{eq:Hamil_Ising_LandT} represents  
the snapshot Hamiltonian of 
$\mathcal{H}(t)=\mathcal{H}_{\mathrm{Ising}}-B^{x}(t)S_{\mathrm{tot}}^{x}$
at some fixed time $t$ corresponding to $B^{x}(t)=B$. 
If the transverse field is not present ($B=0$), 
the spins are completely localized and there is no dispersion 
since the Hamiltonian only contains $S^{z}$. 
The elementary excitation corresponds to a single spin flip, 
which has a gap $J+H$. 
In the presence of a nonzero transverse field, 
this flipped spin can propagate and transform into a magnon. 
The DSF shown in Fig.~\ref{fig:Ising_high_suscep} 
represents the magnon dispersion. 
In Figs.~\ref{fig:Ising_high_suscep}(a)-\ref{fig:Ising_high_suscep}(c), 
a weak intensity is seen at 
twice of the energy of the lowest band (single-magnon dispersion). 
This corresponds to the two-magnon band. 
Since the single-magnon band has a cosine structure 
$E_{1}(q)=c_{1}+c_{2}\cos(q)$, 
the two-magnon band can be represented as 
\begin{align}
 E_{2}(q)=&2c_{1}+c_{2}[\cos(q')+\cos(q-q')]\nonumber\\
   =&2c_{1}+2c_{2}\cos\Big(\frac{q}{2}\Big)
     \cos\Big(\frac{q-2q'}{2}\Big)
\quad(0\leq q'\leq 2\pi)
\nonumber
\end{align}
by considering the momentum conservation, 
and we obtain 
\begin{align}
 2c_{1}-2c_{2}|\cos(q/2)|\leq E_{2}(q)\leq 
 2c_{1}+2c_{2}|\cos(q/2)|.
\label{eq:TwoMagnonBand}
\end{align}
This feature of the two-magnon band is observed more evidently 
when the longitudinal field $H$ is weak 
as mentioned in Appendix~\ref{sec:IsingWeakField}.

%%%%%%%%%% Fig : Evolution of mag with high Hz %%%%%%%%%%
\begin{figure}[t]
\includegraphics[width=0.48\textwidth]{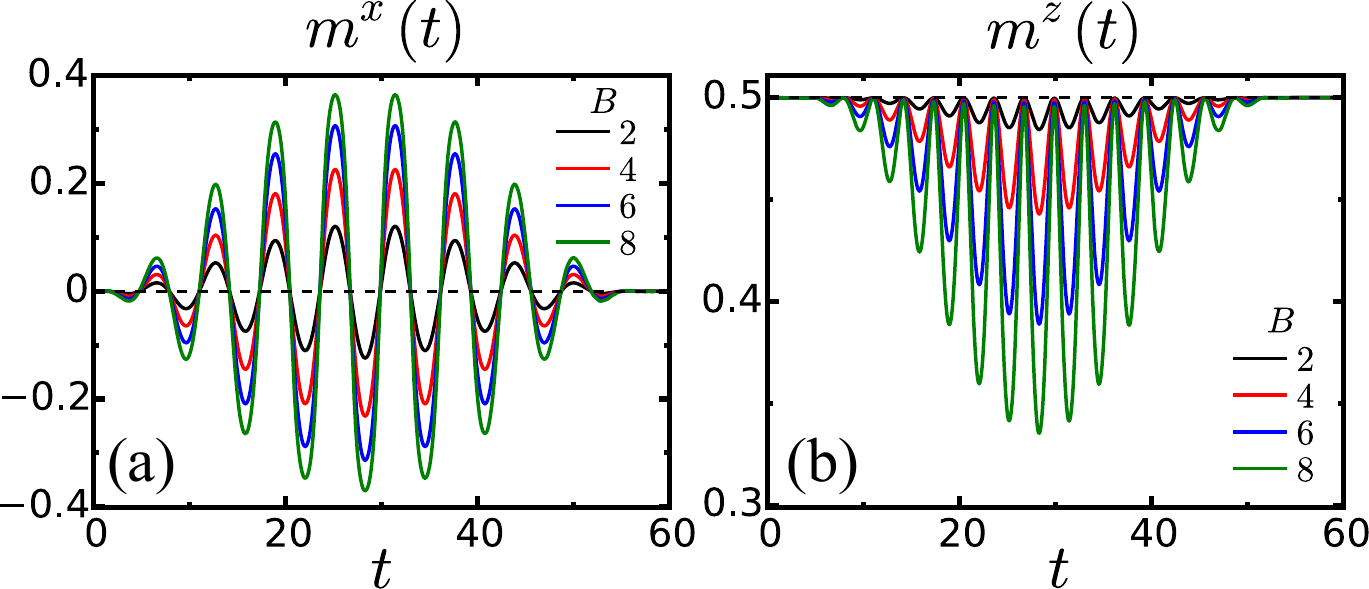}
\caption{Time evolution of (a) $m^{x}$ and (b) $m^{z}$ 
calculated by iTEBD 
for the Ising model with $J=2$ and $H=6$. 
}
\label{fig:Ising_high_mag}
\end{figure}
%%%%%%%%%%%%%%%%%%%%

In Fig.~\ref{fig:Ising_high_mag}, 
we show the time evolution of $m^{x}$ and $m^{z}$ 
for the Hamiltonian 
$\mathcal{H}(t)=\mathcal{H}_{\mathrm{Ising}}-B^{x}(t)S_{\mathrm{tot}}^{x}$
with different values of the laser amplitude $B=2,4,6,8$. 
As the numerical method, we use 
the iTEBD~\cite{Vidal2007PRL}, 
which utilizes a matrix product state (MPS) representation.
This method enables the simulation of infinite size systems, 
i.e., without finite-size effects, 
by assuming the translational invariance of the system. 
In this paper, we take the matrix dimension of the MPS as 100 
and the time evolution is performed by the fourth-order Trotter decomposition 
with the time step $\Delta t=0.05$. 
The shape of the time evolving $m^{x}$ is 
similar to that of the applied 
laser magnetic field (Eq.~\eqref{eq:laser_mag_field}) 
for all values of the laser amplitude. 
The value of $m^{z}$ drops when $|m^{x}|$ grows, 
but otherwise the magnetization in the $z$ direction recovers to $m^{z}=1/2$.
This demonstrates that the state of the system closely follows the ground state 
of the instantaneous Hamiltonian at each time. 
In other words, the laser frequency is slow enough 
for an adiabatic time evolution of the magnetization.  
In the present case of $H\gg J$, 
the gap is large ($\gg \Omega$) even for $B=0$, 
and it increases monotonically with increasing $B$ 
(Fig.~\ref{fig:Ising_high_suscep}). 
Thus, transitions to excited states through the Landau-Zener process are suppressed 
and the state remains in the snapshot ground state. 
However, if $B$ is further increased, 
the chain will eventually be disordered after the laser application,
similarly to what is shown in Fig.~\ref{fig:Ising_low_mag}(b) 
in Appendix~\ref{sec:IsingWeakField}. 

%%%%%%%%%% Fig : HHG with high Hz %%%%%%%%%%
\begin{figure}[t]
\includegraphics[width=0.48\textwidth]{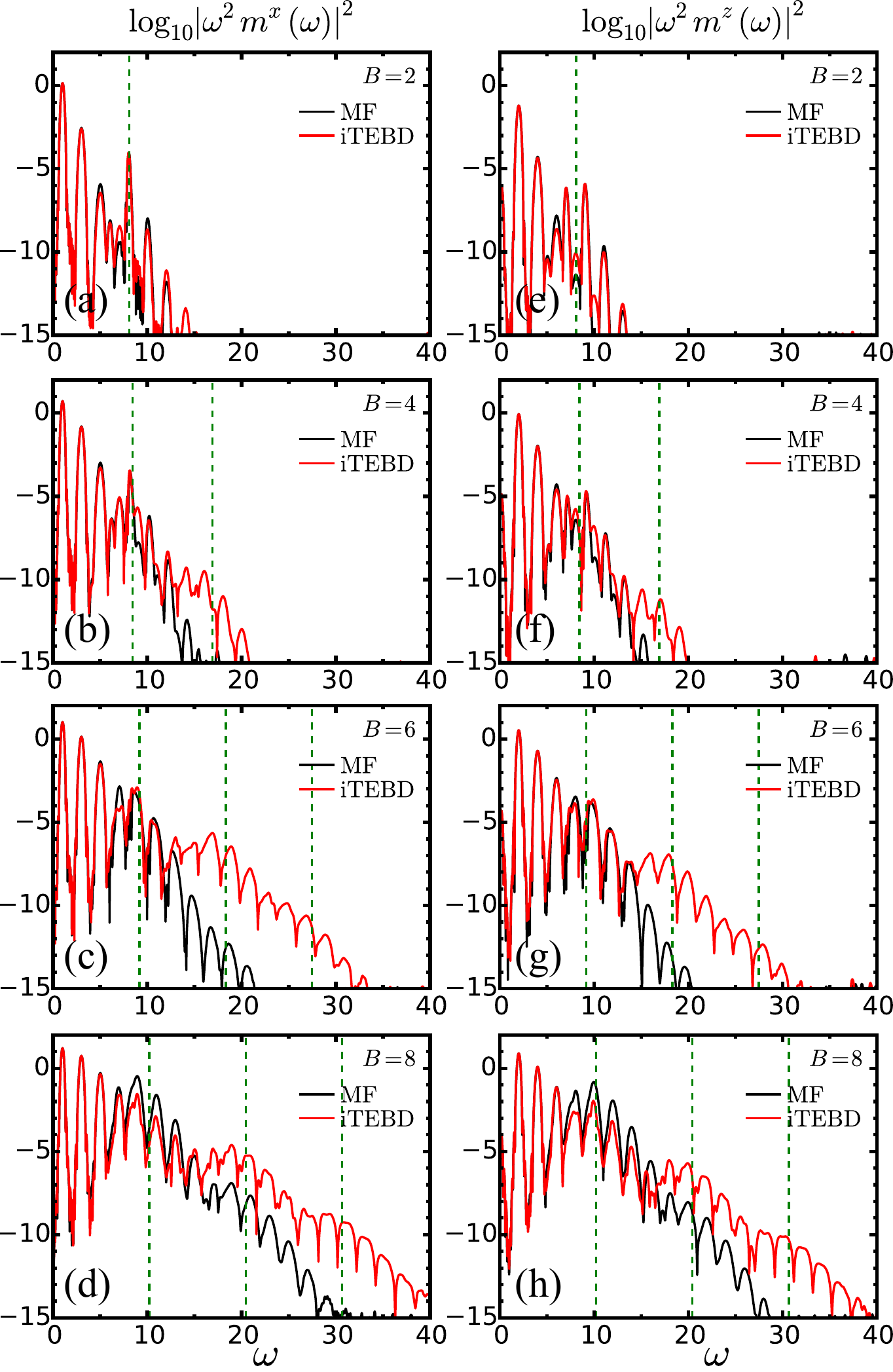}
\caption{The radiation power from 
(a)-(d) $m^{x}$ and (e)-(h) $m^{z}$ 
in the Ising model with $J=2$ and $H=6$. 
The dashed lines correspond to the mass of a magnon at $q=0$ 
(times integer). 
}
\label{fig:Ising_high_HHG}
\end{figure}
%%%%%%%%%%%%%%%%%%%%

\begin{figure}[t]
\includegraphics[width=0.48\textwidth]{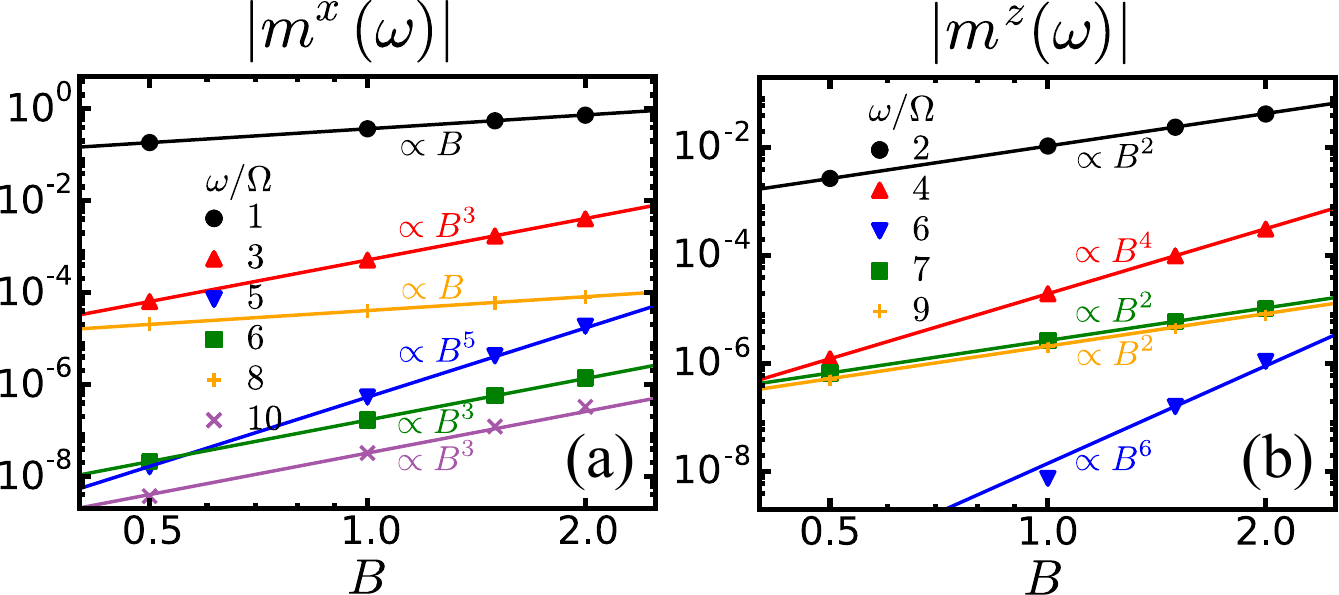}
\caption{Scaling of the magnitude of the Fourier components for the magnetizations 
(a) $|m^{x}(\omega)|$ and (b) $|m^{z}(\omega)|$ 
in the region of small $B$, 
in the Ising model with $J=2$ and $H=6$. 
The power of $B$ agrees well with the prediction 
from the time-dependent perturbation theory. 
}
\label{fig:Ising_high_Bscaling}
\end{figure}

To investigate the HHG, 
we plot $|\omega^{2}m^{x}(\omega)|^{2}$ and 
$|\omega^{2}m^{z}(\omega)|^{2}$ 
on a logarithmic scale in Fig.~\ref{fig:Ising_high_HHG}. 
These spectra were obtained by first differentiating $m^{x(z)}(t)$ numerically as 
${m''}^{x(z)}(t)=[m^{x(z)}(t+\Delta t)+m^{x(z)}(t-\Delta t)
-2m^{x(z)}(t)]/(\Delta t)^{2}$, 
where $\Delta t$ is the time step, 
and then performing the Fourier transform. 
In the Fourier transform, we apply the Blackman window 
$W_{\mathrm{B}}(t)=0.42-0.5\cos(2\pi t/T_{\mathrm{f}})
+0.08\cos(4\pi t/T_{\mathrm{f}})$ 
$(0<t<T_{\mathrm{f}})$ and 
$W_{\mathrm{B}}(t)=0$ (otherwise). 
The result in Fig.~\ref{fig:Ising_high_HHG} 
clearly demonstrates the HHG for all values of $B$ 
in both magnetization components $m^{x}$ and $m^{z}$. 
Since the system satisfies the symmetry Eq.~\eqref{eq:symmetry}, 
$m^{x}(\omega)$ and $m^{z}(\omega)$ become 0 
at $\omega=2n\Omega$ and $\omega=(2n+1)\Omega$ 
($n$ is an integer), respectively, for steady states. 
Although the presented results are for the transient case, 
the magnitudes of $|\omega^{2}m^{x}(\omega)|^{2}$ and 
$|\omega^{2}m^{z}(\omega)|^{2}$ drop 
at $\omega=2n\Omega$ and $\omega=(2n+1)\Omega$, respectively.
An exception occurs when $B$ is small and $\omega$ around the value 
corresponding to the excitation gap, 
as can be seen in Figs.~\ref{fig:Ising_high_HHG}(a), 
\ref{fig:Ising_high_HHG}(b), \ref{fig:Ising_high_HHG}(e), 
and \ref{fig:Ising_high_HHG}(f) 
where the spectra exhibit peaks at $\omega=6,8,10$ in $m^{x}(\omega)$ 
and at $\omega=7,9$ in $m^{z}(\omega)$ for $B=2,4$. 
Since we consider the application of a laser pulse, 
the system is in a transient regime and 
does not reach a nonequilibrium steady state. 
Hence the conditions Eqs.~\eqref{eq:SteadyCondMx} and \eqref{eq:SteadyCondMz} 
are not necessarily satisfied. 
The result for the peak position of the HHG spectra 
is supported by time-dependent perturbation theory 
(Appendix~\ref{sec:TdepPerturb}). 
The peaks resulting from the perturbation theory are located 
at $\omega=H+J$ for $m^{x}$ and at $\omega=H+J\pm\Omega$ for $m^{z}$, 
i.e., they can appear at an arbitrary frequency 
(not necessarily an integer multiple of $\Omega$) 
depending on the values of $H$ and $J$. 
The validity of the time-dependent perturbation theory 
is also confirmed by the scaling of the radiation intensity 
with the laser amplitude $B$. 
In Fig.~\ref{fig:Ising_high_Bscaling}, 
we plot $|m^{x}(\omega)|$ and $|m^{z}(\omega)|$ 
in the region of small $B$. 
$|m^{x}(\omega)|$ and $|m^{z}(\omega)|$ at 
$\omega=n\Omega$ scale as $B^{n}$, while at  
$\omega=H+J\pm n\Omega$ they scale as $B^{n+1}$, 
which agrees with the prediction from the perturbation theory 
presented in Appendix~\ref{sec:TdepPerturb}. 
This result indicates that $B\leq 2$ is in the perturbative regime. 

%%%%%%%%%% Fig : Intensity scaling by B with high Hz %%%%%%%%%%
%%%%%%%%%%%%%%%%%%%%

In Fig.~\ref{fig:Ising_high_HHG}, 
when the field strength is sufficiently large, 
we can identify a frequency above which the intensity drops rapidly 
as well as multiple plateau structures.
We can connect these cut-off energies with the excitation structures of 
the snapshot Hamiltonians, 
in particular those with the maximum value of $B$.
In the dispersion relation obtained from the data 
in Fig.~\ref{fig:Ising_high_suscep}, 
the energy has a minimum (maximum) at $q=0$ ($q=\pi$), 
and the excitation gap corresponds to the mass of a magnon at $q=0$. 
We see that the intensity of $|\omega^{2}m^{x}(\omega)|^{2}$ and 
$|\omega^{2}m^{z}(\omega)|^{2}$ drops above the energies  
corresponding to integer multiples of the magnon mass at $q=0$, 
as indicated by the dashed lines in Fig.~\ref{fig:Ising_high_HHG}. 
This result suggests that for sufficiently large laser field amplitude, 
there occurs a spontaneous annihilation of $n(=1,2,3,\ldots)$ magnons, 
which leads to the emission of light 
with the frequency $n\Delta(B(t))$ at time $t$, 
where $\Delta$ represents the single-magnon energy gap. 
This situation is analogous to electron-hole or doublon-holon 
recombination in electron systems 
such as Mott insulators~\cite{Vampa2014PRL,Murakami2018PRL}, 
where the radiation originates primarily from the interband transitions. 

Further insights can be obtained from a subcycle analysis.
The subcycle Fourier transform of the magnetic moment is defined as 
\begin{align}
 m_{\mathrm{sub}}^{x(z)}(\omega;t_{*})\equiv
   \int dt e^{i\omega t}m^{x(z)}(t)W_{\mathrm{G}}(t;t_{*}),
\label{eq:SubcycFourier}
\end{align}
where 
$W_{\mathrm{G}}(t;t_{*})=\exp\big[-\frac{(t-t_{*})^{2}}{2\sigma^{2}}\big]$ 
($\sigma=T_{\mathrm{per}}/8$) 
is a Gaussian window function. 
(An alternative way to compute time-dependent spectra 
is the wavelet analysis. 
We discuss the result of the wavelet analysis and the difference to 
the window Fourier transform in Appendix~\ref{sec:Wavelet}.) 
In Fig.~\ref{fig:Ising_high_subcyc}, 
we show the subcycle radiation spectrum 
$\log_{10}|\omega^{2}m_{\mathrm{sub}}^{x(z)}(\omega;t_{*})|^{2}$ 
for $B=8$ as a colormap 
and the multiple magnon excitation energies  
of the snapshot Hamiltonian at $t_{*}$
by the solid lines. 
In the low-energy region ($\omega<10$), although 
$|\omega^{2}m_{\mathrm{sub}}^{x(z)}(\omega;t_{*})|^{2}$ 
does not much depend on $t_{*}$, 
one can roughly identify an enhanced HHG signal following the one-magnon energy. 
This is due to the fact that the single-magnon band changes only little 
as a function of the transverse field 
(see Fig.~\ref{fig:Ising_high_suscep}). 
On the other hand, in the high-energy region, 
a high intensity signal is produced when the magnetic field is strong.
In particular, we can clearly identify an enhanced HHG signal tracking 
the two-magnon and three-magnon lines, 
both in the radiation produced by the $x$ and $z$ magnetization components. 
These observations support the interpretation that the plateaus and their thresholds 
in the spin HHG originate from the annihilation of magnons.

%%%%%%%%%% Fig : subcycle analysis with high Hz %%%%%%%%%%
\begin{figure}[t]
\includegraphics[width=0.48\textwidth]{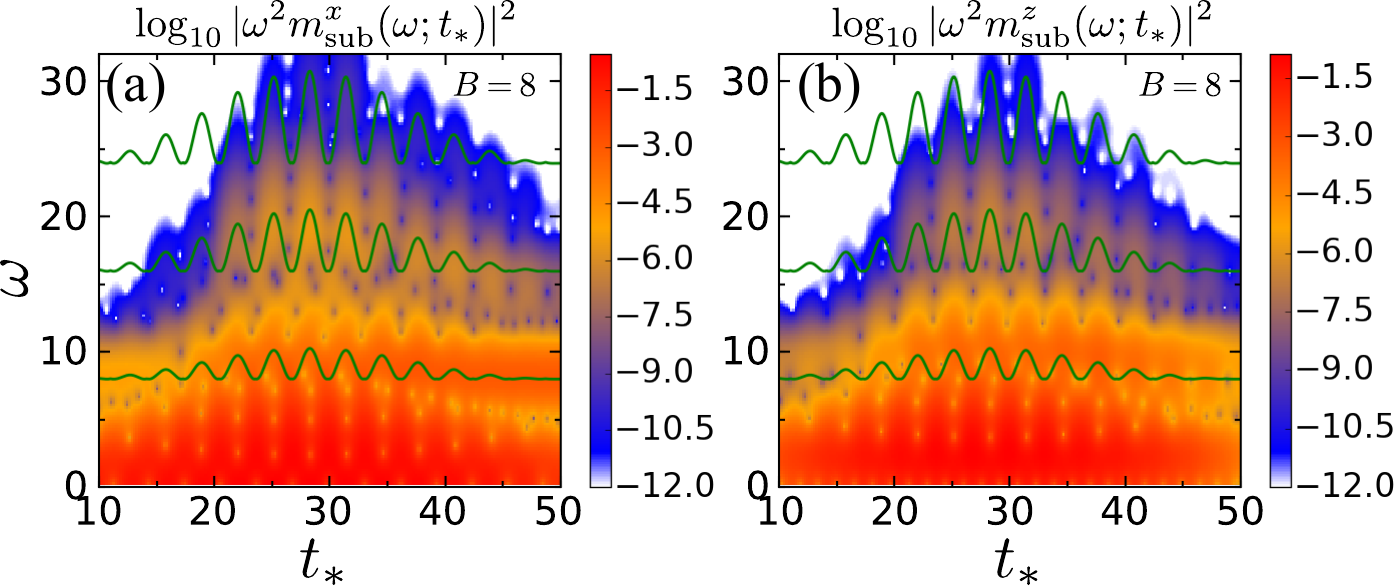}
\caption{Colormap of the subcycle radiation spectrum 
(a) $\log_{10}|\omega^{2}m_{\mathrm{sub}}^{x}(\omega;t_{*})|^{2}$ and 
(b) $\log_{10}|\omega^{2}m_{\mathrm{sub}}^{z}(\omega;t_{*})|^{2}$ 
for the Ising model with $J=2$, $H=6$, and $B=8$. 
The solid lines indicate the energies of one, two, and three magnons 
at the corresponding $B(t_{*})$.
}
\label{fig:Ising_high_subcyc}
\end{figure}
%%%%%%%%%%%%%%%%%%%%

We note that our discussion of the spin HHG so far has been based on 
the eigenstates or the energy structure of the snapshot Hamiltonians, 
as has been done for electronic systems 
using the Houston basis~\cite{Wu2015} or assuming 
a slowly changing field~\cite{Higuchi2014,Murakami2018PRL}.
To be more specific, let us expand the wave function as
$|\Psi(t)\rangle=\sum_{n}\alpha_{n}(t)|\Phi_{n}(B(t))\rangle$, 
where $|\Phi_{n}(B(t))\rangle$ is an eigenstate of the snapshot Hamiltonian 
with the eigenenergy $E_{n}(B(t))$, and express
the magnetization as 
\begin{align}
 M^{x(z)}(t)=\sum_{m,n} \alpha_{m}^{*}(t)\alpha_{n}(t) 
   \langle \Phi_{m}(B(t))|S^{x(z)} |\Phi_{n}(B(t))\rangle.
\label{eq:MagExpand}
\end{align}
We can then classify the contributions to the magnetization dynamics 
according to the character of 
$|\Phi_{m}(B(t))\rangle$ and $|\Phi_{n}(B(t))\rangle$.
The time dependence of the coefficients $\alpha_{n}$ follows from
\begin{align}
 i\partial_{t} \alpha_{n}(t)
   =&E_{n}(B(t))\alpha_{n}(t)\nonumber\\
     &-i\sum_{m\neq n} 
       (\partial_{t} B(t)) F_{nm} (B(t)) \alpha_{m}(t),
\label{eq:coeff}
\end{align}
where $F_{nm} (B)=\langle \Phi_{n}(B)| \partial_{B} |\Phi_{m}(B)\rangle$. 
If the variation of $B(t)$ (with excitation frequency $\Omega$) is slow enough, 
$\partial_{t} B(t)$ is small and $E_{n}(B(t))$ can be approximated as a constant 
for a certain time interval. 
Hence the second term on the right hand side of Eq.~\eqref{eq:coeff} 
can be neglected and we can write $\alpha_{n}(t)\propto e^{-iE_{n}(B(t_{*}))t}$ 
for $t$ around $t_{*}$. 
If these approximations hold and the time-dependence of $|\Phi_n(B(t))\rangle$ 
(and hence that of $ \langle \Phi_{m}(B(t))|S^{x(z)} |\Phi_{n}(B(t))\rangle$)
is also small enough, the main contribution 
to $M^{x(z)}(t)$ [Eq.~\eqref{eq:MagExpand}] 
is proportional to $e^{-i[E_n(B(t_{*}))-E_m(B(t_{*}))]t}$ 
for $t$ around $t_{*}$, 
which oscillates with (multiple) magnon energies.
If $|\Phi_{n}(B)\rangle$ and $|\Phi_{m}(B)\rangle$ differ by $l$ magnons, 
the radiation can be interpreted as originating from an $l$-magnon annihilation. 
However, in practice, there may be contributions 
from the second term on the right hand side of Eq.~(\ref{eq:coeff}) 
and the time-dependence of $|\Phi_{n}(B(t))\rangle$,
which leads to deviations from the simple magnon picture. 
Furthermore, the magnetization curve of the ground states 
for the Hamiltonian Eq.~\eqref{eq:Hamil_Ising_LandT} 
is a nonlinear function of $B$. 
Since $m^{x}$ for the ground state with the field $B$ 
is an odd function, we see, 
by replacing $B$ in this equation by $B\cos(\Omega t)$, 
that the Fourier component of $n\Omega$ 
(with $n$ an odd integer) appears in $m^{x}(\omega)$ 
and its leading order is $B^{n}$.
This partially explains the appearance of well-defined frequency components 
even in the energy region lower than the excitation gap seen 
in Fig.~\ref{fig:Ising_high_HHG}. 

The above results suggest that for the parameters chosen in this study, 
the magnon picture is essentially valid  
and the dynamics is described in terms of well-ordered magnetic moments, 
i.e. the effect of quantum fluctuations is small. 
To confirm this point, we perform a tdMF analysis. 
The approximation 
$\sum_{j}S_{j}^{z}S_{j+1}^{z}\simeq 2m^{z}\sum_{j}S_{j}^{z}-N{m^{z}}^{2}$ 
leads to the tdMF Hamiltonian  
\begin{align}
 \tilde{\mathcal{H}}_{\mathrm{Ising}}(t)
   =-2Jm^{z}(t)S^{z}-HS^{z}-B(t)S^{x},
\label{eq:Hamil_IsingMF}
\end{align}
where $m^{z}(t)\equiv\langle S^{z}(t)\rangle$. 
We solve the Schr\"odinger equation with 
the Hamiltonian~\eqref{eq:Hamil_IsingMF} 
by the fifth order Runge-Kutta method with the Cash-Karp parameters, 
and calculate the dynamics of 
$m^{x}(t)\equiv\langle S^{x}(t)\rangle$ and $m^{z}(t)$. 
The discretized time step is $\Delta t=0.05$. 
The result is also shown in Fig.~\ref{fig:Ising_high_HHG}. 
The curves of $|\omega^{2}m^{x}(\omega)|^{2}$ and 
$|\omega^{2}m^{z}(\omega)|^{2}$ calculated 
by the single spin dynamics agree well with 
those calculated by iTEBD up to the first HHG threshold. 
Note that there is no rescaling of the results 
and the agreement is quantitative. 
The deviations become larger above the first threshold. 
This indicates that correlations between magnons beyond mean-field theory 
are essential for the spontaneous recombination of multiple magnons. 

%%%%%%%%%% Fig : Floquet subbands with high Hz %%%%%%%%%%
\begin{figure}[t]
\includegraphics[width=0.48\textwidth]{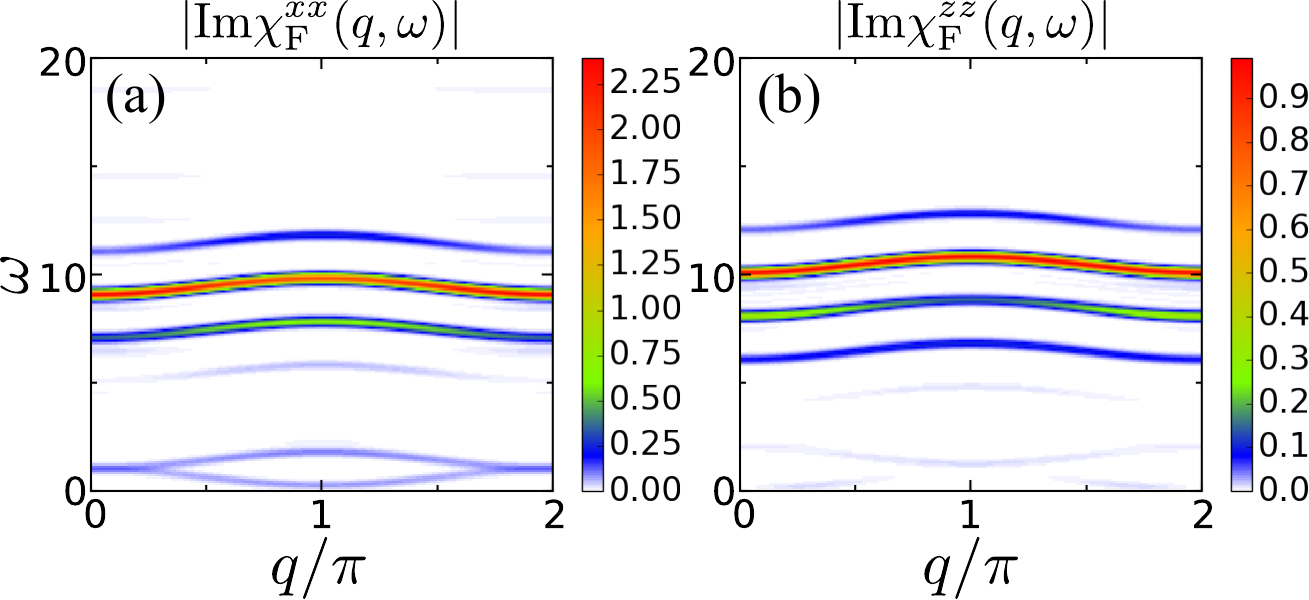}
\caption{The Floquet DSF 
(a) $|\mathrm{Im}\chi_{\mathrm{F}}^{xx}(q,\omega)|$ and 
(b) $|\mathrm{Im}\chi_{\mathrm{F}}^{zz}(q,\omega)|$ 
for the Ising model with $J=2$, $H=6$, and $B=8$. 
}
\label{fig:Ising_high_Floq}
\end{figure}
%%%%%%%%%%%%%%%%%%%%

Another useful perspective on HHG can be obtained 
from the Floquet picture~\cite{Ikeda2018PRA,Murakami2018PRB}. 
The spectrum in the Floquet theory is derived from 
the Floquet DSF $|\chi_{\mathrm{F}}^{\alpha\beta}(q,\omega)|$, 
which is calculated in a similar way as $|\chi^{\alpha\beta}(q,\omega)|$. 
Let us consider the time-dependent Hamiltonian 
$\mathcal{H}(t;\alpha_{0})=\mathcal{H}_{0}
-B\sin(\Omega t+\alpha_{0})S_{\mathrm{tot}}^{x}$ 
and represent the ground state of $\mathcal{H}(0;\alpha_{0})$ by 
$|\Psi(0;\alpha_{0})\rangle$, 
where $\alpha_{0}$ is the phase shift. 
We calculate the Floquet retarded correlation function 
\begin{align}
 \chi_{\mathrm{F}}^{\alpha\beta}(r,t;\alpha_{0})
   =-i\vartheta(t)\langle \Psi(0;\alpha_{0})|
[S_{r}^{\alpha}(t;\alpha_{0}),S_{0}^{\beta}]|\Psi(0;\alpha_{0})\rangle
\end{align}
[cf.~Eq.~\eqref{eq:RetCorrel}], 
where 
$S_{r}^{\alpha}(t;\alpha_{0})=U^{-1}(t;\alpha_{0})S_{0}^{\alpha}U(t;\alpha_{0})$ 
and 
$U(t;\alpha_{0})=\mathcal{T}\int_{0}^{t}dt' e^{-i\mathcal{H}(t';\alpha_{0})t'}$. 
The DSF $\chi_{\mathrm{F}}^{\alpha\beta}(q,\omega;\alpha_{0})$ 
is defined as the Fourier transform of this correlation function, 
and we take the average relative to the phase shift $\alpha_{0}$ 
over a single cycle as 
\begin{align}
 \chi_{\mathrm{F}}^{\alpha\beta}(q,\omega)
   =\langle
     \chi_{\mathrm{F}}^{\alpha\beta}(q,\omega;\alpha_{0})
     \rangle_{\alpha_{0}}. 
\nonumber
\end{align}
Here we take $\alpha_{0}=n\pi/8$ ($n=0,1,\ldots,15$). 
In Fig.~\ref{fig:Ising_high_Floq}, 
we show the Floquet DSF 
$|\mathrm{Im}\chi_{\mathrm{F}}^{xx(zz)}(q,\omega)|$ 
for $B=8$. 
We can see the appearance of Floquet subbands
with an energy splitting of $2\Omega$ rather than $\Omega$. 
The subbands of 
$|\mathrm{Im}\chi_{\mathrm{F}}^{xx}(q,\omega)|$ are located at 
$(\mathrm{one\;magnon\;band})\pm(\mathrm{odd\;integer})\Omega$ 
while those of 
$|\mathrm{Im}\chi_{\mathrm{F}}^{zz}(q,\omega)|$ are located at 
$(\mathrm{one\;magnon\;band})\pm(\mathrm{even \;integer})\Omega$. 
In $|\mathrm{Im}\chi_{\mathrm{F}}^{xx}(q,\omega)|$, 
the Floquet subbands of the negative energy magnon dispersion 
appear around $\omega\simeq1$. 
These Floquet DSFs suggest that we can also interpret the high-harmonic peaks
with energy below the magnon mass in terms of transitions 
between Floquet sidebands of the magnon spectrum. 

In this section, we have focused on the model with strong 
longitudinal field ($H\gg J$) and large gap. 
With decreasing $H$, the gap decreases and the HHG behavior changes. 
We discuss the results of a weak $H$ model ($J=4$, $H=2$) 
in Appendix~\ref{sec:IsingWeakField}. 
The emergence of HHG plateaus with a close relation 
to magnon energies can also be observed there.

\section{HHG in XXZ models}
\label{sec:XXZweakJxy}

%%%%%%%%%% Fig : susceptibility with weak Jxy %%%%%%%%%%
\begin{figure}[t]
\includegraphics[width=0.48\textwidth]{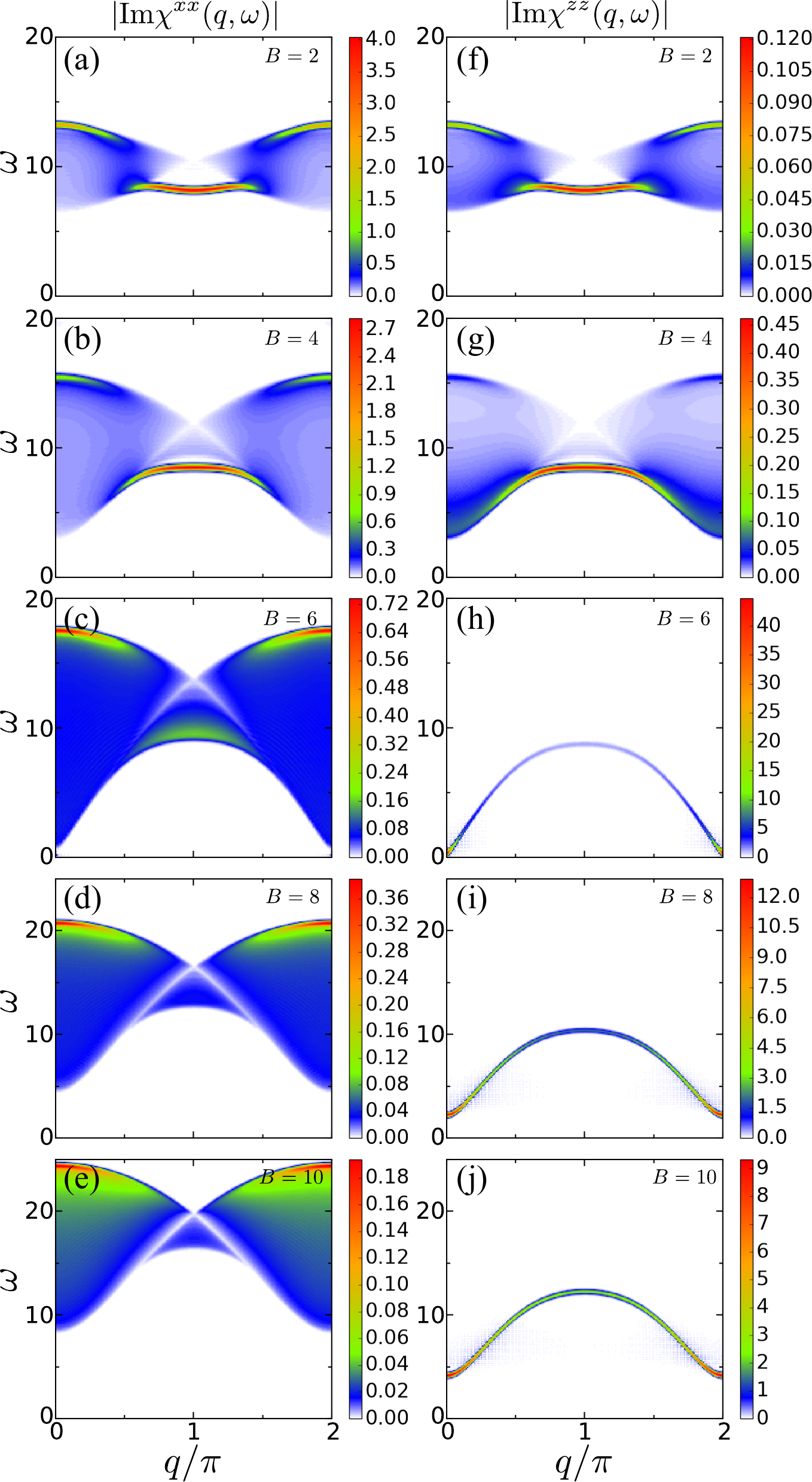}
\caption{DSFs 
(a)-(e) $|\mathrm{Im}\chi^{xx}(q,\omega)|$ and 
(f)-(j) $|\mathrm{Im}\chi^{zz}(q,\omega)|$ 
for the XXZ model with $J_{xy}=2$ and $J_{z}=10$. 
}
\label{fig:XXZ_weakJxy_suscep}
\end{figure}
%%%%%%%%%%%%%%%%%%%%

In this section we consider another fundamental model 
of quantum magnets, the ferromagnetic XXZ model. 
The spin Hamiltonian is 
\begin{align}
 \mathcal{H}_{\mathrm{XXZ}}
   =\sum_{j}[J_{xy}(S_{j}^{x}S_{j+1}^{x}+S_{j}^{y}S_{j+1}^{y})
     -J_{z}S_{j}^{z}S_{j+1}^{z}],
\label{eq:XXZModel}
\end{align}
where $J_{z}>J_{xy}>0$. 
The difference from the Ising model is the term
$S_{j}^{x}S_{j+1}^{x}+S_{j}^{y}S_{j+1}^{y}
=\frac{1}{2}(S_{j}^{+}S_{j+1}^{-}+S_{j}^{-}S_{j+1}^{+})$, 
which acts as a kinetic term for the magnons. 
Note that the spins are completely frozen 
in the Ising model without laser. 
The ground state of Eq.~\eqref{eq:XXZModel} is a
ferromagnetic state for $J_{z}>J_{xy}>0$ 
while it is a gapless Luttinger liquid 
for $|J_{z}|<|J_{xy}|$~\cite{Giamarchi2004Book}. 
The low energy excitations of Eq.~\eqref{eq:XXZModel} are magnons 
with dispersion $E(q)=J_{xy}\cos(q)+J_{z}$. 
Since the Hamiltonian Eq.~\eqref{eq:XXZModel} does not include 
the longitudinal static field $HS_{\mathrm{tot}}^{z}$, 
the system has a $\mathbf{Z}_{2}$ symmetry, 
and thus a quantum phase transition can be induced 
by applying a transverse field. 

Here we consider the case where $J_{xy}$ is weak, $J_{xy}\ll J_{z}$, 
which is relevant for the modeling of quasi-one-dimensional 
magnetic insulators such as $\mathrm{CoNb_{2}O_{6}}$~\cite{Coldea2010Science}. 
The parameters are set to $J_{xy}=2$, $J_{z}=10$, and $\Omega=1$. 
We take both $J_{xy}$ and $J_{z}$ to be larger than $\Omega$ 
so that the HHG plateau contains several harmonics. 
For the analysis of the model with strong $J_{xy}$ 
($J_{xy}\lesssim J_{z}$), 
see Appendix~\ref{sec:XXZStrongJxy}. 
In Fig.~\ref{fig:XXZ_weakJxy_suscep}, 
we show the DSF in the ground state of the XXZ model with a transverse field $B$, 
which corresponds to the snapshot Hamiltonian 
of the system under laser irradiation, 
\begin{align}
 \mathcal{H}
   =\mathcal{H}_{\mathrm{XXZ}}
     -BS_{\mathrm{tot}}^{x}.
\label{eq:XXZ_Transverse}
\end{align}
In contrast to the case of the Ising model, 
the low-energy excitation spectrum is continuous 
due to the existence of the kinetic term. 
The lower bound of the dispersion at $q=0$ decreases with increasing $B$. 
The gap closes and a phase transition happens at $B_{\mathrm{c}}\simeq 6$. 
Before the transition ($B<B_{\mathrm{c}}$), 
$\chi^{xx}$ shows a stronger intensity than $\chi^{zz}$, 
because the spins are primarily aligned 
in the $z$ direction in the ground state. 
When $B$ is small enough, 
the DSF has a strong intensity near the one magnon dispersion for $B=0$ 
(i.e., $E(q)=J_{xy}\cos(q)+J_{z}$), and in particular 
the strongest intensity is found at $q=\pi$. 
On the other hand, after the transition ($B>B_\mathrm{c}$), 
the intensity of $\chi^{zz}$ becomes much stronger than $\chi^{xx}$, 
because the spins are mainly aligned 
in the $x$ direction in the ground state, 
and the strongest intensity is observed at $q=0$. 
The dispersion captured by $\chi^{zz}$ is sharp, 
and it can be interpreted as a single-magnon band in terms of 
the spin wave theory (see Appendix~\ref{sec:SpinWave}). 
We also note that the upper bound of the continuous dispersion at $q=0$ 
captured by $\chi^{xx}$ corresponds to a two-magnon state 
since its energy is twice the excitation energy at $q=\pi$ 
captured by $\chi^{zz}$ for $B>B_{\mathrm{c}}$ . 

%%%%%%%%%% Fig : Evolution of mag with weak Jxy %%%%%%%%%%
\begin{figure}[t]
\includegraphics[width=0.48\textwidth]{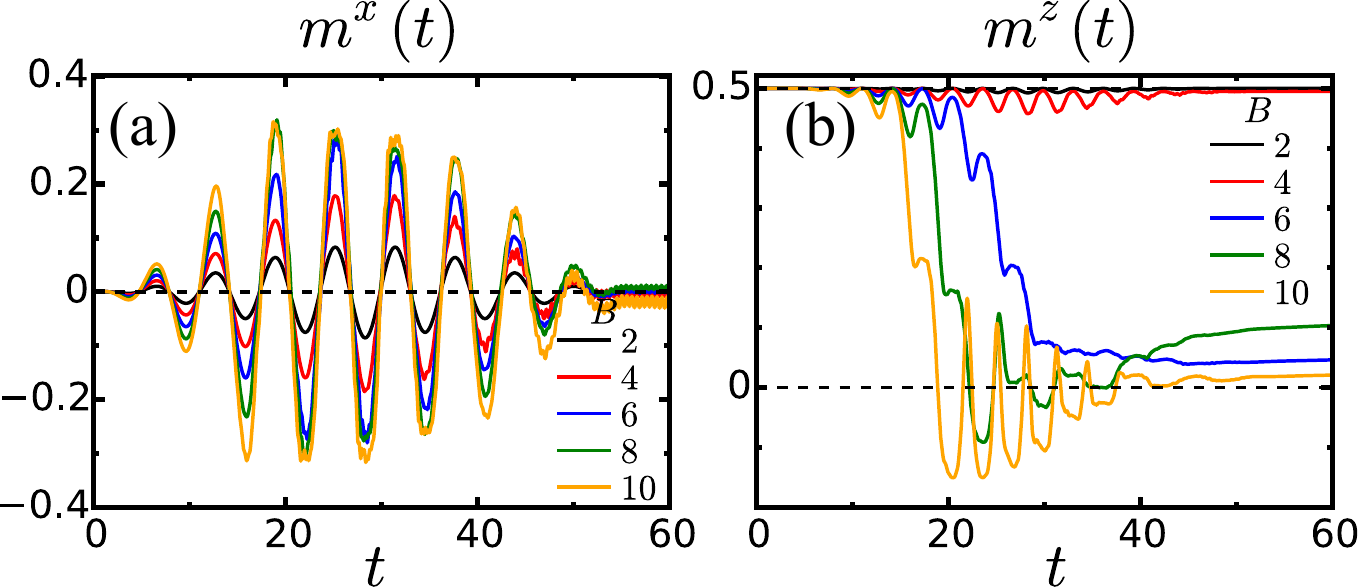}
\caption{Time evolution of (a) $m^{x}$ and (b) $m^{z}$ 
for the XXZ model with $J_{xy}=2$ and $J_{z}=10$. 
}
\label{fig:XXZ_weakJxy_mag}
\end{figure}
%%%%%%%%%%%%%%%%%%%%

The time evolution of $m^{x}$ and $m^{z}$ 
calculated by iTEBD is shown in Fig.~\ref{fig:XXZ_weakJxy_mag}. 
The time evolution of $m^{x}$ essentially tracks  
the laser magnetic field Eq.~\eqref{eq:laser_mag_field} for small $B$, 
but the shape changes especially near the peaks of the intensity 
as $B$ is increased. 
Higher frequency components than $\Omega$ 
appear near the peaks, and these contribute to the HHG 
(see the sub-cycle analysis below). 
The time evolution of $m^{z}$ drastically changes its behavior 
depending on whether $B$ is smaller or larger than $B_{\mathrm{c}}$. 
For $B<B_{\mathrm{c}}$, the magnitude of $m^{z}$ decreases 
when the laser intensity is strong, 
otherwise $m^{z}\simeq 1/2$, 
which demonstrates that the state follows the ground state 
of the snapshot Hamiltonian, i.e., the time evolution is almost adiabatic. 
However, for $B>B_{\mathrm{c}}$, $m^{z}$ suddenly decreases from $1/2$, 
which shows that the system makes transitions 
to excited states of the snapshot Hamiltonian. 

%%%%%%%%%% Fig : HHG with weak Jxy %%%%%%%%%%
\begin{figure}[t]
\includegraphics[width=0.48\textwidth]{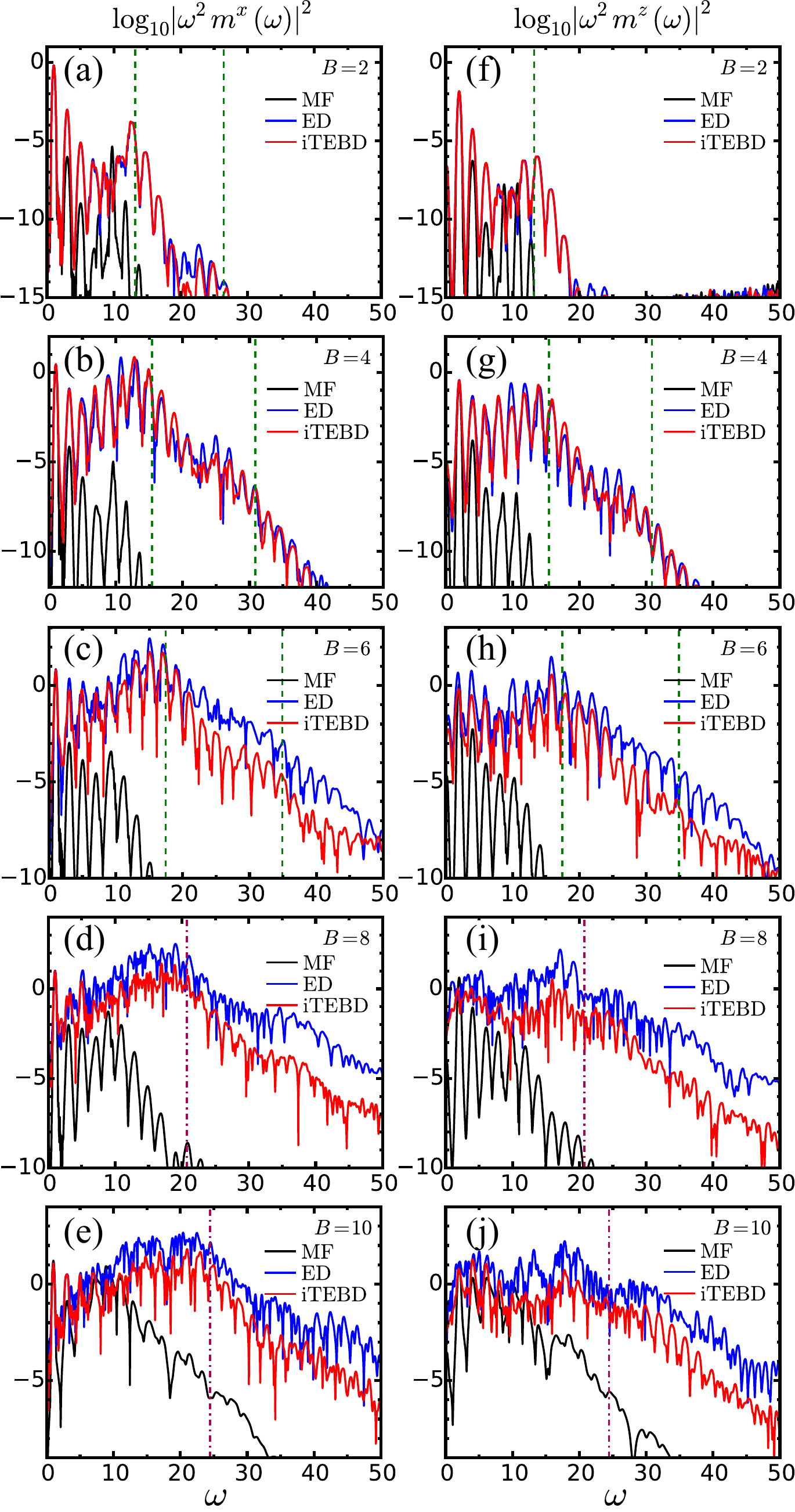}
\caption{HHG from (a)-(e) $m^{x}$ and (f)-(j) $m^{z}$ 
for the XXZ model with $J_{xy}=2$ and $J_{z}=10$. 
Dashed lines correspond to $\Delta_{q=0}$ (times integer) 
and dashed-dotted lines correspond to $2\tilde{\Delta}_{q=\pi}$. 
}
\label{fig:XXZ_weakJxy_HHG}
\end{figure}
%%%%%%%%%%%%%%%%%%%%

The HHG spectra $|\omega^{2}m^{x}(\omega)|^{2}$ and 
$|\omega^{2}m^{z}(\omega)|^{2}$ are shown 
in Fig.~\ref{fig:XXZ_weakJxy_HHG}. 
Here the same Blackman window is used as in the Ising case. 
The HHG structure is clear for the weak field $B$ 
while it is noisier after the transition. 
Since the system satisfies the symmetry Eq.~\eqref{eq:symmetry}, 
the magnitudes of $|\omega^{2}m^{x}(\omega)|^{2}$ and 
$|\omega^{2}m^{z}(\omega)|^{2}$ drop 
at $\omega=2n\Omega$ and $\omega=(2n+1)\Omega$, respectively, 
except that $m^{x}(\omega)$ has a peak 
and $m^{z}(\omega)$ has a dip around $\omega=12$ for $B=2$. 
This energy corresponds to the upper bound of the single-magnon band $J_{xy}+J_{z}$, 
and we can explain the peaks at $\omega=J_{xy}+J_{z}$ 
for $m^{x}(\omega)$ and at $\omega=J_{xy}+J_{z}\pm\Omega$ for $m^{z}(\omega)$ 
in the small $B$ region in terms of the time-dependent perturbation theory 
as shown in Appendix~\ref{sec:TdepPerturb}. 

%%%%%%%%%% Fig : subcycle analysis with weak Jxy %%%%%%%%%%
\begin{figure}[t]
\includegraphics[width=0.48\textwidth]{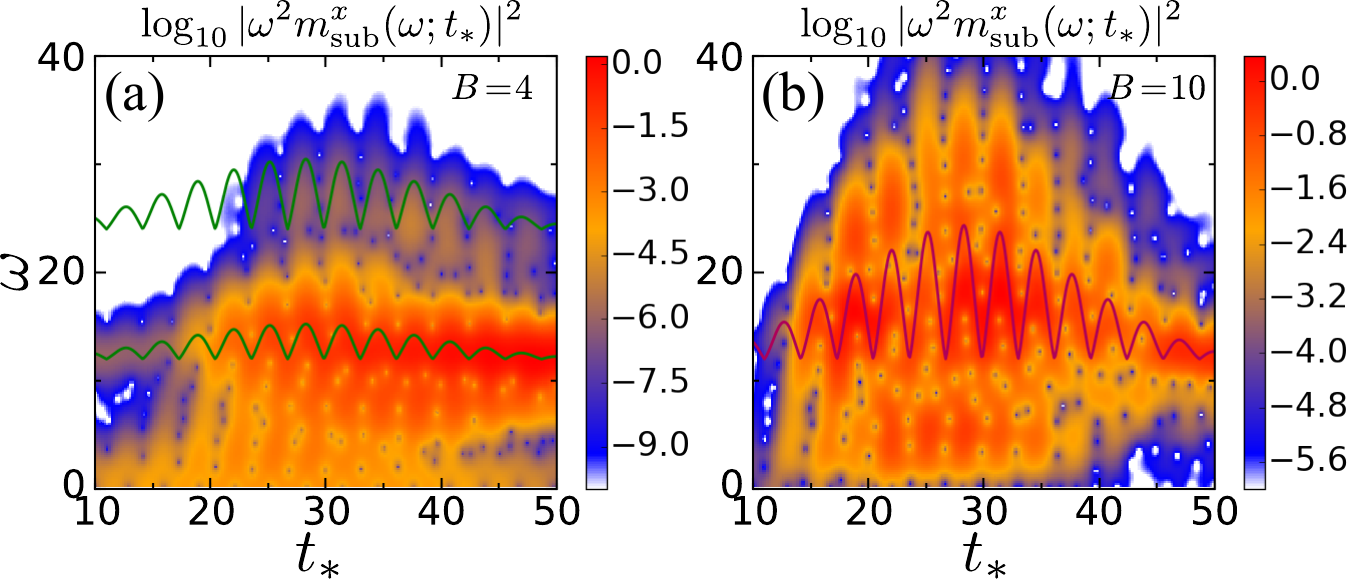}
\caption{Colormap of the subcycle radiation spectrum 
$\log_{10}|\omega^{2}m_{\mathrm{sub}}^{x}(\omega;t_{*})|^{2}$
for the XXZ model 
with $J_{xy}=2$ and $J_{z}=10$ 
under the laser field (a) $B=4$ and (b) $B=10$. 
The solid lines show the single-magnon and two-magnon modes 
of the snapshot Hamiltonian at time $t_*$ 
in (a) and the two-magnon mode in (b). 
}
\label{fig:XXZ_weakJxy_subcyc}
\end{figure}
%%%%%%%%%%%%%%%%%%%%

As we increase $B$ and leave the perturbative regime, 
plateau structures develop in the low-energy region. 
Again we can connect these cut-off energies (threshold energies)
with the spin excitation structure. 
As depicted in Fig.~\ref{fig:XXZ_weakJxy_HHG}, 
for $B<B_{\mathrm{c}}$, the threshold of the HHG plateau 
corresponds to $\Delta_{q=0}$, which is the upper bound of the  
dispersion obtained from $\chi^{xx}$ at $q=0$. 
For $B>B_{\mathrm{c}}$, 
the threshold of the first HHG plateau 
is determined by $2\tilde{\Delta}_{q=\pi}$, 
where $\tilde{\Delta}_{q=\pi}$ is the excitation gap corresponding 
to $\chi^{zz}$ at $q=\pi$. 
This energy scale is not very apparent in $|\omega^{2}m^{z}(\omega)|^{2}$ 
but we can see that $|\omega^{2}m^{x}(\omega)|^{2}$ is larger than 
$|\omega^{2}m^{z}(\omega)|^{2}$ by several orders  
near the threshold energy 
[dashed-dotted lines in Figs.~\ref{fig:XXZ_weakJxy_HHG}(d), 
\ref{fig:XXZ_weakJxy_HHG}(e), \ref{fig:XXZ_weakJxy_HHG}(i), and 
\ref{fig:XXZ_weakJxy_HHG}(j)]
and dominates the HHG. 
Note that for $B>B_{\mathrm{c}}$, the spins are mostly aligned 
in the $x$ direction in the ground state and 
$S^{z}$ works as a spin-flip (magnon generation) operator.
Even though $\Delta_{q=0}=2\tilde{\Delta}_{q=\pi}$ and 
this mode can also be excited by the $S^{x}$ operator,
the intensity of $\chi^{zz}$ is much larger than that of $\chi^{xx}$ 
as seen in Fig.~\ref{fig:XXZ_weakJxy_suscep}. 
Thus it is more natural to regard it as a two-magnon process. 

In the same way as we have done for the Ising model, 
we can obtain further insight into the origin of the HHG 
by performing a subcycle analysis for the XXZ model. 
In Fig.~\ref{fig:XXZ_weakJxy_subcyc}, 
we show the subcycle radiation spectrum Eq.~\eqref{eq:SubcycFourier} 
for $B=4$ and $B=10$ (below and above the critical field, respectively). 
In the case of $B=4$ [Fig.~\ref{fig:XXZ_weakJxy_subcyc}(a)], 
the strong intensity in the HHG signal 
follows the single-magnon and two-magnon excitation energy 
($\Delta_{q=0}$ and $2\Delta_{q=0}$) 
of the snapshot Hamiltonian at each time, 
which suggests that again the threshold can be associated with 
the annihilation of multiple magnons at $q=0$ for $B<B_{\mathrm{c}}$. 
In the case of $B=10$ [Fig.~\ref{fig:XXZ_weakJxy_subcyc}(b)], 
the strong intensity in the HHG signal 
follows the two-magnon excitation energy 
($2\tilde{\Delta}_{q=\pi}$). 
There is also some additional intensity in the energy range 
$\omega=25$ -- 40 in Fig.~\ref{fig:XXZ_weakJxy_subcyc}(b), 
which may correspond to higher order excitations 
such as four magnon processes. 

%%%%%%%%%% Fig : Overlap calculation by ED with weak Jxy %%%%%%%%%%
\begin{figure}[t]
\includegraphics[width=0.48\textwidth]{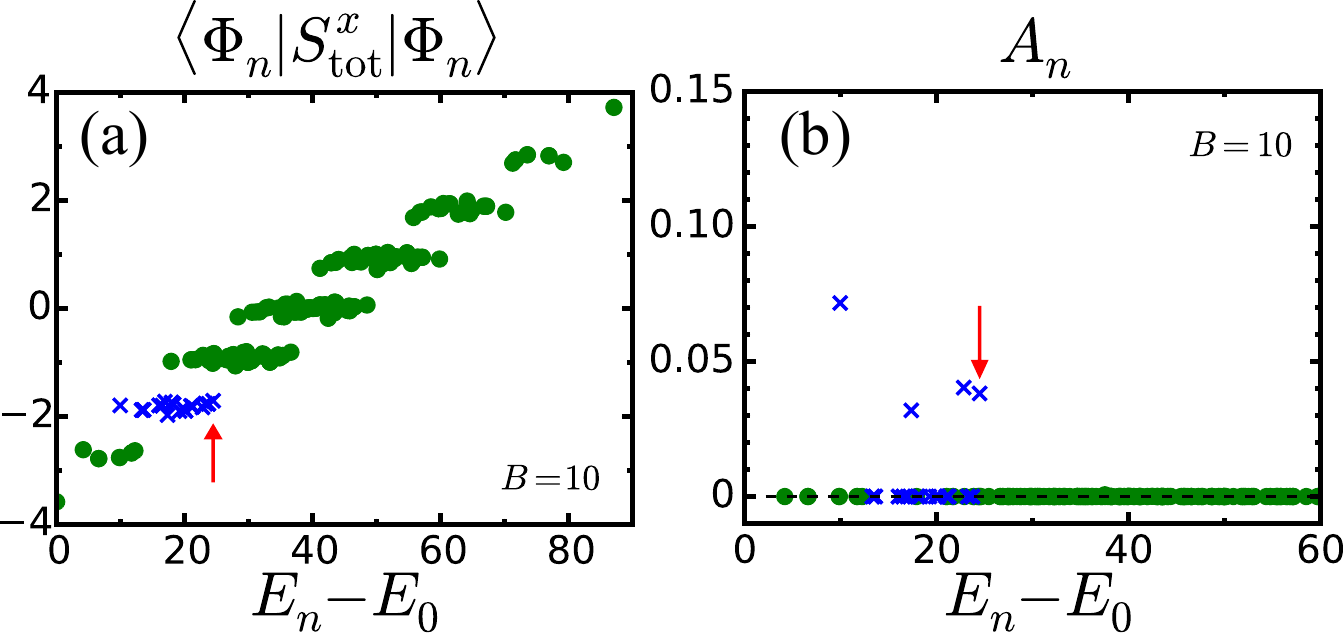}
\caption{(a) $\langle\Phi_{n}|S_{\mathrm{tot}}^{x}|\Phi_{n}\rangle$ 
and (b) $A_{n}$ [Eq.~\eqref{eq:SxWeightWeakXXZ}] 
calculated by ED for the model~\eqref{eq:XXZ_Transverse} 
with $J_{xy}=2$, $J_{z}=10$, and $B=10$. 
The arrows show the energy $E_{n}-E_{0}=24.5$. 
Cross marks are used for the states in the 
$\langle\Phi_{n}|S_{\mathrm{tot}}^{x}|\Phi_{n}\rangle\simeq -2$ 
sector to demonstrate that the contribution to the HHG signal comes 
mainly from this sector. 
}
\label{fig:XXZ_weakJxy_ED}
\end{figure}
%%%%%%%%%%%%%%%%%%%%

To confirm that the threshold of the HHG plateaus 
corresponds to the magnetic excitation structure,  
especially magnon modes, 
we perform an ED calculation for a system of $N=8$ sites. 
The system size is small, 
but the ED calculations reproduce quantitatively
the behavior of the HHG spectra for small $B$ 
as can be seen in Fig.~\ref{fig:XXZ_weakJxy_HHG}. 
Although there is a quantitative deviation from the iTEBD results 
in the case of strong $B$, 
the HHG signals show a qualitative agreement. 
In particular, the threshold energy of the first plateau 
is the same for ED and iTEBD. 
We denote the eigenstates of the snapshot Hamiltonian 
at the time when the laser intensity takes the maximum 
($t_{\mathrm{peak}}=\pi N_{\mathrm{cyc}}/\Omega$) 
by $|\Phi_{n}\rangle$ 
and their eigenenergies by $E_{n}$. 
In Fig.~\ref{fig:XXZ_weakJxy_ED}(a), 
we show 
$\langle\Phi_{n}|S_{\mathrm{tot}}^{x}|\Phi_{n}\rangle$ 
calculated by ED for large $B$. 
In the present model, 
even though $S_{\mathrm{tot}}^{x}$ is not a conserved quantity, 
the spins basically align in the $S^{x}$ direction in the ground state
for large $B$ and the expectation values 
$\langle\Phi_{n}|S_{\mathrm{tot}}^{x}|\Phi_{n}\rangle$ 
are almost discretized and distributed around integer values. 
The expectation values near $-2$ are highlighted with cross markers in Fig.~\ref{fig:XXZ_weakJxy_ED}.
From Fig.~\ref{fig:XXZ_weakJxy_ED}(a), 
the energy threshold of the first HHG plateau 
corresponds to the upper bound of the 
$\langle\Phi_{n}|S_{\mathrm{tot}}^{x}|\Phi_{n}\rangle\simeq -2$ 
sector ($E_{n}-E_{0}=24.5$). 
Since the ground state is in the 
$\langle\Phi_{n}|S_{\mathrm{tot}}^{x}|\Phi_{n}\rangle\simeq -4$ 
sector, two spins are flipped, i.e., 
two magnons are generated. 
In Fig.~\ref{fig:XXZ_weakJxy_ED}(b), 
we plot the quantity 
\begin{align}
 A_{n}=|\alpha_{n}^{*}\alpha_{0}
   \langle \Phi_{n}|S_{\mathrm{tot}}^{x} |\Phi_{0}\rangle|,
\label{eq:SxWeightWeakXXZ}
\end{align}
where $\alpha_{n}\equiv\langle\Phi_{n}|\Psi(t_{\mathrm{peak}})\rangle$ 
represents the overlap between the state at $t=t_{\mathrm{peak}}$ 
and the $n$-th excited state $|\Phi_{n}\rangle$ of the snapshot Hamiltonian 
($|\Phi_{0}\rangle$ is the ground state). 
This quantity is directly related to $m^{x}$ 
through Eq.~\eqref{eq:MagExpand}. 
We see that there is a strong intensity 
at the energy $E_{n}-E_{0}=24.5$, 
which agrees with the threshold energy 
in Fig.~\ref{fig:XXZ_weakJxy_HHG}(e).  
Hence we can conclude that 
the threshold of the first HHG plateau is dictated by 
the two-magnon mode $2\tilde{\Delta}_{q=\pi}$. 
In addition, Fig.~\ref{fig:XXZ_weakJxy_ED}(b) suggests that 
the contribution to the HHG signal mainly comes from the two-magnon sector ($\langle\Phi_{n}|S_{\mathrm{tot}}^{x}|\Phi_{n}\rangle\simeq -2$).

Further insight into the HHG signal with large $B$ can be obtained by rewriting the Hamiltonian. 
Since the spin alignment axis is $S^{x}$ 
in the case of very strong laser field $B$, 
the magnon creation and annihilation operators correspond to  
$\tilde{S}^{\pm}=S^{y}\pm iS^{z}$. 
Using these operators, the Hamiltonian 
(Eq.~\eqref{eq:Hamil_tdep} with Eq.~\eqref{eq:XXZModel}) becomes
\begin{align}
 \mathcal{H}
   &=J_{xy}\sum_{j}S_{j}^{x}S_{j+1}^{x}+
     \frac{J_{xy}-J_{z}}{4}\sum_{j}(\tilde{S}_{j}^{+}\tilde{S}_{j+1}^{-}
     +\tilde{S}_{j}^{-}\tilde{S}_{j+1}^{+})\nonumber\\
   &+\frac{J_{xy}+J_{z}}{4}\sum_{j}(\tilde{S}_{j}^{+}\tilde{S}_{j+1}^{+}
     +\tilde{S}_{j}^{-}\tilde{S}_{j+1}^{-})
     -B(t)S_{\mathrm{tot}}^{x}.
\label{eq:HamilSxBasis}
\end{align}
The $\tilde{S}_{j}^{+}\tilde{S}_{j+1}^{+}+\tilde{S}_{j}^{-}\tilde{S}_{j+1}^{-}$ 
term creates and annihilates magnons (at large $B$) in pairs. 
In the Hamiltonian Eq.~\eqref{eq:HamilSxBasis}, 
the Hilbert space is separated into the sectors with
$S_{\mathrm{tot}}^{x}=(\mathrm{even\;integer})$ and 
$S_{\mathrm{tot}}^{x}=(\mathrm{odd \;integer})$ 
since the parity of the magnon number is a conserved quantity. 
Hence the state remains in the same sector during the time evolution. 
The initial state is the ferromagnetic state, 
which corresponds to a Schr\"odinger cat state in the $S^{x}$ basis,  
\begin{align}
 \otimes_{j}|\!\uparrow\rangle_{j}
   =\otimes_{j}\frac{|S^{x}=1/2\rangle_{j}+|S^{x}=-1/2\rangle_{j}}{\sqrt{2}}.
\nonumber
\end{align}
Thus, this state has weight in both 
$S_{\mathrm{tot}}^{x}=(\mathrm{even\;integer})$ and 
$S_{\mathrm{tot}}^{x}=(\mathrm{odd \;integer})$ sectors. 
$M_{x}$ has nonzero expectation values for states within the same sector,  
$\langle\Psi_{\mathrm{even}}|S_{\mathrm{tot}}^{x}|\Psi_{\mathrm{even}}\rangle
+\langle\Psi_{\mathrm{ odd}}|S_{\mathrm{tot}}^{x}|\Psi_{\mathrm{ odd}}\rangle$, 
while $M_{z}$ has nonzero expectation values 
for the states between different sectors, 
$\langle\Psi_{\mathrm{ odd}}|S_{\mathrm{tot}}^{z}|\Psi_{\mathrm{even}}\rangle
+\langle\Psi_{\mathrm{even}}|S_{\mathrm{tot}}^{z}|\Psi_{\mathrm{ odd}}\rangle$. 
This expression explains why the two-magnon mode is evident in $m^{x}(\omega)$ 
[Figs.~\ref{fig:XXZ_weakJxy_HHG}(d) and \ref{fig:XXZ_weakJxy_HHG}(e)]
while it is not apparent in $m^{z}(\omega)$ 
[Figs.~\ref{fig:XXZ_weakJxy_HHG}(i) and \ref{fig:XXZ_weakJxy_HHG}(j)]. 

We also analyze the dynamics of this system 
by means of the tdMF theory. 
The mean field Hamiltonian is 
\begin{align}
 \tilde{\mathcal{H}}_{\mathrm{XXZ}}(t)
   =&2J_{xy}(m^{x}(t)S^{x}+m^{y}(t)S^{y})
     -2J_{z}m^{z}(t)S^{z}\nonumber\\
   &-B(t)S^{x}.
\label{eq:Hamil_XXZMF}
\end{align}
The radiation power spectrum calculated 
by the Hamiltonian Eq.~\eqref{eq:Hamil_XXZMF} 
is shown in Fig.~\ref{fig:XXZ_weakJxy_HHG}. 
The tdMF result shows a peak or plateau structure in the HHG spectrum, 
but quantitatively it deviates strongly 
from the iTEBD and ED results,  
in contrast to the case of the Ising model. 
This is due to the strong quantum fluctuations induced by 
the $S_{j}^{+}S_{j+1}^{-}+S_{j}^{-}S_{j+1}^{+}$ term, 
and implies that the tdMF theory does not provide a good description 
of the XXZ model.

\section{Summary and discussions}
\label{sec:Summary}

In this paper, we studied HHG in quantum spin systems 
driven by a laser magnetic field. 
When the laser is applied to magnetic insulators, 
it drives the magnetic dipole which generates electromagnetic radiation with power 
proportional to $|\omega^{2}M(\omega)|^{2}$. 
We considered two specific but fundamental quantum spin chain models, 
the Ising model with static longitudinal field and the XXZ model. 
In both cases, when the magnetic field is strong enough, the spin HHG shows a (multiple-)plateau structure,
which is associated with the annihilation of (multiple) magnons.

To be more specific, in the Ising model case, 
the excitation gap does not close in the presence of a transverse field 
since the $\mathbf{Z}_{2}$ symmetry is explicitly broken. 
When the laser amplitude is weak enough, 
the time-dependent perturbation theory is valid, 
which explains the appearance of a peak around the frequency $J+H$. 
With increasing laser amplitude, 
the shape of the HHG spectrum changes from a peak structure 
to a plateau structure.
The subcycle analysis suggests that the HHG originates from the annihilation of magnons. 
The cutoff energies, above which the radiation intensity drops, 
correspond to integer multiples of the single-magnon excitation energy at $q=0$.
Since the magnetic field is stronger than the interaction, 
the tdMF theory provides a quantitative description. 

In the XXZ model without longitudinal field, 
the system has a $\mathbf{Z}_{2}$ symmetry and 
a phase transition happens at a critical value of the transverse field. 
The structure of the HHG spectrum changes 
depending on whether the peak amplitude of the laser magnetic field 
is below or above the critical field. 
Similarly to the Ising case, when the laser amplitude is small, 
the time-dependent perturbation theory is valid  
and explains the appearance of a peak around the frequency $J_{xy}+J_{z}$. 
As the laser amplitude increases, 
the peak structure transforms into a plateau structure. 
The cutoff energy of this plateau corresponds 
to the single-magnon mass at $q=0$ below the critical field. 
When the laser amplitude is larger than the critical field, 
the threshold is determined by the two-magnon excitation at $q=\pi$.
The subcycle analysis and the ED analysis suggest that also in the XXZ model case, 
the annihilation of magnons leads to the HHG signal.
The tdMF approach is not effective in this model 
due to the quantum fluctuation caused by the $J_{xy}$ term.

Now let us discuss the similarities and differences 
between the HHG from spin systems and 
that from insulating electron systems 
such as semiconductors and Mott insulators~\cite{Vampa2014PRL, Vampa2015PRB,Murakami2018PRL}. 
In the latter case, a periodic electric field creates charge carriers 
(electrons and holes in semiconductors, 
and doublons and holons in Mott insulators) 
and these carriers move around in response to the applied electric field. 
The HHG originates from the dynamics of these charge carriers, 
which can be separated into the interband and intraband current.
The interband current corresponds to the creation and recombination of charge carriers,
while the intraband current represents the contribution 
from hopping processes which do not change the number of charge carriers, 
i.e. where the carriers remain in the same conduction/valence or Hubbard band. 
In contrast, in the spin systems, the magnetic field can excite 
magnetic excitations (magnons) but there is no preferable direction to move 
since the homogeneous magnetic field, unlike the electric field, does not 
produce a spatially dependent potential. 
Hence, the HHG signal originating from the dynamics of the magnetization is analogous to the interband current, 
while there is no counterpart to the intraband current. 
Our finding that the spin HHG is associated with the annihilation of magnons
is reminiscent of the electron HHG which is dominated by the recombination of charge carriers \cite{Vampa2014PRL,Murakami2018PRL}.

Experimentally, the HHG from spins excited by time-periodic magnetic fields 
can be realized by choosing large gap insulating materials, 
and by taking advantage of metamaterials 
to selectively enhance the magnetic field~\cite{Mukai2014APL}. 
For example, $\mathrm{CoNb_{2}O_{6}}$~\cite{Coldea2010Science} 
can be represented as a ferromagnetic XXZ chain with $J_{xy}\ll J_{z}$, 
and therefore the discussion in Sec.~\ref{sec:XXZweakJxy} is relevant for this material,  
while examples of Ising magnets (Sec.~\ref{sec:Ising}) 
such as $\mathrm{Dy(C_{2}H_{5}SO_{4})_{3}\cdot 9H_{2}O}$, 
$\mathrm{LiTbF_{4}}$ and $\mathrm{LiHoF_{4}}$ are discussed in~Ref.~\cite{Wolf2000BJP}. 
For $\mathrm{CoNb_{2}O_{6}}$, since the value of $J_{z}(=10)$ is 
1.94 meV~\cite{Coldea2010Science}, 
the energy unit is 
$0.194\;\mathrm{meV}=1.67\;\mathrm{T}=2\pi\times 0.0469\;\mathrm{THz}$
by noting that $g\mu_{\mathrm{B}}B$ and $\hbar\Omega$ 
have the dimension of energy, 
where $g\simeq 2$ is the Land\'e $g$ factor for electron spins, 
$\mu_{\mathrm{B}}=0.0579\;\mathrm{meV/T}$ is the Bohr magneton, 
and $\hbar=6.58\times 10^{-13}\;\mathrm{meV}\cdot\mathrm{s}$. 
$\Omega=1$ and $B=2$ thus correspond to 
$\Omega=2\pi\times 0.0469\;\mathrm{THz}$ and 
$B=3.34\;\mathrm{T}$, respectively. 

Our results demonstrate the possibility of generating high-harmonic signals in spin systems,
which may be utilized for new laser sources in the THz regime or to obtain information about the magnetic excitations of these spin systems under strong fields.
In the present work, we focused on one-dimensional ferromagnets 
but the fact that the tdMF results show a similar HHG spectrum strongly suggests 
that the HHG signal can also be produced in higher dimensional magnets. 
Radiation from the magnetic dipole 
should be possible also in ferrimagnets and antiferromagnets. 
Although the total magnetization is zero in antiferromagnets, 
the laser magnetic field produces a net magnetization 
and a HHG signal can be expected. 
Since there exist various kinds of quantum spin systems, 
studying these other types of magnetic insulators is 
an interesting direction for future research.

\acknowledgements

We acknowledge fruitful discussions with Takashi Oka, Tran Trung Luu and Gabriel Aeppli. 
ST is supported by the Swiss National Science Foundation 
under Division II and ImPact project (No. 2015-PM12-05-01) 
from the Japan Science and Technology Agency. 
YM and PW are supported by ERC Consolidator Grant No. 724103.

\appendix

\section{Time-dependent perturbation theory}
\label{sec:TdepPerturb}

In this appendix we analyze the spin system in the presence of a laser field  
using the time-dependent perturbation theory. 
The Hamiltonian is 
\begin{align}
 \mathcal{H}(t)=\mathcal{H}_{\mathrm{spin}}+V(t),
\nonumber
\end{align}
where $V(t)$ represents the laser-matter interaction 
which is assumed here for simplicity to have the form 
\begin{align}
 V(t)=
\left\{
\begin{array}{ll}
 0 & (t<0) \\
 -BS_{\mathrm{tot}}^{x}\sin(\Omega t) & (t\geq 0)
\end{array}
\right. .
\label{eq:SimpleLaser}
\end{align}
We switch to the interaction picture. 
The state and operator are represented as 
$|\Psi(t)\rangle_{\mathrm{I}}
=e^{+i\mathcal{H}_{\mathrm{spin}}t}|\Psi(t)\rangle$ 
and 
$\mathcal{O}_{\mathrm{I}}=e^{i\mathcal{H}_{\mathrm{spin}}t}
\mathcal{O}e^{-i\mathcal{H}_{\mathrm{spin}}t}$, respectively, 
where $|\Psi(t)\rangle$ and $\mathcal{O}$ are 
the state and operator in the Schr\"odinger picture. 
The equation of motion becomes 
\begin{align}
 i\frac{d}{dt}|\Psi(t)\rangle_{\mathrm{I}}
   &=V_{\mathrm{I}}(t)|\Psi(t)\rangle_{\mathrm{I}},
\label{eq:EOM1_InterPic}\\
 \frac{d}{dt}\mathcal{O}_{\mathrm{I}}
   &=i[\mathcal{H}_{\mathrm{spin}},\mathcal{O}_{\mathrm{I}}],
\nonumber
\end{align}
where $V_{\mathrm{I}}(t)=e^{i\mathcal{H}_{\mathrm{spin}}t}
V(t)e^{-i\mathcal{H}_{\mathrm{spin}}t}$. 
From Eq.~\eqref{eq:EOM1_InterPic}, 
we derive 
\begin{widetext}
\begin{align}
 |\Psi(t)\rangle_{\mathrm{I}}
   =\Big(1+\sum_{n=1}^{\infty}(-i)^{n}
     \int_{0}^{t}dt_{1}\int_{0}^{t_{1}}dt_{2}\cdots
     \int_{0}^{t_{n-1}}dt_{n}
     \times V_{\mathrm{I}}(t_{1})\cdots
     V_{\mathrm{I}}(t_{n})\Big)
     |\Psi(0)\rangle_{\mathrm{I}}.
\label{eq:StateTdepPerturbExpand}
\end{align}

We denote the eigenenergy and eigenstate of 
$\mathcal{H}_{\mathrm{spin}}$ by 
$E_{n}$ and $|\varphi_{n}\rangle$, respectively. 
Let us expand $|\Psi(t)\rangle_{\mathrm{I}}$ 
in the basis of $|\varphi_{n}\rangle$, 
\begin{align}
 |\Psi(t)\rangle_{\mathrm{I}}
   =\sum_{n}c_{n}(t)|\varphi_{n}\rangle.
\label{eq:StateExpand}
\end{align}
We substitute~\eqref{eq:StateExpand} 
into~\eqref{eq:StateTdepPerturbExpand} 
and take the inner product with $\langle\varphi_{n}|$, 
to obtain 
\begin{align}
 c_{n}(t)=c_{n}(0)-i\sum_{m}
   \int_{0}^{t}dt_{1}V_{nm}(t_{1})c_{m}(0)
     -\sum_{m,l}\int_{0}^{t}dt_{1}\int_{0}^{t_{1}}dt_{2}
     V_{nl}(t_{1})V_{lm}(t_{2})c_{m}(0)+\cdots,
\label{eq:CoeffPerturb}
\end{align}
where 
$V_{nm}(t)=\langle \varphi_{n}|V_{\mathrm{I}}(t)|\varphi_{m}\rangle
=e^{-i(E_{m}-E_{n})t}\langle \varphi_{n}|V(t)|\varphi_{m}\rangle.$
In the present case, 
\begin{align}
 V_{nm}(t)=-Be^{-i(E_{m}-E_{n})t}\sin(\Omega t)
   \langle \varphi_{n}|S_{\mathrm{tot}}^{x}|\varphi_{m}\rangle
\nonumber
\end{align}
for $t\geq 0$. 
At $t=0$, the system is in the ground state 
$c_{0}(0)=1$ and $c_{n}(0)=0$ ($n\geq 1$), 
thus Eq.~\eqref{eq:CoeffPerturb} becomes 
\begin{align}
 c_{n}(t)=c_{n}(0)-i\int_{0}^{t}dt_{1}V_{n0}(t_{1})
     -\sum_{l}\int_{0}^{t}dt_{1}\int_{0}^{t_{1}}dt_{2}
     V_{nl}(t_{1})V_{l0}(t_{2})+\cdots.
\nonumber
\end{align}
Physical observables are calculated as 
\begin{align}
 \langle\mathcal{O}\rangle
   &={}_{\mathrm{I}}\langle\Psi(t)|\mathcal{O}_{\mathrm{I}}
   |\Psi(t)\rangle_{\mathrm{I}}\nonumber\\
   &=\sum_{m,n}c_{m}^{*}(t)c_{n}(t)
     \langle\varphi_{m}|e^{i\mathcal{H}_{\mathrm{spin}}t}
     \mathcal{O}e^{-i\mathcal{H}_{\mathrm{spin}}t}|\varphi_{n}\rangle
   =\sum_{m,n}c_{m}^{*}(t)c_{n}(t)e^{i(E_{m}-E_{n})t}
     \langle\varphi_{m}|\mathcal{O}|\varphi_{n}\rangle,
\nonumber
\end{align}
and specifically for the magnetization as 
\begin{align}
 M_{(x,y,z)}
   =\langle S_{\mathrm{tot}}^{(x,y,z)}\rangle
   =\sum_{m,n}c_{m}^{*}(t)c_{n}(t)e^{i(E_{m}-E_{n})t}
     \langle\varphi_{m}|S_{\mathrm{tot}}^{(x,y,z)}|\varphi_{n}\rangle.
\label{eq:Mag_TDPT}
\end{align}

First we consider the Ising model 
$\mathcal{H}_{\mathrm{spin}}=\mathcal{H}_{\mathrm{Ising}}$ 
[Eq.~\eqref{eq:IsingModel} in the main text]. 
The ground state is the configuration 
with all spins up 
$|\varphi_{0}\rangle=|\!\!\!\uparrow\uparrow\cdots\uparrow\rangle$ 
and the first excited states $|\varphi_{n}\rangle$ ($n=1,\ldots,N$) 
are single spin flipped states 
$|\varphi_{n}\rangle=S_{n}^{-}|\varphi_{0}\rangle$. 
Since the excitation gap is $E_{n}-E_{0}=H+J$ ($n=1,\ldots,N$),
we can calculate 
\begin{align}
 c_{0}(t)&=1-\sum_{l}\int_{0}^{t}dt_{1}\int_{0}^{t_{1}}dt_{2}
     V_{0l}(t_{1})V_{l0}(t_{2})+\mathcal{O}(B^{4})\nonumber\\
 &=1-\frac{NB^{2}}{4}
   \Big[\frac{2(H+J)t}{4i\{(H+J)^{2}-\Omega^{2}\}}
     -\frac{e^{-2i\Omega t}-1}{8\Omega(H+J-\Omega)}
     +\frac{e^{ 2i\Omega t}-1}{8\Omega(H+J+\Omega)}\nonumber\\
 &\qquad-\frac{\Omega}{(H+J)^{2}-\Omega^{2}}\Big\{
   \frac{e^{-i(H+J-\Omega)t}-1}{2(H+J-\Omega)}
   -\frac{e^{-i(H+J+\Omega)t}-1}{2(H+J+\Omega)}\Big\}\Big]
     +\mathcal{O}(B^{4}),\nonumber\\
 c_{n}(t)&=-i\int_{0}^{t}dt_{1}V_{n0}(t_{1})
   =-\frac{iB}{4}\Big[\frac{e^{i(H+J+\Omega)t}-1}{H+J+\Omega}
     -\frac{e^{i(H+J-\Omega)t}-1}{H+J-\Omega}\Big]+\mathcal{O}(B^{3})
   \quad(n=1,\ldots,N).
\nonumber
\end{align}
Hence the magnetization~\eqref{eq:Mag_TDPT} becomes 
\begin{align}
 M_{x}&=\sum_{n=1}^{N}\frac{1}{2}c_{n}^{*}(t)c_{0}(t)e^{i(H+J)t}
     +\mathrm{c.c.}+\cdots\nonumber\\
   &=\frac{NB(H+J)}{2\{(H+J)^{2}-\Omega^{2}\}}\sin(\Omega t)
     -\frac{NB\Omega}{2\{(H+J)^{2}-\Omega^{2}\}}\sin[(H+J)t]
     +\mathcal{O}(B^{3}),\nonumber\\
 M_{z}&=\frac{N}{2}c_{0}^{*}(t)c_{0}(t)
     +\sum_{n=1}^{N}\frac{N-2}{2}c_{n}^{*}(t)c_{n}(t)
     +\cdots\nonumber\\
   &=\frac{N}{2}-\frac{NB^{2}}{8}\Big[
     \frac{(H+J)^{2}+3\Omega^{2}}{\{(H+J)^{2}-\Omega^{2}\}^{2}}
     -\frac{\cos(2\Omega t)}{(H+J)^{2}-\Omega^{2}}
     -\frac{2\Omega\cos[(H+J-\Omega)t]}{(H+J+\Omega)(H+J-\Omega)^{2}}\nonumber\\
    &\qquad+\frac{2\Omega\cos[(H+J+\Omega)t]}{(H+J+\Omega)^{2}(H+J-\Omega)}\Big]
     +\mathcal{O}(B^{4}).
\nonumber
\end{align}
For $M_{x}$, the order $B$ term contains 
components with frequency $\Omega$ and $H+J$, 
while for $M_{z}$, the order $B^{2}$ term contains components 
with frequency $2\Omega$ and $H+J\pm\Omega$. 
The full calculation of the $\mathcal{O}(B^{3})$ terms is difficult, 
but we can see that the $c_{n}^{*}(t)c_{0}(t)e^{i(H+J)t}$ term 
contains $e^{3i\Omega t}$ and $e^{i(H+J\pm 2\Omega)t}$. 
Thus, we can surmise that for $M_{x}$, 
the leading order of frequency $n\Omega$ is $B^{n}$ ($n$: odd) 
and that of frequency $H+J\pm n\Omega$ is $B^{n+1}$ ($n$: even) 
while for $M_{z}$, 
the leading order of frequency $n\Omega$ is $B^{n}$ ($n$: even) 
and that of frequency $H+J\pm n\Omega$ is $B^{n+1}$ ($n$: odd). 

Next we consider the XXZ model 
$\mathcal{H}_{\mathrm{spin}}=\mathcal{H}_{\mathrm{XXZ}}$ 
[Eq.~\eqref{eq:XXZModel} in the main text], 
where the laser field is again assumed 
to be Eq.~\eqref{eq:SimpleLaser}. 
The ground state of $\mathcal{H}_{\mathrm{XXZ}}$ is 
the fully polarized ferromagnetic state 
$|\varphi_{0}\rangle=|\!\uparrow\uparrow\cdots\uparrow\rangle$ 
and its eigenenergy is $E_{0}=-\frac{NJ_{z}}{4}$. 
Due to the symmetry breaking, 
$|\!\downarrow\downarrow\cdots\downarrow\rangle$ is also a ground state, 
but we assume that the initial state is $|\varphi_{0}\rangle$. 
The low-energy excited states are single-magnon states 
$|\varphi_{n}\rangle=\frac{1}{\sqrt{N}}\sum_{j=1}^{N}e^{i\frac{2\pi n}{N}j}
S_{j}^{-}|\varphi_{0}\rangle$ ($n=1,\ldots,N$) 
and their eigenenergy is 
$E_{n}=-\frac{(N-4)J_{z}}{4}+J_{xy}\cos(\frac{2\pi n}{N})$ 
($n=1,\ldots,N$). 
Noting that 
\begin{align}
 \langle\varphi_{n}|S_{\mathrm{tot}}^{x}|\varphi_{0}\rangle
   =\frac{1}{2\sqrt{N}}\sum_{j=1}^{N}e^{-i\frac{2\pi n}{N}j}
   =\frac{\sqrt{N}}{2}\delta_{nN},
\nonumber
\end{align}
we obtain
\begin{align}
 V_{n0}(t)=-\frac{\sqrt{N}B}{2}\delta_{nN}
   e^{i(J_{xy}+J_{z})t}\sin(\Omega t)
\nonumber
\end{align}
for $n=1,\ldots,N$, 
where $\delta_{nN}$ is the Kronecker delta. 
Thus we derive 
\begin{align}
 c_{0}(t)&=1-\sum_{l}\int_{0}^{t}dt_{1}\int_{0}^{t_{1}}dt_{2}
     V_{0l}(t_{1})V_{l0}(t_{2})+\mathcal{O}(B^{4})
   =1-\int_{0}^{t}dt_{1}\int_{0}^{t_{1}}dt_{2}
     V_{0N}(t_{1})V_{N0}(t_{2})+\mathcal{O}(B^{4})\nonumber\\
 &=1-\frac{NB^{2}}{4}
   \Big[\frac{2(J_{xy}+J_{z})t}{4i\{(J_{xy}+J_{z})^{2}-\Omega^{2}\}}
     -\frac{e^{-2i\Omega t}-1}{8\Omega(J_{xy}+J_{z}-\Omega)}
     +\frac{e^{ 2i\Omega t}-1}{8\Omega(J_{xy}+J_{z}+\Omega)}+\nonumber\\
 &\qquad -\frac{\Omega}{(J_{xy}+J_{z})^{2}-\Omega^{2}}\Big\{
   \frac{e^{-i(J_{xy}+J_{z}-\Omega)t}-1}{2(J_{xy}+J_{z}-\Omega)}
   -\frac{e^{-i(J_{xy}+J_{z}+\Omega)t}-1}{2(J_{xy}+J_{z}+\Omega)}\Big\}\Big]
   +\mathcal{O}(B^{4}),\nonumber\\
 c_{1}(t)&=\cdots=c_{N-1}(t)=\mathcal{O}(B^{3}),\nonumber\\
 c_{N}(t)&=-i\int_{0}^{t}dt_{1}V_{N0}(t_{1})\nonumber\\
   &=-\frac{i\sqrt{N}B}{4}\Big[\frac{e^{i(J_{xy}+J_{z}+\Omega)t}-1}{J_{xy}+J_{z}+\Omega}
     -\frac{e^{i(J_{xy}+J_{z}-\Omega)t}-1}{J_{xy}+J_{z}-\Omega}\Big]
     +\mathcal{O}(B^{3}).
\nonumber
\end{align}
Therefore the magnetization~\eqref{eq:Mag_TDPT} becomes 
\begin{align}
 M_{x}&=\frac{\sqrt{N}}{2}c_{N}^{*}(t)c_{0}(t)e^{i(J_{xy}+J_{z})t}
     +\mathrm{c.c.}+\cdots\nonumber\\
   &=\frac{NB(J_{xy}+J_{z})}{2\{(J_{xy}+J_{z})^{2}-\Omega^{2}\}}\sin(\Omega t)
     -\frac{NB\Omega}{2\{(J_{xy}+J_{z})^{2}-\Omega^{2}\}}\sin[(J_{xy}+J_{z})t]
     +\mathcal{O}(B^{3}),\\
 M_{z}&=\frac{N}{2}c_{0}^{*}(t)c_{0}(t)
     +\frac{N-2}{2}c_{N}^{*}(t)c_{N}(t)
     +\cdots\nonumber\\
   &=\frac{N}{2}-\frac{NB^{2}}{8}\Big[
     \frac{(J_{xy}+J_{z})^{2}+3\Omega^{2}}{\{(J_{xy}+J_{z})^{2}-\Omega^{2}\}^{2}}
     -\frac{\cos(2\Omega t)}{(J_{xy}+J_{z})^{2}-\Omega^{2}}\nonumber\\
     &-\frac{2\Omega\cos[(J_{xy}+J_{z}-\Omega)t]}{(J_{xy}+J_{z}+\Omega)(J_{xy}+J_{z}-\Omega)^{2}}
     +\frac{2\Omega\cos[(J_{xy}+J_{z}+\Omega)t]}{(J_{xy}+J_{z}+\Omega)^{2}(J_{xy}+J_{z}-\Omega)}\Big]
     +\mathcal{O}(B^{4}).
\end{align}
\end{widetext}
Similarly to the case of the Ising model, 
we can surmise that for $M_{x}$, 
the leading order of frequency $n\Omega$ is $B^{n}$ ($n$: odd) 
and that of frequency $J_{xy}+J_{z}\pm n\Omega$ is $B^{n+1}$ ($n$: even) 
while for $M_{z}$, 
the leading order of frequency $n\Omega$ is $B^{n}$ ($n$: even) 
and that of frequency $J_{xy}+J_{z}\pm n\Omega$ is $B^{n+1}$ ($n$: odd).

\section{Spin wave theory}
\label{sec:SpinWave}

%%%%%%%%%% Fig : Band structure from spin wave thoery %%%%%%%%%%
\begin{figure}[t]
\includegraphics[width=0.48\textwidth]{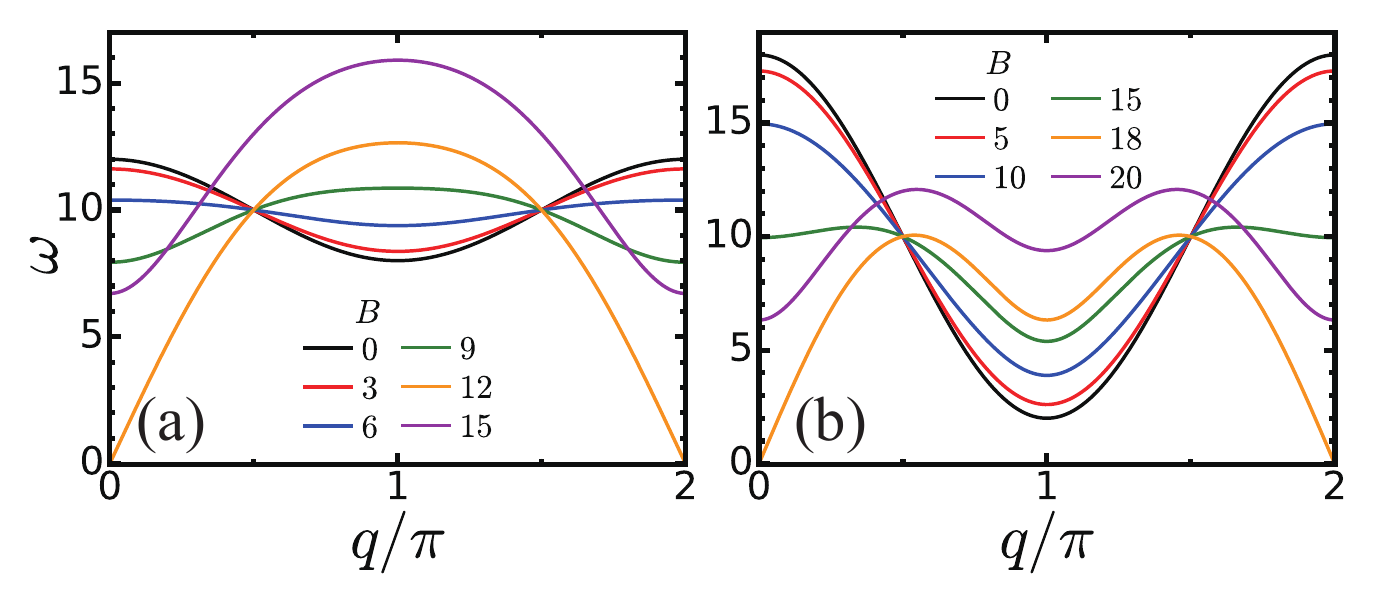}
\caption{The magnon band structure from the spin wave theory 
for the XXZ model with (a) $J_{xy}=2$, $J_{z}=10$ 
and (b) $J_{xy}=8$, $J_{z}=10$. 
}
\label{fig:SpinWave}
\end{figure}
%%%%%%%%%%%%%%%%%%%%

We consider the system 
\begin{equation}
 \mathcal{H}=J_{xy}\sum_{j}(S_{j}^{x}S_{j+1}^{x}+S_{j}^{y}S_{j+1}^{y})
   -J_{z}\sum_{j}S_{j}^{z}S_{j+1}^{z}
   -B\sum_{j}S_{j}^{x},
\nonumber
\end{equation}
where $J_{z}>J_{xy}>0$ with general spin-$S$. 
The number of sites is $N$ and we consider periodic boundary conditions.

First, let us determine the classical ground state. 
For $B=0$ (large $B$), the spin is polarized along the $S^{z}$ ($S^{x}$) axis, 
thus we can assume that the direction of the spins is in the $xz$ plane, 
$S_{j}=S(\sin\phi,0,\cos\phi)$. 
The energy is 
\begin{align}
 E=&NS^{2}(J_{xy}\sin^{2}\phi-J_{z}\cos^{2}\phi)-NSB\sin\phi\nonumber\\
   =&NS^{2}(J_{xy}+J_{z})\Big[\sin\phi-\frac{B}{2S(J_{xy}+J_{z})}\Big]^{2},
\label{eq:Egs}
\end{align}
where the constant term is neglected. 
The configuration minimizing $E$ is 
\begin{equation}
\begin{split}
 \sin\phi=&\frac{B}{2S(J_{xy}+J_{z})}\quad
   (0\leq B\leq 2S(J_{xy}+J_{z}))\\
 \phi=&\pi/2\quad (B>2S(J_{xy}+J_{z}).
\end{split}
\label{eq:Emin_cond}
\end{equation}

We introduce new spin axes $\tilde{S}_{j}^{x,y,z}$ as 
$S_{j}^{x}=\cos\phi\tilde{S}_{j}^{x}+\sin\phi\tilde{S}_{j}^{z}$, 
$S_{j}^{y}=\tilde{S}_{j}^{y}$ and 
$S_{j}^{z}=-\sin\phi\tilde{S}_{j}^{x}+\cos\phi\tilde{S}_{j}^{z}$, 
so that the spin is polarized along the $\tilde{S}_{j}^{z}$ axis. 
Next we perform the Holstein-Primakoff transformation,
\begin{align}
 &\tilde{S}_{j}^{z}=S-n_{j},\nonumber\\
 &\tilde{S}_{j}^{x}+i\tilde{S}_{j}^{y}
   =\sqrt{2S}\Big(1-\frac{n_{j}}{2S}\Big)^{1/2}a_{j},\nonumber\\
 &\tilde{S}_{j}^{x}-i\tilde{S}_{j}^{y}
   =\sqrt{2S}a_{j}^{\dagger}\Big(1-\frac{n_{j}}{2S}\Big)^{1/2},\nonumber
\end{align}
where $a_{j}$ and $a_{j}^{\dagger}$ are 
annihilation and creation operators for bosons (magnons), 
and $n_{j}\equiv a_{j}^{\dagger}a_{j}$ is the number operator. 
Expanding in powers of $1/S$ and 
retaining terms up to second order in $a_{j}$ and $a_{j}^{\dagger}$ 
yields an expression of the Hamiltonian in terms of magnon operators, 
\begin{align}
 \mathcal{H}&=\frac{S}{2}\sum_{j}
     [(J_{xy}\cos^{2}\phi-J_{z}\sin^{2}\phi-J_{xy})
      (a_{j}a_{j+1}+a_{j}^{\dagger}a_{j+1}^{\dagger})\nonumber\\
     &+(J_{xy}\cos^{2}\phi-J_{z}\sin^{2}\phi+J_{xy})
      (a_{j}a_{j+1}^{\dagger}+a_{j}^{\dagger}a_{j+1})\nonumber\\
   &+(J_{xy}\sin^{2}\phi-J_{z}\cos^{2}\phi)(S^{2}-2Sn_{j})].
\nonumber
\end{align}
The first order term of $a_{j}$, $a_{j}^{\dagger}$ vanishes 
if one imposes the condition~\eqref{eq:Emin_cond}. 
After the Fourier transform 
$a_{k}=\frac{1}{\sqrt{N}}\sum_{j}e^{-ikj}a_{j}$, 
$a_{k}^{\dagger}=\frac{1}{\sqrt{N}}\sum_{j}e^{ikj}a_{j}^{\dagger}$, 
we obtain 
\begin{align}
 \mathcal{H}=&\sum_{k>0}
     [f(k,\phi)(a_{k}a_{-k}+a_{k}^{\dagger}a_{-k}^{\dagger})
     +g(k,\phi)(n_{k}+n_{-k})]\nonumber\\
   &+f(k,\phi)(a_{0}^{2}+(a_{0}^{\dagger})^{2})+g(0,\phi)n_{0}
   +E_{\mathrm{CL}},
\nonumber
\end{align}
where
\begin{align}
 f(k,\phi)=&S(J_{xy}\cos^{2}\phi-J_{z}\sin^{2}\phi-J_{xy})\cos k,\nonumber\\
 g(k,\phi)=&S(J_{xy}\cos^{2}\phi-J_{z}\sin^{2}\phi+J_{xy})\cos k\nonumber\\
   &-2S(J_{xy}\sin^{2}\phi-J_{z}\cos^{2}\phi)+B\sin\phi,
\nonumber
\end{align} 
and $E_{\mathrm{CL}}=NS^{2}(J_{xy}\sin^{2}\phi-J_{z}\cos^{2}\phi)-NSB\sin\phi$ 
is the classical ground state energy (see Eq.~\eqref{eq:Egs}). Note that $g(k,\phi)=g(-k,\phi)$.
We then perform the Bogoliubov transformation 
$b_{k}=a_{k}\cosh\theta_{k}+a_{-k}^{\dagger}\sinh\theta_{k}$,
$a_{k}=b_{k}\cosh\theta_{k}-b_{-k}^{\dagger}\sinh\theta_{k}$,
where 
$\tanh 2\theta_{k}=\frac{f(k,\phi)}{g(k,\phi)}$ 
($\theta_{k}=\theta_{-k}$). 
Finally the Hamiltonian becomes 
\begin{align}
 \mathcal{H}-E_{\mathrm{CL}}
   =&\sum_{k}[-f(k,\phi)\sinh 2\theta_{k}
       +g(k,\phi)\cosh 2\theta_{k}]n_{k}\nonumber\\
   &+E_{\mathrm{QC}},
\nonumber
\end{align}
where 
\begin{align}
 E_{\mathrm{QC}}=&\frac{N}{2\pi}\int_{-\pi}^{\pi}dk
   \Big[-\frac{1}{2}f(k,\phi)\sinh 2\theta_{k}\nonumber\\
   &\qquad+\frac{1}{2}g(k,\phi)(\cosh 2\theta_{k}-1)\Big]
\nonumber
\end{align}
is the quantum correction to 
the classical ground state energy which is a constant. 

The magnon band structure from the spin wave theory 
$-f(k,\phi)\sinh 2\theta_{k}+g(k,\phi)\cosh 2\theta_{k}$ 
is shown in Fig.~\ref{fig:SpinWave}. 
The excitation gap closes and the transition happens at 
$B=J_{xy}+J_{z}$. 

\section{Additional analysis of models with different parameters}

\subsection{Ising model with weak static field}
\label{sec:IsingWeakField}

%%%%%%%%%% Fig : susceptibility with low Hz %%%%%%%%%%
\begin{figure}[t]
\includegraphics[width=0.48\textwidth]{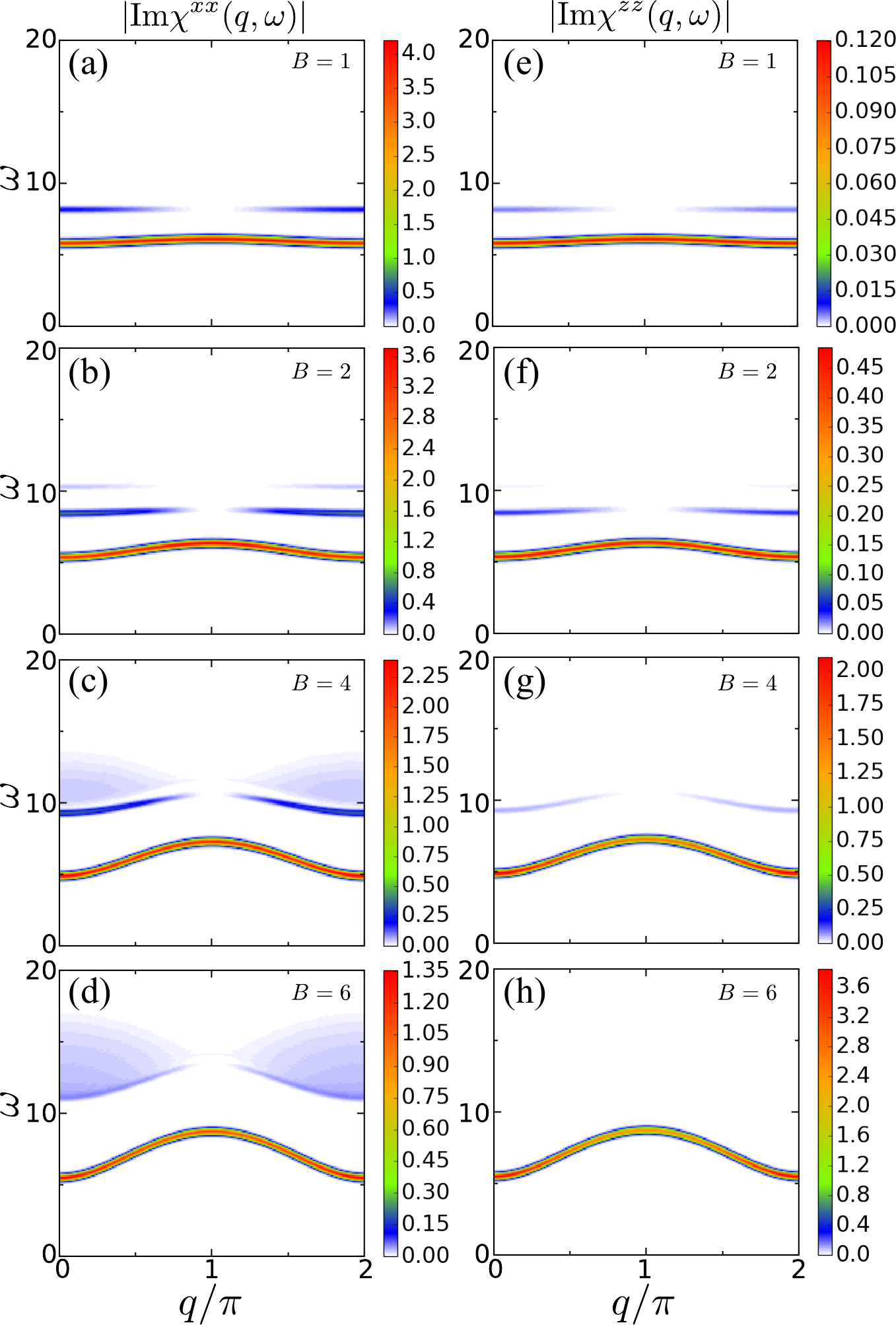}
\caption{DSFs 
(a)-(d) $|\mathrm{Im}\chi^{xx}(q,\omega)|$ and 
(e)-(h) $|\mathrm{Im}\chi^{zz}(q,\omega)|$ 
for the Ising model with $J=4$ and $H=2$. 
}
\label{fig:Ising_low_suscep}
\end{figure}
%%%%%%%%%%%%%%%%%%%%

In the main text, we considered the Ising model 
with a strong longitudinal field $H\gg J$. 
In this subsection, we study how the radiation spectrum and 
the excitation structure are changed 
if the static field is weak $H\ll J$. 
The parameters are set to $J=4$, $H=2$, and $\Omega=1$. 

The DSFs $|\mathrm{Im}\chi^{xx}(q,\omega)|$ and 
$|\mathrm{Im}\chi^{zz}(q,\omega)|$ 
for the ground states of Eq.~\eqref{eq:Hamil_Ising_LandT} 
are shown in Fig.~\ref{fig:Ising_low_suscep}. 
The low energy excitation is again a magnon 
and the shape of the dispersion is similar to the high field case, 
but the size of the excitation gap decreases at first 
with the introduction of $B$ and then increases. 
This behavior is caused by the weak $\mathbf{Z}_{2}$ symmetry breaking 
due to the small longitudinal field $H$. 
For $H=0$, the $\mathbf{Z}_{2}$ symmetry is recovered 
and a gap closing (i.e., a quantum phase transition) happens 
at the critical field $B=B_{\mathrm{c}}$. 
For $H>0$,
the $\mathbf{Z}_{2}$ symmetry is explicitly broken 
and the excitation gap opens at $B_{\mathrm{c}}$. 
However the gap size is small for $H \ll J$, 
and thus the gap size becomes a nonmonotonous function of $B$. 
The continuous spectrum corresponding 
to the two-magnon mode [Eq.~\eqref{eq:TwoMagnonBand} in the main text]
appears more evidently in 
Figs.~\ref{fig:Ising_low_suscep}(b)-\ref{fig:Ising_low_suscep}(d) 
compared with the strong field case. 
We can see an additional excitation between the single-magnon band 
and the two-magnon continuum, which is a two-magnon bound state. 
When $B$ is small, the energy of this state 
(two-spin flips on nearest neighbor sites) is $\simeq J+2H(=8)$. 
With increasing $B$, this bound state is strongly hybridized 
with the two-magnon continuum and is finally merged into it. 

%%%%%%%%%% Fig : Evolution of mag with low Hz %%%%%%%%%%
\begin{figure}[t]
\includegraphics[width=0.48\textwidth]{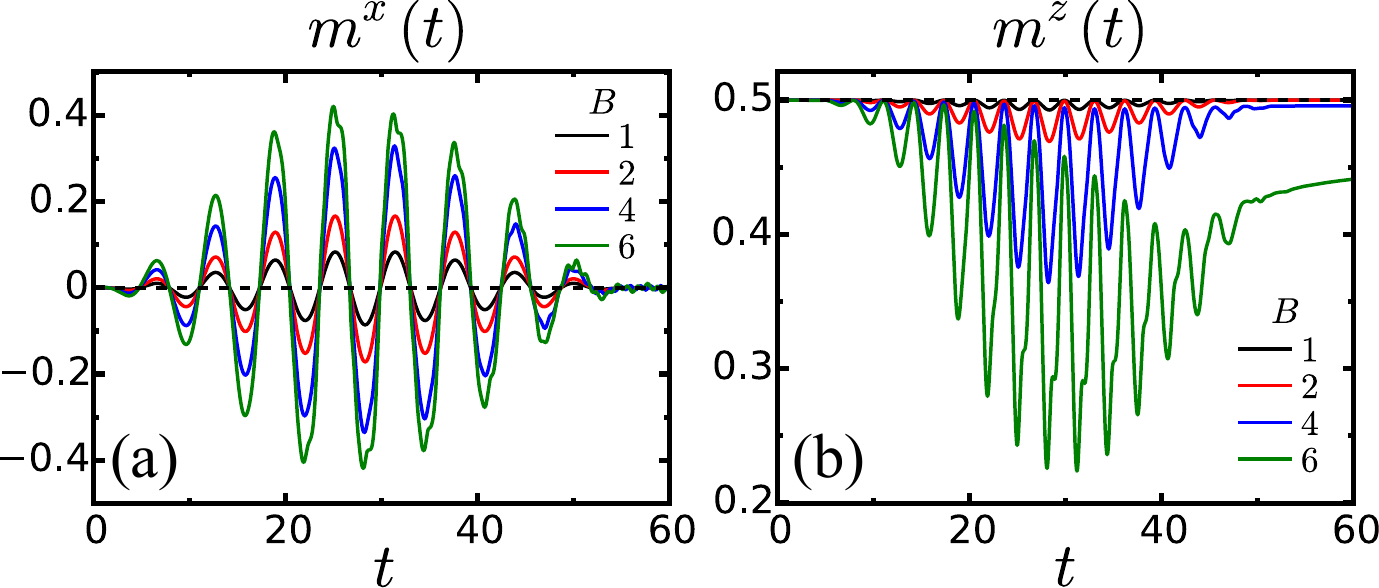}
\caption{Time evolution of (a) $m^{x}$ and (b) $m^{z}$ 
for the Ising model with $J=4$ and $H=2$ calculated by iTEBD. 
}
\label{fig:Ising_low_mag}
\end{figure}
%%%%%%%%%%%%%%%%%%%%

%%%%%%%%%% Fig : HHG with low Hz %%%%%%%%%%
\begin{figure}[t] 
\includegraphics[width=0.48\textwidth]{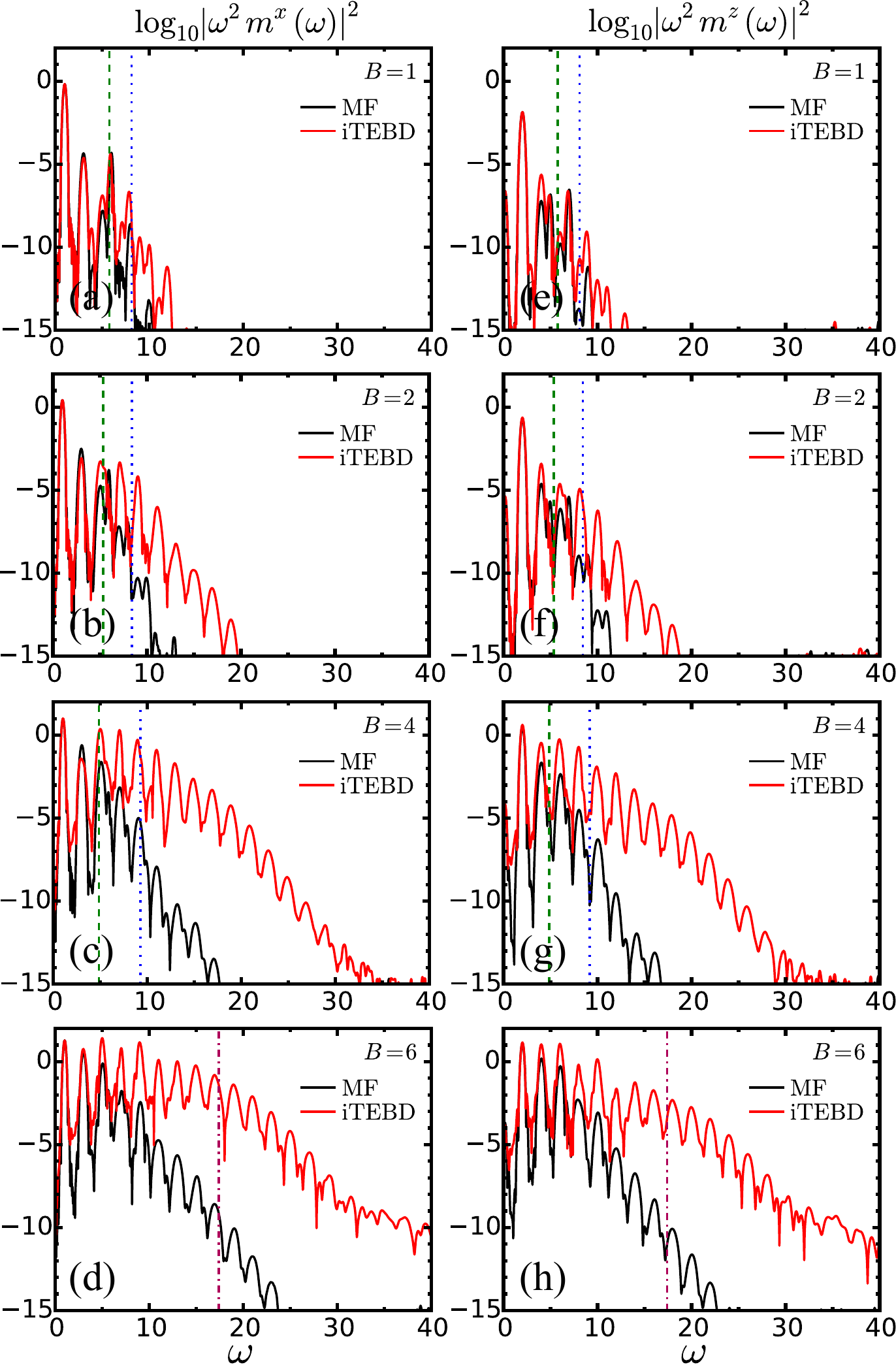}
\caption{Radiation power from (a)-(d) $m^{x}$ and (e)-(h) $m^{z}$ 
for the Ising model with $J=4$ and $H=2$. 
The dashed and dotted lines represent the mass of the single-magnon 
at $q=0$ and that of the two-magnon bound state, respectively. 
The dashed-dotted line corresponds to 
the energy of two magnons at $q=\pi$. 
The crossover of the threshold energy from the former to the latter 
occurs with increasing $B$. 
}
\label{fig:Ising_low_HHG}
\end{figure}
%%%%%%%%%%%%%%%%%%%%

In Fig.~\ref{fig:Ising_low_mag}, 
we show the time evolution of $m^{x}$ and $m^{z}$. 
For $B=6$, the shape of $m^{x}(t)$ is clearly different
from the sinusoidal curve of the laser field 
[Eq.~\eqref{eq:laser_mag_field} in the main text] 
especially near the peaks, which gives rise to a strong HHG signal. 
In contrast to the case of strong static fields, 
the final value of $m^{z}$ deviates from the value $1/2$ for large $B$. 
(This deviation  will also happen in the strong longitudinal field case 
for large $B/H$.) 
As discussed above, the gap of the system first decreases and  
then increases as a function of $B$, 
while the Landau-Zener tunneling happens mainly 
near the minimum of the gap. 
Thus, transitions to excited states of the snapshot Hamiltonian occur, 
which results in the drop of the final value of $m^{z}$ from 1/2. 

In order to investigate the HHG, 
we show $|\omega^{2}m^{x}(\omega)|^{2}$ 
and $|\omega^{2}m^{z}(\omega)|^{2}$ in Fig.~\ref{fig:Ising_low_HHG}. 
When $B$ is small, the behavior of the radiation spectrum is 
similar to the case of high static field. 
The intensity of $|\omega^{2}m^{x}(\omega)|^{2}$ 
and $|\omega^{2}m^{z}(\omega)|^{2}$ generically peaks at 
$\omega=(2n+1)\Omega$ and $\omega=2n\Omega$ ($n$: integer), respectively, 
but at $\omega=J+H=6$, $|\omega^{2}m^{x}(\omega)|^{2}$ 
exhibits a local maximum and $|\omega^{2}m^{z}(\omega)|^{2}$ shows a dip. 
This is consistent with the time-dependent perturbation theory. 
When $B$ becomes larger, a plateau structure appears in the HHG signal  
and its threshold corresponds to 
the single-magnon excitation energy 
(the dashed lines in Fig.~\ref{fig:Ising_low_HHG}) 
or the energy of the two-magnon bound state 
(the dotted lines in Fig.~\ref{fig:Ising_low_HHG}). 
As $B$ is further increased, the threshold of the HHG plateau 
changes from the single-magnon energy to twice of the magnon energy at $q=\pi$ 
(dashed-dotted lines in Fig.~\ref{fig:Ising_low_HHG}). 
This behavior is similar to the XXZ model with small $J_{xy}$ 
(see Sec.~\ref{sec:XXZweakJxy}), 
where the threshold corresponds to the single-magnon energy 
before the transition and the two-magnon energy after the transition. 
In the present case, due to the existence of the longitudinal field $H$, 
the change of the threshold energy scale is not a transition 
but a crossover. 

We also show the analysis by the tdMF theory with the Hamiltonian 
Eq.~\eqref{eq:Hamil_IsingMF} in Fig.~\ref{fig:Ising_low_HHG}. 
For the weak laser amplitude $B$, 
the agreement between the iTEBD and tdMF theories is quantitatively good. 
When $B$ becomes large, the spectra start to deviate 
above the single-magnon energy but the threshold of the HHG plateau
is almost the same ($B=2$). 
For $B=6$ [Fig.~\ref{fig:Ising_low_HHG}(d) and \ref{fig:Ising_low_HHG}(h)], 
the HHG signal calculated by the tdMF theory 
becomes less prominent above the single-magnon energy ($\omega\simeq 5$) 
while the threshold is the two-magnon energy for iTEBD. 
This result implies that the tdMF theory can reproduce 
the single-magnon dynamics but fails to capture multiple-magnon processes. 

%%%%%%%%%% Fig : subcycle analysis with low Hz %%%%%%%%%%
\begin{figure}[t]
\includegraphics[width=0.48\textwidth]{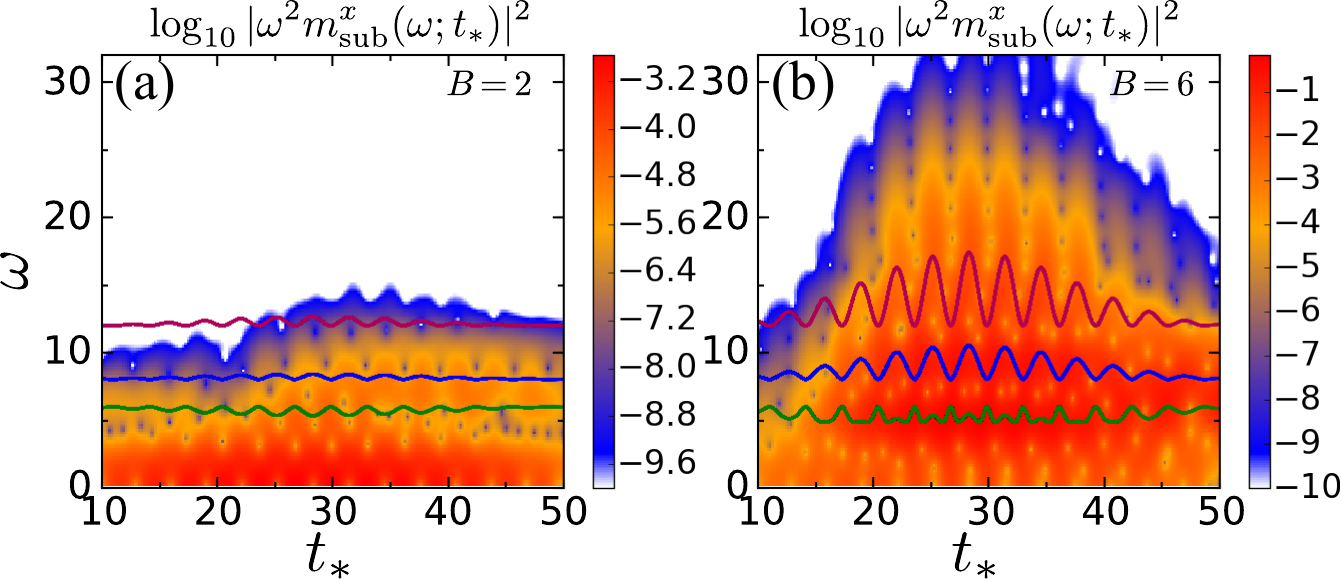}
\caption{Colormap of the subcycle radiation spectrum 
$\log_{10}|\omega^{2}m_{\mathrm{sub}}^{x}(\omega;t_{*})|^{2}$ 
for the Ising model with $J=4$ and $H=2$ 
under the (a) weak transverse field $B=2$ and 
(b) strong transverse  field $B=6$. 
The solid lines show the single-magnon, two-magnon-bound state, 
and two-magnon (with $q=\pi$) state from bottom to top, respectively, 
of the snapshot Hamiltonian at time $t_*$. 
}
\label{fig:Ising_low_subcyc}
\end{figure}
%%%%%%%%%%%%%%%%%%%%

In Fig.~\ref{fig:Ising_low_subcyc}, 
we show the subcycle radiation spectrum 
$\log_{10}|\omega^{2}m_{\mathrm{sub}}^{x}(\omega;t_{*})|^{2}$ 
for $B=2$ and $B=6$. 
Green and purple solid lines show the energy of 
the single-magnon at $q=0$ and of two magnons at $q=\pi$ 
for the snapshot Hamiltonian, respectively. 
In Fig.~\ref{fig:Ising_low_subcyc}(a), 
some intensity exists between the two lines, 
which may be associated with the two-magnon bound state 
seen in Fig.~\ref{fig:Ising_low_suscep}(a) 
and \ref{fig:Ising_low_suscep}(b) (around $\omega=8$\,--\,9). 
For large laser field amplitude ($B=6$), 
the intensity around the two-magnon energy becomes prominent. 
This subcycle analysis supports the interpretation that the crossover of the HHG signal 
is caused by a change of the dynamics from a single-magnon 
to a two-magnon dominated process.

\subsection{XXZ model with strong $J_{xy}$}
\label{sec:XXZStrongJxy}

%%%%%%%%%% Fig : susceptibility with strong Jxy %%%%%%%%%%
\begin{figure}[t]
\includegraphics[width=0.48\textwidth]{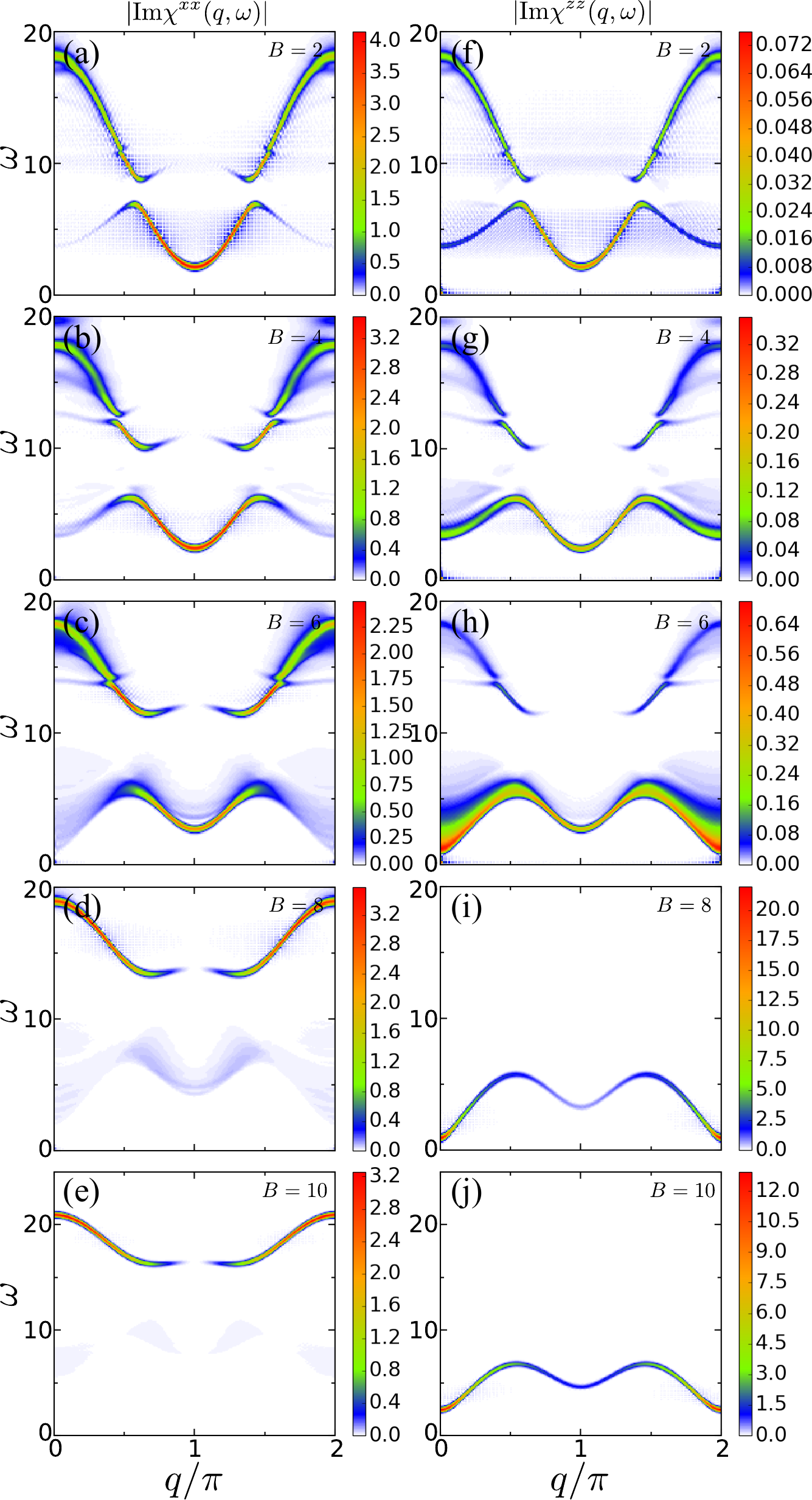}
\caption{DSFs 
(a)-(e) $|\mathrm{Im}\chi^{xx}(q,\omega)|$ and 
(f)-(j) $|\mathrm{Im}\chi^{zz}(q,\omega)|$ 
for the XXZ model with $J_{xy}=8$ and $J_{z}=10$. 
}
\label{fig:XXZ_strongJxy_suscep}
\end{figure}
%%%%%%%%%%%%%%%%%%%%

%%%%%%%%%% Fig : Evolution of mag with strong Jxy %%%%%%%%%%
\begin{figure}[t]
\includegraphics[width=0.48\textwidth]{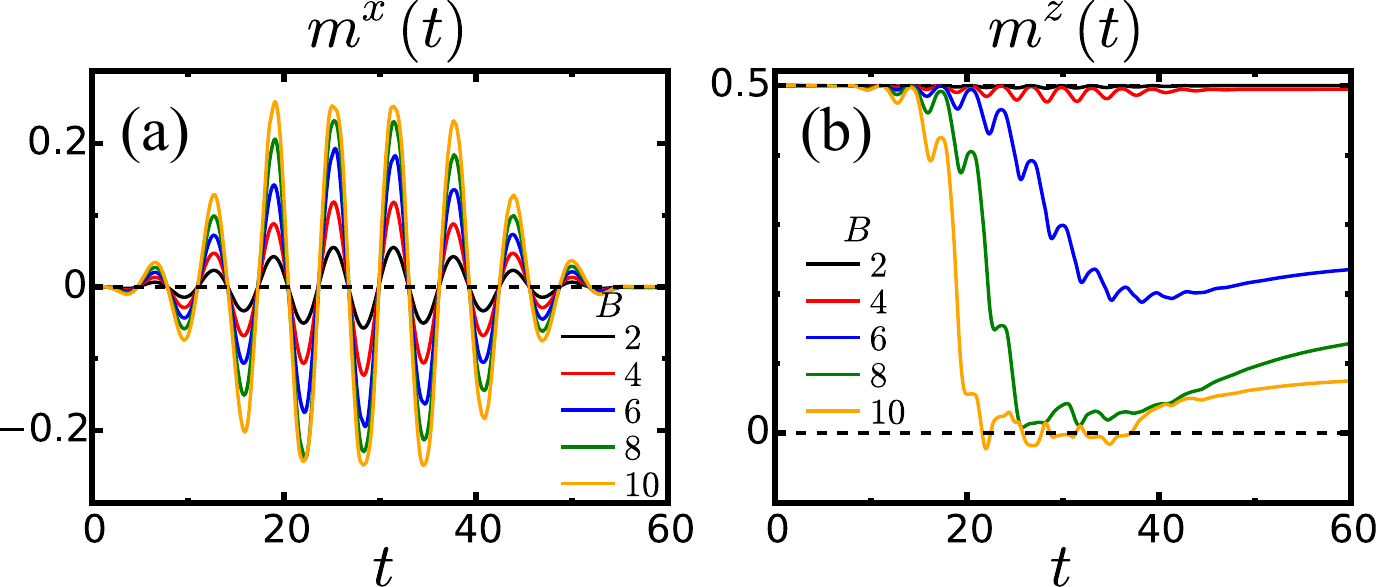}
\caption{Time evolution of (a) $m^{x}$ and (b) $m^{z}$ 
for the XXZ model with $J_{xy}=8$ and $J_{z}=10$.
}
\label{fig:XXZ_strongJxy_mag}
\end{figure}
%%%%%%%%%%%%%%%%%%%%

We next consider the XXZ model with $J_{xy}$ stronger than that in the main text, i.e. $J_{xy}\lesssim J_{z}$. 
The parameters are set to $J_{xy}=8$, $J_{z}=10$, and $\Omega=1$. 
In Fig.~\ref{fig:XXZ_strongJxy_suscep}, 
we show the DSF of the Hamiltonian Eq.~\eqref{eq:XXZ_Transverse} (in the main text). 
The single-magnon dispersion ($E(q)=J_{xy}\cos(q)+J_{z}$ for $B=0$) 
splits by the introduction of the $B$ field. 
The lower bound of the spectrum at $q=0$ decreases with increasing $B$, 
and the gap closes at $B_{\mathrm{c}}\simeq 6$, where a phase transition happens.   
After the transition, 
the intensity of $\chi^{zz}$ is stronger than $\chi^{xx}$, 
but both are still comparable. 
The dispersion captured by $\chi^{zz}$ has a dip around $q=\pi$,  
a property which is reproduced by the spin wave theory 
(see Appendix~\ref{sec:SpinWave}). 
However, the relation $\tilde{\Delta}_{q=0}=2\tilde{\Delta}_{q=\pi}$ 
(for $B>B_{\mathrm{c}}$) does not hold 
in contrast to the weak $J_{xy}$ case. 

%%%%%%%%%% Fig : HHG with strong Jxy %%%%%%%%%%
\begin{figure}[t]
\includegraphics[width=0.48\textwidth]{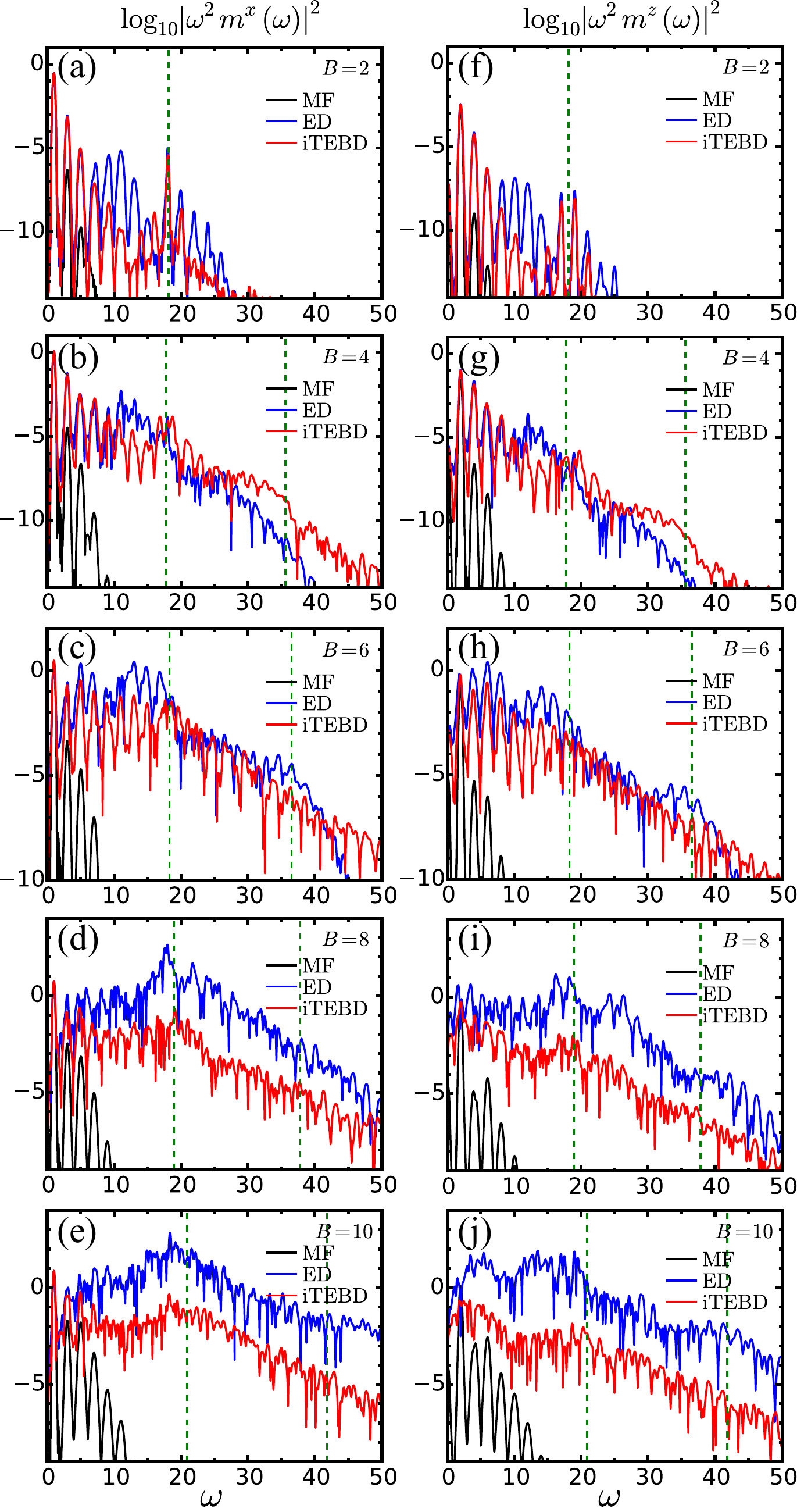}
\caption{HHG from (a)-(e) $m^{x}$ and (f)-(j) $m^{z}$ 
for the XXZ model with $J_{xy}=8$ and $J_{z}=10$. 
Dashed lines correspond to $m_{q=0}$ (times integer). 
}
\label{fig:XXZ_strongJxy_HHG}
\end{figure}
%%%%%%%%%%%%%%%%%%%%

The time evolution of $m^{x}$ and $m^{z}$ 
calculated by iTEBD is shown in Fig.~\ref{fig:XXZ_strongJxy_mag}. 
The behavior of $m^{x}(t)$ and $m^{z}(t)$ is similar 
to that in the weak $J_{xy}$ case. 
The time evolution of $m^{z}$ is different
depending on whether $B$ is smaller or larger than $B_{\mathrm{c}}$. 
In particular, $m^{z}$ suddenly decreases from $1/2$, 
when $B$ exceeds $B_{\mathrm{c}}$. 

In Fig.~\ref{fig:XXZ_strongJxy_HHG}, 
we show the HHG spectra $|\omega^{2}m^{x}(\omega)|^{2}$ and 
$|\omega^{2}m^{z}(\omega)|^{2}$. 
When $B$ is small, 
the result is again described by the time-dependent perturbation theory 
(Appendix~\ref{sec:TdepPerturb}), 
and there is a peak at $\omega=J_{xy}+J_{z}=18$ for $m^{x}(\omega)$ 
and $\omega=J_{xy}+J_{z}\pm\Omega$ for $m^{z}(\omega)$ in the case of $B=2$. 
In contrast to the weak $J_{xy}$ case, 
the threshold of the plateau corresponds to $m_{q=0}$ 
for both $B<B_{\mathrm{c}}$ and $B>B_{\mathrm{c}}$. 
Since there is a dip around $q=\pi$ for $B>B_{\mathrm{c}}$
as is seen from the dispersions in 
Figs.~\ref{fig:XXZ_strongJxy_suscep}(i) and 
\ref{fig:XXZ_strongJxy_suscep}(j),
the relation $\Delta_{q=0}=2\tilde{\Delta}_{q=\pi}$ does not hold. 
Hence the energy scale of the threshold of the plateau 
corresponds to the mode excited by the operator $S_{\mathrm{tot}}^{x}$. 

%%%%%%%%%% Fig : subcycle analysis with strong Jxy %%%%%%%%%%
\begin{figure}[t]
\includegraphics[width=0.48\textwidth]{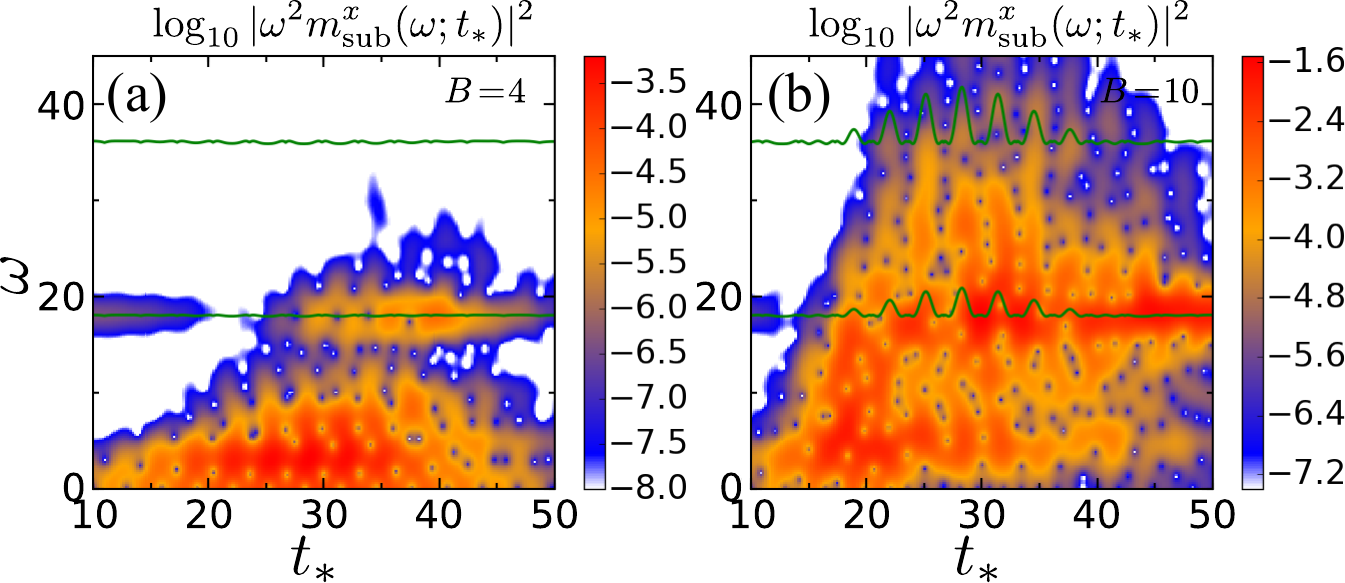}
\caption{Colormap of the subcycle radiation spectrum 
$\log_{10}|\omega^{2}m_{\mathrm{sub}}^{x}(\omega;t_{*})|^{2}$
for the XXZ model with $J_{xy}=8$ and $J_{z}=10$ 
under the laser field (a) $B=4$ and (b) $B=10$. 
The solid lines show the single-magnon, 
and two-magnon modes of the snapshot Hamiltonian at time $t_*$. 
}
\label{fig:XXZ_strongJxy_subcyc}
\end{figure}
%%%%%%%%%%%%%%%%%%%%

Figure~\ref{fig:XXZ_strongJxy_subcyc} shows the subcycle radiation spectrum 
Eq.~\eqref{eq:SubcycFourier} (in the main text) 
for $B=4$ and $B=10$ (below and above the critical field, respectively). 
In the case of $B=4$ [Fig.~\ref{fig:XXZ_strongJxy_subcyc}(a)], 
the strong intensity in the HHG signal 
follows the single-magnon excitation energy $\Delta_{q=0}$ 
of the snapshot Hamiltonian, 
which indicates that the threshold is related to 
the annihilation of single magnons at $q=0$ for $B<B_{\mathrm{c}}$. 
In the case of $B=10$ [Fig.~\ref{fig:XXZ_strongJxy_subcyc}(b)], 
the strong intensity in the HHG signal 
still roughly follows the energy of $\Delta_{q=0}$ and $2\Delta_{q=0}$. 

%%%%%%%%%% Fig : Overlap calculation by ED with strong Jxy %%%%%%%%%%
\begin{figure}[t]
\includegraphics[width=0.48\textwidth]{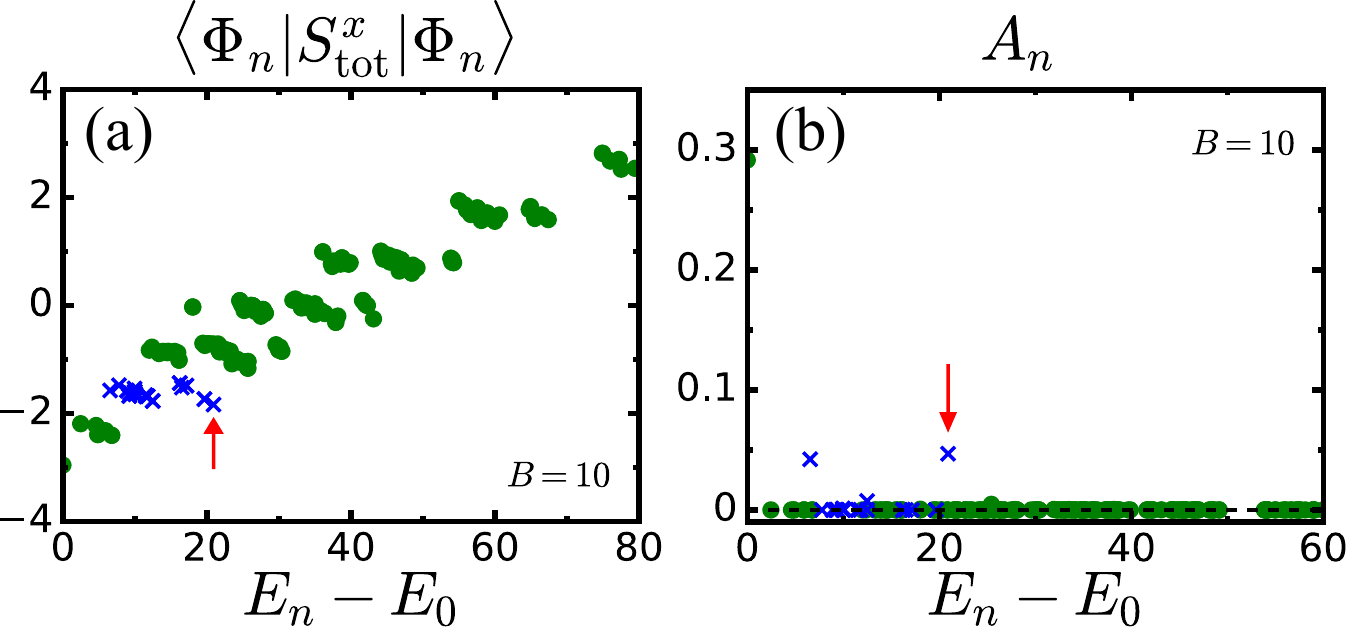}
\caption{(a) $\langle\Phi_{n}|S_{\mathrm{tot}}^{x}|\Phi_{n}\rangle$ 
and (b) $A_{n}$ [Eq.~\eqref{eq:SxWeightWeakXXZ}] 
calculated by ED for the model Eq.~\eqref{eq:XXZ_Transverse} 
(in the main text) with $J_{xy}=8$, $J_{z}=10$, and $B=10$. 
The arrows show the energy $E_{n}-E_{0}=20.9$. 
Cross marks are used for the states in the 
$\langle\Phi_{n}|S_{\mathrm{tot}}^{x}|\Phi_{n}\rangle\simeq -1.5$ 
sector to demonstrate that the contribution to the HHG signal comes 
mainly from this sector. 
}
\label{fig:XXZ_strongJxy_ED}
\end{figure}
%%%%%%%%%%%%%%%%%%%%

To obtain more information on the relation 
between the HHG spectra and the excitation structure, 
we perform ED calculations for a system with $N=8$ sites. 
Although there is a quantitative deviation from the iTEBD results, 
the ED calculations qualitatively reproduce the behavior of 
the HHG spectra, especially the peaks and plateaus as shown 
in Fig.~\ref{fig:XXZ_strongJxy_HHG}. 
In Fig.~\ref{fig:XXZ_strongJxy_ED}(a), 
we show $\langle\Phi_{n}|S_{\mathrm{tot}}^{x}|\Phi_{n}\rangle$ 
calculated for the eigenstates of the snapshot Hamiltonian 
at $t=t_{\mathrm{peak}}$. 
Although the discretization of 
$\langle\Phi_{n}|S_{\mathrm{tot}}^{x}|\Phi_{n}\rangle$ 
is not as clear as in the weak $J_{xy}$ case 
and the values are not necessarily close to integers, 
the eigenstates can be roughly classified into sectors. 
In Fig.~\ref{fig:XXZ_strongJxy_ED}(b), 
we plot the quantity 
$A_{n}=|\alpha_{n}^{*}\alpha_{0}\langle \Phi_{n}|S_{\mathrm{tot}}^{x} |\Phi_{0}\rangle|$
($\alpha_{n}\equiv\langle\Phi_{n}|\Psi(t_{\mathrm{peak}})\rangle$) 
[Eq.~\eqref{eq:SxWeightWeakXXZ} in the main text]. 
We see that there is a strong intensity 
at the energy $E_{n}-E_{0}=20.9$, 
which agrees with the threshold energy 
in Figs.~\ref{fig:XXZ_strongJxy_HHG}(e) 
and \ref{fig:XXZ_strongJxy_HHG}(j). 
The eigenstate at $E_{n}-E_{0}=20.9$ belongs to the 
$\langle\Phi_{n}|S_{\mathrm{tot}}^{x}|\Phi_{n}\rangle\simeq -1.5$ sector 
(depicted by the cross marks in Fig.~\ref{fig:XXZ_strongJxy_ED}), 
and this sector is connected to the 
$\langle\Phi_{n}|S_{\mathrm{tot}}^{x}|\Phi_{n}\rangle\simeq -2$ sector 
in the weak $J_{xy}$ case. 
As is seen from Eq.~\eqref{eq:HamilSxBasis} in the main text, 
the hybridization between two sectors characterized 
by different eigenvalues of $S_{\mathrm{tot}}^{x}$ 
is caused by the term 
$\frac{J_{xy}+J_{z}}{4}\sum_{j}
(\tilde{S}_{j}^{+}\tilde{S}_{j+1}^{+}+\tilde{S}_{j}^{-}\tilde{S}_{j+1}^{-})$, 
which becomes stronger as $J_{xy}$ is increased. 
This strong hybridization explains the results that 
the values of 
$\langle\Phi_{n}|S_{\mathrm{tot}}^{x}|\Phi_{n}\rangle$ 
deviate from integer and that the state with the energy $E_{n}-E_{0}=20.9$ 
is strongly excited by the $S_{\mathrm{tot}}^{x}$ operator. 
By recalling that $\Delta_{q=0}$ is not equal to 
$2\tilde{\Delta}_{q=\pi}$, 
this excitation of $E_{n}-E_{0}=20.9$ cannot be regarded as 
two free magnons created by the $S^{z}$ operator, 
which implies that magnon-magnon interaction effects are important. 

The radiation power spectrum calculated by 
the tdMF Hamiltonian Eq.~\eqref{eq:Hamil_XXZMF} (in the main text)
is also shown in Fig.~\ref{fig:XXZ_strongJxy_HHG}. 
The plateau structure of the radiation spectrum does not appear 
in the tdMF analysis and the high harmonic signals 
decay exponentially as the frequency becomes larger. 
Due to the strong quantum fluctuations induced by $J_{xy}=8$, 
the tdMF theory does not give a good description in this case. 

\section{Subtraction of linear response}

%%%%%%%%%% Fig : linear response subtracted %%%%%%%%%%
\begin{figure}[t]
\includegraphics[width=0.48\textwidth]{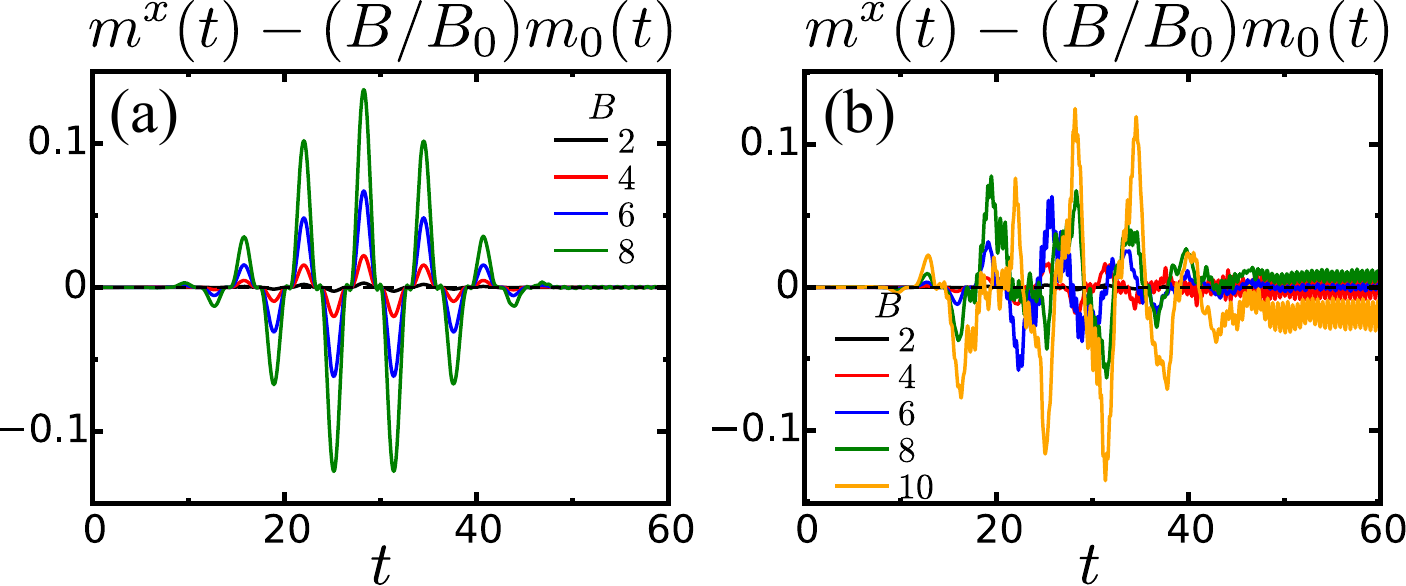}
\caption{Time evolution of $m^{x}$ 
for (a) the Ising model with $J=2$ and $H=6$ 
and (b) the XXZ model with $J_{xy}=2$ and $J_{z}=10$ 
after the subtraction of the linear response component 
$(B/B_{0})m_{0}(t)$.}
\label{fig:mag_subt}
\end{figure}
%%%%%%%%%%%%%%%%%%%%

In this section, we show the time evolution of $m^{x}$ 
after the subtraction of the linear response component 
in order to illustrate the origin of high harmonic generation. 
First we calculate the magnetization dynamics 
$m_{0}(t)\equiv \langle S^{x}(t)\rangle$ 
under the weak laser field~\eqref{eq:laser_mag_field} with $B=0.2$ 
(i.e., in the linear response regime). 
Then we subtract this linear response component from $m^{x}(t)$. 
In Fig.~\ref{fig:mag_subt}, $m^{x}(t)-(B/B_{0})m_{0}(t)$ is shown 
for both the Ising model with $J=2$ and $H=6$ 
and the XXZ model with $J_{xy}=2$ and $J_{z}=10$. 
In both models, the discrepancy from the linear response 
becomes large near the maxima of the laser field amplitude, 
and the nonlinear component increases  
with increasing laser intensity. 
In particular, $m^{x}(t)-(B/B_{0})m_{0}(t)$ has a non-sinusoidal shape, 
which is a manifestation of strong nonlinearity.

\section{Wavelet analysis}
\label{sec:Wavelet}

%%%%%%%%%% Fig : Wavelet analysis %%%%%%%%%%
\begin{figure}[t]
\includegraphics[width=0.48\textwidth]{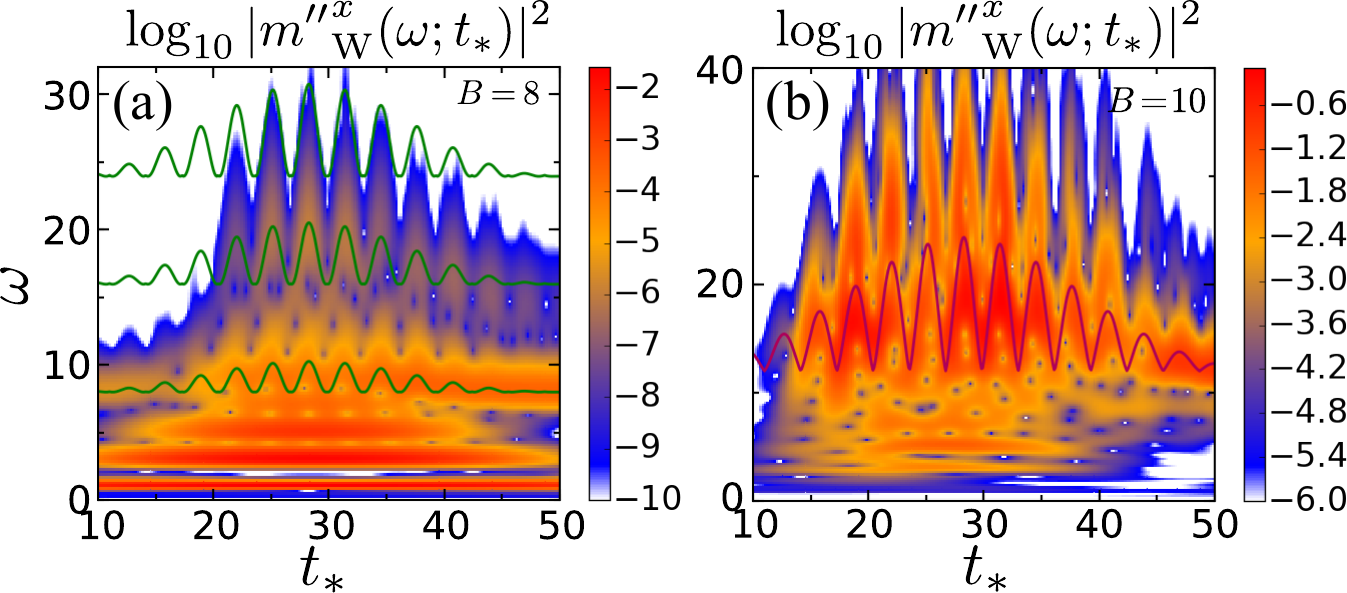}
\caption{Colormap of the wavelet radiation spectrum 
$\log_{10}|{m''}_{\mathrm{W}}^{x}(\omega;t_{*})|^{2}$ 
for the Ising model with $J=4$, $H=2$, and $B=8$ (panel (a))
and the XXZ model with $J_{xy}=2$, $J_{z}=10$, and $B=10$ (panel (b)). 
The solid lines show (a) the (multiple) single-magnon modes and 
(b) the two-magnon (with $q=\pi$) mode 
of the snapshot Hamiltonian at time $t_*$. 
}
\label{fig:Wavelet}
\end{figure}
%%%%%%%%%%%%%%%%%%%%

In this section, we show the time-resolved radiation spectra obtained by a wavelet analysis. 
This approach is similar to the subcycle analysis, 
but in contrast to the latter, the time and energy resolution depends on $\omega$. 
In the low energy regime, the time ($t_{*}$) resolution is low 
and the energy ($\omega$) resolution is high, 
while it is the opposite in the high energy regime. 
The wavelet transform for the second derivative of the magnetic moment 
is defined as 
\begin{align}
 {m''}_{\mathrm{W}}^{x}(\omega;t_{*})\equiv
   \int dt \frac{d^{2}m^{x}(t)}{dt^{2}}
     \omega F_{\mathrm{W}}(\omega(t-t_{*})),
\nonumber
\end{align}
where 
\begin{align}
 F_{\mathrm{W}}(x)
   =\frac{1}{\sqrt{2\pi}\sigma}
     e^{ix}e^{-\frac{x^{2}}{2\sigma^{2}}}
\quad(\sigma=10)
\nonumber
\end{align}
is a mother function for the Gabor wavelet. 
In Fig.~\ref{fig:Wavelet}, 
we show the wavelet spectrum 
$|{m''}_{\mathrm{W}}^{x}(\omega;t_{*})|^{2}$ 
for the Ising model with $J=4$, $H=2$, and $B=8$ 
and the XXZ model with $J_{xy}=2$, $J_{z}=10$, and $B=10$. 
In the low-energy region, 
the signal is smeared out in the time direction 
while there are clearly resolved peaks  
along the $\omega$ direction. 
In the high energy region, on the other hand, 
the structures are smeared out along the $\omega$ axis,  
while the time evolution of the spectral features can be well captured.


\begin{thebibliography}{67}%
\makeatletter
\providecommand \@ifxundefined [1]{%
 \@ifx{#1\undefined}
}%
\providecommand \@ifnum [1]{%
 \ifnum #1\expandafter \@firstoftwo
 \else \expandafter \@secondoftwo
 \fi
}%
\providecommand \@ifx [1]{%
 \ifx #1\expandafter \@firstoftwo
 \else \expandafter \@secondoftwo
 \fi
}%
\providecommand \natexlab [1]{#1}%
\providecommand \enquote  [1]{``#1''}%
\providecommand \bibnamefont  [1]{#1}%
\providecommand \bibfnamefont [1]{#1}%
\providecommand \citenamefont [1]{#1}%
\providecommand \href@noop [0]{\@secondoftwo}%
\providecommand \href [0]{\begingroup \@sanitize@url \@href}%
\providecommand \@href[1]{\@@startlink{#1}\@@href}%
\providecommand \@@href[1]{\endgroup#1\@@endlink}%
\providecommand \@sanitize@url [0]{\catcode `\\12\catcode `\$12\catcode
  `\&12\catcode `\#12\catcode `\^12\catcode `\_12\catcode `\%12\relax}%
\providecommand \@@startlink[1]{}%
\providecommand \@@endlink[0]{}%
\providecommand \url  [0]{\begingroup\@sanitize@url \@url }%
\providecommand \@url [1]{\endgroup\@href {#1}{\urlprefix }}%
\providecommand \urlprefix  [0]{URL }%
\providecommand \Eprint [0]{\href }%
\providecommand \doibase [0]{http://dx.doi.org/}%
\providecommand \selectlanguage [0]{\@gobble}%
\providecommand \bibinfo  [0]{\@secondoftwo}%
\providecommand \bibfield  [0]{\@secondoftwo}%
\providecommand \translation [1]{[#1]}%
\providecommand \BibitemOpen [0]{}%
\providecommand \bibitemStop [0]{}%
\providecommand \bibitemNoStop [0]{.\EOS\space}%
\providecommand \EOS [0]{\spacefactor3000\relax}%
\providecommand \BibitemShut  [1]{\csname bibitem#1\endcsname}%
\let\auto@bib@innerbib\@empty
%</preamble>
\bibitem [{\citenamefont {Matsunaga}\ \emph {et~al.}(2014)\citenamefont
  {Matsunaga}, \citenamefont {Tsuji}, \citenamefont {Fujita}, \citenamefont
  {Sugioka}, \citenamefont {Makise}, \citenamefont {Uzawa}, \citenamefont
  {Terai}, \citenamefont {Wang}, \citenamefont {Aoki},\ and\ \citenamefont
  {Shimano}}]{Matsunaga2014Science}%
  \BibitemOpen
  \bibfield  {author} {\bibinfo {author} {\bibfnamefont {R.}~\bibnamefont
  {Matsunaga}}, \bibinfo {author} {\bibfnamefont {N.}~\bibnamefont {Tsuji}},
  \bibinfo {author} {\bibfnamefont {H.}~\bibnamefont {Fujita}}, \bibinfo
  {author} {\bibfnamefont {A.}~\bibnamefont {Sugioka}}, \bibinfo {author}
  {\bibfnamefont {K.}~\bibnamefont {Makise}}, \bibinfo {author} {\bibfnamefont
  {Y.}~\bibnamefont {Uzawa}}, \bibinfo {author} {\bibfnamefont
  {H.}~\bibnamefont {Terai}}, \bibinfo {author} {\bibfnamefont
  {Z.}~\bibnamefont {Wang}}, \bibinfo {author} {\bibfnamefont {H.}~\bibnamefont
  {Aoki}}, \ and\ \bibinfo {author} {\bibfnamefont {R.}~\bibnamefont
  {Shimano}},\ }\href {\doibase 10.1126/science.1254697} {\bibfield  {journal}
  {\bibinfo  {journal} {Science}\ }\textbf {\bibinfo {volume} {345}},\ \bibinfo
  {pages} {1145} (\bibinfo {year} {2014})}\BibitemShut {NoStop}%
\bibitem [{\citenamefont {Lu}\ \emph {et~al.}(2017)\citenamefont {Lu},
  \citenamefont {Li}, \citenamefont {Hwang}, \citenamefont {Ofori-Okai},
  \citenamefont {Kurihara}, \citenamefont {Suemoto},\ and\ \citenamefont
  {Nelson}}]{Lu2017PRL}%
  \BibitemOpen
  \bibfield  {author} {\bibinfo {author} {\bibfnamefont {J.}~\bibnamefont
  {Lu}}, \bibinfo {author} {\bibfnamefont {X.}~\bibnamefont {Li}}, \bibinfo
  {author} {\bibfnamefont {H.~Y.}\ \bibnamefont {Hwang}}, \bibinfo {author}
  {\bibfnamefont {B.~K.}\ \bibnamefont {Ofori-Okai}}, \bibinfo {author}
  {\bibfnamefont {T.}~\bibnamefont {Kurihara}}, \bibinfo {author}
  {\bibfnamefont {T.}~\bibnamefont {Suemoto}}, \ and\ \bibinfo {author}
  {\bibfnamefont {K.~A.}\ \bibnamefont {Nelson}},\ }\href {\doibase
  10.1103/PhysRevLett.118.207204} {\bibfield  {journal} {\bibinfo  {journal}
  {Phys. Rev. Lett.}\ }\textbf {\bibinfo {volume} {118}},\ \bibinfo {pages}
  {207204} (\bibinfo {year} {2017})}\BibitemShut {NoStop}%
\bibitem [{\citenamefont {Kirilyuk}\ \emph {et~al.}(2010)\citenamefont
  {Kirilyuk}, \citenamefont {Kimel},\ and\ \citenamefont
  {Rasing}}]{Kirilyuk2010RMP}%
  \BibitemOpen
  \bibfield  {author} {\bibinfo {author} {\bibfnamefont {A.}~\bibnamefont
  {Kirilyuk}}, \bibinfo {author} {\bibfnamefont {A.~V.}\ \bibnamefont {Kimel}},
  \ and\ \bibinfo {author} {\bibfnamefont {T.}~\bibnamefont {Rasing}},\ }\href
  {\doibase 10.1103/RevModPhys.82.2731} {\bibfield  {journal} {\bibinfo
  {journal} {Rev. Mod. Phys.}\ }\textbf {\bibinfo {volume} {82}},\ \bibinfo
  {pages} {2731} (\bibinfo {year} {2010})}\BibitemShut {NoStop}%
\bibitem [{\citenamefont {Kampfrath}\ \emph {et~al.}(2013)\citenamefont
  {Kampfrath}, \citenamefont {Tanaka},\ and\ \citenamefont
  {Nelson}}]{Kampfrath2013NatPhoton}%
  \BibitemOpen
  \bibfield  {author} {\bibinfo {author} {\bibfnamefont {T.}~\bibnamefont
  {Kampfrath}}, \bibinfo {author} {\bibfnamefont {K.}~\bibnamefont {Tanaka}}, \
  and\ \bibinfo {author} {\bibfnamefont {K.~A.}\ \bibnamefont {Nelson}},\
  }\href {https://doi.org/10.1038/nphoton.2013.184} {\bibfield  {journal}
  {\bibinfo  {journal} {Nat. Photon.}\ }\textbf {\bibinfo {volume} {7}},\
  \bibinfo {pages} {680} (\bibinfo {year} {2013})}\BibitemShut {NoStop}%
\bibitem [{\citenamefont {Nasu}(2004)}]{Nasu2004Book}%
  \BibitemOpen
  \bibfield  {author} {\bibinfo {author} {\bibfnamefont {K.}~\bibnamefont
  {Nasu}},\ }\href@noop {} {\emph {\bibinfo {title} {Photoinduced phase
  transitions}}}\ (\bibinfo  {publisher} {World Scientific},\ \bibinfo
  {address} {Singapore},\ \bibinfo {year} {2004})\BibitemShut {NoStop}%
\bibitem [{\citenamefont {Oka}\ and\ \citenamefont {Aoki}(2009)}]{Oka2009PRB}%
  \BibitemOpen
  \bibfield  {author} {\bibinfo {author} {\bibfnamefont {T.}~\bibnamefont
  {Oka}}\ and\ \bibinfo {author} {\bibfnamefont {H.}~\bibnamefont {Aoki}},\
  }\href {\doibase 10.1103/PhysRevB.79.081406} {\bibfield  {journal} {\bibinfo
  {journal} {Phys. Rev. B}\ }\textbf {\bibinfo {volume} {79}},\ \bibinfo
  {pages} {081406} (\bibinfo {year} {2009})}\BibitemShut {NoStop}%
\bibitem [{\citenamefont {Lindner}\ \emph {et~al.}(2011)\citenamefont
  {Lindner}, \citenamefont {Refael},\ and\ \citenamefont
  {Galitski}}]{Lindner2011NatPhys}%
  \BibitemOpen
  \bibfield  {author} {\bibinfo {author} {\bibfnamefont {N.~H.}\ \bibnamefont
  {Lindner}}, \bibinfo {author} {\bibfnamefont {G.}~\bibnamefont {Refael}}, \
  and\ \bibinfo {author} {\bibfnamefont {V.}~\bibnamefont {Galitski}},\ }\href
  {https://doi.org/10.1038/nphys1926} {\bibfield  {journal} {\bibinfo
  {journal} {Nat. Phys.}\ }\textbf {\bibinfo {volume} {7}},\ \bibinfo {pages}
  {490} (\bibinfo {year} {2011})}\BibitemShut {NoStop}%
\bibitem [{\citenamefont {Krausz}\ and\ \citenamefont
  {Ivanov}(2009)}]{Krausz2009RMP}%
  \BibitemOpen
  \bibfield  {author} {\bibinfo {author} {\bibfnamefont {F.}~\bibnamefont
  {Krausz}}\ and\ \bibinfo {author} {\bibfnamefont {M.}~\bibnamefont
  {Ivanov}},\ }\href {\doibase 10.1103/RevModPhys.81.163} {\bibfield  {journal}
  {\bibinfo  {journal} {Rev. Mod. Phys.}\ }\textbf {\bibinfo {volume} {81}},\
  \bibinfo {pages} {163} (\bibinfo {year} {2009})}\BibitemShut {NoStop}%
\bibitem [{\citenamefont {Cavalieri}\ \emph {et~al.}(2007)\citenamefont
  {Cavalieri}, \citenamefont {M{\"u}ller}, \citenamefont {Uphues},
  \citenamefont {Yakovlev}, \citenamefont {Baltu{\v{s}}ka}, \citenamefont
  {Horvath}, \citenamefont {Schmidt}, \citenamefont {Bl{\"u}mel}, \citenamefont
  {Holzwarth}, \citenamefont {Hendel}, \citenamefont {Drescher}, \citenamefont
  {Kleineberg}, \citenamefont {Echenique}, \citenamefont {Kienberger},
  \citenamefont {Krausz},\ and\ \citenamefont {Heinzmann}}]{Cavalieri2007}%
  \BibitemOpen
  \bibfield  {author} {\bibinfo {author} {\bibfnamefont {A.~L.}\ \bibnamefont
  {Cavalieri}}, \bibinfo {author} {\bibfnamefont {N.}~\bibnamefont
  {M{\"u}ller}}, \bibinfo {author} {\bibfnamefont {T.}~\bibnamefont {Uphues}},
  \bibinfo {author} {\bibfnamefont {V.~S.}\ \bibnamefont {Yakovlev}}, \bibinfo
  {author} {\bibfnamefont {A.}~\bibnamefont {Baltu{\v{s}}ka}}, \bibinfo
  {author} {\bibfnamefont {B.}~\bibnamefont {Horvath}}, \bibinfo {author}
  {\bibfnamefont {B.}~\bibnamefont {Schmidt}}, \bibinfo {author} {\bibfnamefont
  {L.}~\bibnamefont {Bl{\"u}mel}}, \bibinfo {author} {\bibfnamefont
  {R.}~\bibnamefont {Holzwarth}}, \bibinfo {author} {\bibfnamefont
  {S.}~\bibnamefont {Hendel}}, \bibinfo {author} {\bibfnamefont
  {M.}~\bibnamefont {Drescher}}, \bibinfo {author} {\bibfnamefont
  {U.}~\bibnamefont {Kleineberg}}, \bibinfo {author} {\bibfnamefont {P.~M.}\
  \bibnamefont {Echenique}}, \bibinfo {author} {\bibfnamefont {R.}~\bibnamefont
  {Kienberger}}, \bibinfo {author} {\bibfnamefont {F.}~\bibnamefont {Krausz}},
  \ and\ \bibinfo {author} {\bibfnamefont {U.}~\bibnamefont {Heinzmann}},\
  }\href {http://dx.doi.org/10.1038/nature06229} {\bibfield  {journal}
  {\bibinfo  {journal} {Nature (London)}\ }\textbf {\bibinfo {volume} {449}},\
  \bibinfo {pages} {1029} (\bibinfo {year} {2007})}\BibitemShut {NoStop}%
\bibitem [{\citenamefont {McPherson}\ \emph {et~al.}(1987)\citenamefont
  {McPherson}, \citenamefont {Gibson}, \citenamefont {Jara}, \citenamefont
  {Johann}, \citenamefont {Luk}, \citenamefont {McIntyre}, \citenamefont
  {Boyer},\ and\ \citenamefont {Rhodes}}]{McPherson1987}%
  \BibitemOpen
  \bibfield  {author} {\bibinfo {author} {\bibfnamefont {A.}~\bibnamefont
  {McPherson}}, \bibinfo {author} {\bibfnamefont {G.}~\bibnamefont {Gibson}},
  \bibinfo {author} {\bibfnamefont {H.}~\bibnamefont {Jara}}, \bibinfo {author}
  {\bibfnamefont {U.}~\bibnamefont {Johann}}, \bibinfo {author} {\bibfnamefont
  {T.~S.}\ \bibnamefont {Luk}}, \bibinfo {author} {\bibfnamefont {I.~A.}\
  \bibnamefont {McIntyre}}, \bibinfo {author} {\bibfnamefont {K.}~\bibnamefont
  {Boyer}}, \ and\ \bibinfo {author} {\bibfnamefont {C.~K.}\ \bibnamefont
  {Rhodes}},\ }\href {\doibase 10.1364/JOSAB.4.000595} {\bibfield  {journal}
  {\bibinfo  {journal} {J. Opt. Soc. Am. B}\ }\textbf {\bibinfo {volume} {4}},\
  \bibinfo {pages} {595} (\bibinfo {year} {1987})}\BibitemShut {NoStop}%
\bibitem [{\citenamefont {Ferray}\ \emph {et~al.}(1988)\citenamefont {Ferray},
  \citenamefont {L'Huillier}, \citenamefont {Li}, \citenamefont {Lompre},
  \citenamefont {Mainfray},\ and\ \citenamefont {Manus}}]{Ferray1988}%
  \BibitemOpen
  \bibfield  {author} {\bibinfo {author} {\bibfnamefont {M.}~\bibnamefont
  {Ferray}}, \bibinfo {author} {\bibfnamefont {A.}~\bibnamefont {L'Huillier}},
  \bibinfo {author} {\bibfnamefont {X.~F.}\ \bibnamefont {Li}}, \bibinfo
  {author} {\bibfnamefont {L.~A.}\ \bibnamefont {Lompre}}, \bibinfo {author}
  {\bibfnamefont {G.}~\bibnamefont {Mainfray}}, \ and\ \bibinfo {author}
  {\bibfnamefont {C.}~\bibnamefont {Manus}},\ }\href
  {http://stacks.iop.org/0953-4075/21/i=3/a=001} {\bibfield  {journal}
  {\bibinfo  {journal} {J. Phys. B: Atom. Mol. Opt. Phys.}\ }\textbf {\bibinfo
  {volume} {21}},\ \bibinfo {pages} {L31} (\bibinfo {year} {1988})}\BibitemShut
  {NoStop}%
\bibitem [{\citenamefont {Corkum}(1993)}]{Corkum1993PRL}%
  \BibitemOpen
  \bibfield  {author} {\bibinfo {author} {\bibfnamefont {P.~B.}\ \bibnamefont
  {Corkum}},\ }\href {\doibase 10.1103/PhysRevLett.71.1994} {\bibfield
  {journal} {\bibinfo  {journal} {Phys. Rev. Lett.}\ }\textbf {\bibinfo
  {volume} {71}},\ \bibinfo {pages} {1994} (\bibinfo {year}
  {1993})}\BibitemShut {NoStop}%
\bibitem [{\citenamefont {Lewenstein}\ \emph {et~al.}(1994)\citenamefont
  {Lewenstein}, \citenamefont {Balcou}, \citenamefont {Ivanov}, \citenamefont
  {L'Huillier},\ and\ \citenamefont {Corkum}}]{Lewenstein1994}%
  \BibitemOpen
  \bibfield  {author} {\bibinfo {author} {\bibfnamefont {M.}~\bibnamefont
  {Lewenstein}}, \bibinfo {author} {\bibfnamefont {P.}~\bibnamefont {Balcou}},
  \bibinfo {author} {\bibfnamefont {M.~Y.}\ \bibnamefont {Ivanov}}, \bibinfo
  {author} {\bibfnamefont {A.}~\bibnamefont {L'Huillier}}, \ and\ \bibinfo
  {author} {\bibfnamefont {P.~B.}\ \bibnamefont {Corkum}},\ }\href {\doibase
  10.1103/PhysRevA.49.2117} {\bibfield  {journal} {\bibinfo  {journal} {Phys.
  Rev. A}\ }\textbf {\bibinfo {volume} {49}},\ \bibinfo {pages} {2117}
  (\bibinfo {year} {1994})}\BibitemShut {NoStop}%
\bibitem [{\citenamefont {Ghimire}\ \emph {et~al.}(2011)\citenamefont
  {Ghimire}, \citenamefont {DiChiara}, \citenamefont {Sistrunk}, \citenamefont
  {Agostini}, \citenamefont {DiMauro},\ and\ \citenamefont
  {Reis}}]{Ghimire2011NatPhys}%
  \BibitemOpen
  \bibfield  {author} {\bibinfo {author} {\bibfnamefont {S.}~\bibnamefont
  {Ghimire}}, \bibinfo {author} {\bibfnamefont {A.~D.}\ \bibnamefont
  {DiChiara}}, \bibinfo {author} {\bibfnamefont {E.}~\bibnamefont {Sistrunk}},
  \bibinfo {author} {\bibfnamefont {P.}~\bibnamefont {Agostini}}, \bibinfo
  {author} {\bibfnamefont {L.~F.}\ \bibnamefont {DiMauro}}, \ and\ \bibinfo
  {author} {\bibfnamefont {D.~A.}\ \bibnamefont {Reis}},\ }\href
  {https://doi.org/10.1038/nphys1847} {\bibfield  {journal} {\bibinfo
  {journal} {Nat. Phys.}\ }\textbf {\bibinfo {volume} {7}},\ \bibinfo {pages}
  {138} (\bibinfo {year} {2011})}\BibitemShut {NoStop}%
\bibitem [{\citenamefont {Schubert}\ \emph {et~al.}(2014)\citenamefont
  {Schubert}, \citenamefont {Hohenleutner}, \citenamefont {Langer},
  \citenamefont {Urbanek}, \citenamefont {Lange}, \citenamefont {Huttner},
  \citenamefont {Golde}, \citenamefont {Meier}, \citenamefont {Kira},
  \citenamefont {Koch},\ and\ \citenamefont {Huber}}]{Schubert2014}%
  \BibitemOpen
  \bibfield  {author} {\bibinfo {author} {\bibfnamefont {O.}~\bibnamefont
  {Schubert}}, \bibinfo {author} {\bibfnamefont {M.}~\bibnamefont
  {Hohenleutner}}, \bibinfo {author} {\bibfnamefont {F.}~\bibnamefont
  {Langer}}, \bibinfo {author} {\bibfnamefont {B.}~\bibnamefont {Urbanek}},
  \bibinfo {author} {\bibfnamefont {C.}~\bibnamefont {Lange}}, \bibinfo
  {author} {\bibfnamefont {U.}~\bibnamefont {Huttner}}, \bibinfo {author}
  {\bibfnamefont {D.}~\bibnamefont {Golde}}, \bibinfo {author} {\bibfnamefont
  {T.}~\bibnamefont {Meier}}, \bibinfo {author} {\bibfnamefont
  {M.}~\bibnamefont {Kira}}, \bibinfo {author} {\bibfnamefont {S.~W.}\
  \bibnamefont {Koch}}, \ and\ \bibinfo {author} {\bibfnamefont
  {R.}~\bibnamefont {Huber}},\ }\href
  {http://dx.doi.org/10.1038/nphoton.2013.349} {\bibfield  {journal} {\bibinfo
  {journal} {Nat. Photon.}\ }\textbf {\bibinfo {volume} {8}},\ \bibinfo {pages}
  {119} (\bibinfo {year} {2014})}\BibitemShut {NoStop}%
\bibitem [{\citenamefont {Luu}\ \emph {et~al.}(2015)\citenamefont {Luu},
  \citenamefont {Garg}, \citenamefont {Kruchinin}, \citenamefont {Moulet},
  \citenamefont {Hassan},\ and\ \citenamefont {Goulielmakis}}]{Luu2015}%
  \BibitemOpen
  \bibfield  {author} {\bibinfo {author} {\bibfnamefont {T.~T.}\ \bibnamefont
  {Luu}}, \bibinfo {author} {\bibfnamefont {M.}~\bibnamefont {Garg}}, \bibinfo
  {author} {\bibfnamefont {S.~Y.}\ \bibnamefont {Kruchinin}}, \bibinfo {author}
  {\bibfnamefont {A.}~\bibnamefont {Moulet}}, \bibinfo {author} {\bibfnamefont
  {M.~T.}\ \bibnamefont {Hassan}}, \ and\ \bibinfo {author} {\bibfnamefont
  {E.}~\bibnamefont {Goulielmakis}},\ }\href
  {http://www.nature.com/doifinder/10.1038/nature14456} {\bibfield  {journal}
  {\bibinfo  {journal} {Nature (London)}\ }\textbf {\bibinfo {volume} {521}},\
  \bibinfo {pages} {498} (\bibinfo {year} {2015})}\BibitemShut {NoStop}%
\bibitem [{\citenamefont {Vampa}\ \emph
  {et~al.}(2015{\natexlab{a}})\citenamefont {Vampa}, \citenamefont {Hammond},
  \citenamefont {Thire}, \citenamefont {Schmidt}, \citenamefont {Legare},
  \citenamefont {McDonald}, \citenamefont {Brabec},\ and\ \citenamefont
  {Corkum}}]{Vampa2015Nature}%
  \BibitemOpen
  \bibfield  {author} {\bibinfo {author} {\bibfnamefont {G.}~\bibnamefont
  {Vampa}}, \bibinfo {author} {\bibfnamefont {T.~J.}\ \bibnamefont {Hammond}},
  \bibinfo {author} {\bibfnamefont {N.}~\bibnamefont {Thire}}, \bibinfo
  {author} {\bibfnamefont {B.~E.}\ \bibnamefont {Schmidt}}, \bibinfo {author}
  {\bibfnamefont {F.}~\bibnamefont {Legare}}, \bibinfo {author} {\bibfnamefont
  {C.~R.}\ \bibnamefont {McDonald}}, \bibinfo {author} {\bibfnamefont
  {T.}~\bibnamefont {Brabec}}, \ and\ \bibinfo {author} {\bibfnamefont {P.~B.}\
  \bibnamefont {Corkum}},\ }\href {http://dx.doi.org/10.1038/nature14517}
  {\bibfield  {journal} {\bibinfo  {journal} {Nature (London)}\ }\textbf
  {\bibinfo {volume} {522}},\ \bibinfo {pages} {462} (\bibinfo {year}
  {2015}{\natexlab{a}})}\BibitemShut {NoStop}%
\bibitem [{\citenamefont {Langer}\ \emph {et~al.}(2016)\citenamefont {Langer},
  \citenamefont {Hohenleutner}, \citenamefont {Schmid}, \citenamefont
  {P{\"o}llmann}, \citenamefont {Nagler}, \citenamefont {Korn}, \citenamefont
  {Sch{\"u}ller}, \citenamefont {Sherwin}, \citenamefont {Huttner},
  \citenamefont {Steiner}, \citenamefont {Koch}, \citenamefont {Kira},\ and\
  \citenamefont {Huber}}]{Langer2016Nature}%
  \BibitemOpen
  \bibfield  {author} {\bibinfo {author} {\bibfnamefont {F.}~\bibnamefont
  {Langer}}, \bibinfo {author} {\bibfnamefont {M.}~\bibnamefont
  {Hohenleutner}}, \bibinfo {author} {\bibfnamefont {C.~P.}\ \bibnamefont
  {Schmid}}, \bibinfo {author} {\bibfnamefont {C.}~\bibnamefont
  {P{\"o}llmann}}, \bibinfo {author} {\bibfnamefont {P.}~\bibnamefont
  {Nagler}}, \bibinfo {author} {\bibfnamefont {T.}~\bibnamefont {Korn}},
  \bibinfo {author} {\bibfnamefont {C.}~\bibnamefont {Sch{\"u}ller}}, \bibinfo
  {author} {\bibfnamefont {M.~S.}\ \bibnamefont {Sherwin}}, \bibinfo {author}
  {\bibfnamefont {U.}~\bibnamefont {Huttner}}, \bibinfo {author} {\bibfnamefont
  {J.~T.}\ \bibnamefont {Steiner}}, \bibinfo {author} {\bibfnamefont {S.~W.}\
  \bibnamefont {Koch}}, \bibinfo {author} {\bibfnamefont {M.}~\bibnamefont
  {Kira}}, \ and\ \bibinfo {author} {\bibfnamefont {R.}~\bibnamefont {Huber}},\
  }\href {http://www.nature.com/doifinder/10.1038/nature17958} {\bibfield
  {journal} {\bibinfo  {journal} {Nature (London)}\ }\textbf {\bibinfo {volume}
  {533}},\ \bibinfo {pages} {225} (\bibinfo {year} {2016})}\BibitemShut
  {NoStop}%
\bibitem [{\citenamefont {Hohenleutner}\ \emph {et~al.}(2015)\citenamefont
  {Hohenleutner}, \citenamefont {Langer}, \citenamefont {Schubert},
  \citenamefont {Knorr}, \citenamefont {Huttner}, \citenamefont {Koch},
  \citenamefont {Kira},\ and\ \citenamefont {Huber}}]{Hohenleutner2015Nature}%
  \BibitemOpen
  \bibfield  {author} {\bibinfo {author} {\bibfnamefont {M.}~\bibnamefont
  {Hohenleutner}}, \bibinfo {author} {\bibfnamefont {F.}~\bibnamefont
  {Langer}}, \bibinfo {author} {\bibfnamefont {O.}~\bibnamefont {Schubert}},
  \bibinfo {author} {\bibfnamefont {M.}~\bibnamefont {Knorr}}, \bibinfo
  {author} {\bibfnamefont {U.}~\bibnamefont {Huttner}}, \bibinfo {author}
  {\bibfnamefont {S.}~\bibnamefont {Koch}}, \bibinfo {author} {\bibfnamefont
  {M.}~\bibnamefont {Kira}}, \ and\ \bibinfo {author} {\bibfnamefont
  {R.}~\bibnamefont {Huber}},\ }\href {https://doi.org/10.1038/nature14652}
  {\bibfield  {journal} {\bibinfo  {journal} {Nature (London)}\ }\textbf
  {\bibinfo {volume} {523}},\ \bibinfo {pages} {572} (\bibinfo {year}
  {2015})}\BibitemShut {NoStop}%
\bibitem [{\citenamefont {Ndabashimiye}\ \emph {et~al.}(2016)\citenamefont
  {Ndabashimiye}, \citenamefont {Ghimire}, \citenamefont {Wu}, \citenamefont
  {Browne}, \citenamefont {Schafer}, \citenamefont {Gaarde},\ and\
  \citenamefont {Reis}}]{Ndabashimiye2016}%
  \BibitemOpen
  \bibfield  {author} {\bibinfo {author} {\bibfnamefont {G.}~\bibnamefont
  {Ndabashimiye}}, \bibinfo {author} {\bibfnamefont {S.}~\bibnamefont
  {Ghimire}}, \bibinfo {author} {\bibfnamefont {M.}~\bibnamefont {Wu}},
  \bibinfo {author} {\bibfnamefont {D.~A.}\ \bibnamefont {Browne}}, \bibinfo
  {author} {\bibfnamefont {K.~J.}\ \bibnamefont {Schafer}}, \bibinfo {author}
  {\bibfnamefont {M.~B.}\ \bibnamefont {Gaarde}}, \ and\ \bibinfo {author}
  {\bibfnamefont {D.~A.}\ \bibnamefont {Reis}},\ }\href
  {http://www.nature.com/doifinder/10.1038/nature17660} {\bibfield  {journal}
  {\bibinfo  {journal} {Nature (London)}\ }\textbf {\bibinfo {volume} {534}},\
  \bibinfo {pages} {520} (\bibinfo {year} {2016})}\BibitemShut {NoStop}%
\bibitem [{\citenamefont {Liu}\ \emph {et~al.}(2017)\citenamefont {Liu},
  \citenamefont {Li}, \citenamefont {You}, \citenamefont {Ghimire},
  \citenamefont {Heinz},\ and\ \citenamefont {Reis}}]{Liu2017}%
  \BibitemOpen
  \bibfield  {author} {\bibinfo {author} {\bibfnamefont {H.}~\bibnamefont
  {Liu}}, \bibinfo {author} {\bibfnamefont {Y.}~\bibnamefont {Li}}, \bibinfo
  {author} {\bibfnamefont {Y.~S.}\ \bibnamefont {You}}, \bibinfo {author}
  {\bibfnamefont {S.}~\bibnamefont {Ghimire}}, \bibinfo {author} {\bibfnamefont
  {T.~F.}\ \bibnamefont {Heinz}}, \ and\ \bibinfo {author} {\bibfnamefont
  {D.~A.}\ \bibnamefont {Reis}},\ }\href
  {https://www.nature.com/articles/nphys3946} {\bibfield  {journal} {\bibinfo
  {journal} {Nat. Phys.}\ }\textbf {\bibinfo {volume} {13}},\ \bibinfo {pages}
  {262} (\bibinfo {year} {2017})}\BibitemShut {NoStop}%
\bibitem [{\citenamefont {You}\ \emph {et~al.}(2017)\citenamefont {You},
  \citenamefont {Reis},\ and\ \citenamefont {Ghimire}}]{You2016}%
  \BibitemOpen
  \bibfield  {author} {\bibinfo {author} {\bibfnamefont {Y.~S.}\ \bibnamefont
  {You}}, \bibinfo {author} {\bibfnamefont {D.~A.}\ \bibnamefont {Reis}}, \
  and\ \bibinfo {author} {\bibfnamefont {S.}~\bibnamefont {Ghimire}},\ }\href
  {http://www.nature.com/doifinder/10.1038/nphys3955} {\bibfield  {journal}
  {\bibinfo  {journal} {Nat. Phys.}\ }\textbf {\bibinfo {volume} {13}},\
  \bibinfo {pages} {345} (\bibinfo {year} {2017})}\BibitemShut {NoStop}%
\bibitem [{\citenamefont {Kaneshima}\ \emph {et~al.}(2018)\citenamefont
  {Kaneshima}, \citenamefont {Shinohara}, \citenamefont {Takeuchi},
  \citenamefont {Ishii}, \citenamefont {Imasaka}, \citenamefont {Kaji},
  \citenamefont {Ashihara}, \citenamefont {Ishikawa},\ and\ \citenamefont
  {Itatani}}]{Kaneshima2018}%
  \BibitemOpen
  \bibfield  {author} {\bibinfo {author} {\bibfnamefont {K.}~\bibnamefont
  {Kaneshima}}, \bibinfo {author} {\bibfnamefont {Y.}~\bibnamefont
  {Shinohara}}, \bibinfo {author} {\bibfnamefont {K.}~\bibnamefont {Takeuchi}},
  \bibinfo {author} {\bibfnamefont {N.}~\bibnamefont {Ishii}}, \bibinfo
  {author} {\bibfnamefont {K.}~\bibnamefont {Imasaka}}, \bibinfo {author}
  {\bibfnamefont {T.}~\bibnamefont {Kaji}}, \bibinfo {author} {\bibfnamefont
  {S.}~\bibnamefont {Ashihara}}, \bibinfo {author} {\bibfnamefont {K.~L.}\
  \bibnamefont {Ishikawa}}, \ and\ \bibinfo {author} {\bibfnamefont
  {J.}~\bibnamefont {Itatani}},\ }\href {\doibase
  10.1103/PhysRevLett.120.243903} {\bibfield  {journal} {\bibinfo  {journal}
  {Phys. Rev. Lett.}\ }\textbf {\bibinfo {volume} {120}},\ \bibinfo {pages}
  {243903} (\bibinfo {year} {2018})}\BibitemShut {NoStop}%
\bibitem [{\citenamefont {Luu}\ and\ \citenamefont
  {W{\"o}rner}(2018)}]{Luu2018}%
  \BibitemOpen
  \bibfield  {author} {\bibinfo {author} {\bibfnamefont {T.~T.}\ \bibnamefont
  {Luu}}\ and\ \bibinfo {author} {\bibfnamefont {H.~J.}\ \bibnamefont
  {W{\"o}rner}},\ }\href {https://www.nature.com/articles/s41467-018-03397-4}
  {\bibfield  {journal} {\bibinfo  {journal} {Nat. Comm.}\ }\textbf {\bibinfo
  {volume} {9}},\ \bibinfo {pages} {916} (\bibinfo {year} {2018})}\BibitemShut
  {NoStop}%
\bibitem [{\citenamefont {Golde}\ \emph {et~al.}(2008)\citenamefont {Golde},
  \citenamefont {Meier},\ and\ \citenamefont {Koch}}]{Golde2008}%
  \BibitemOpen
  \bibfield  {author} {\bibinfo {author} {\bibfnamefont {D.}~\bibnamefont
  {Golde}}, \bibinfo {author} {\bibfnamefont {T.}~\bibnamefont {Meier}}, \ and\
  \bibinfo {author} {\bibfnamefont {S.~W.}\ \bibnamefont {Koch}},\ }\href
  {\doibase 10.1103/PhysRevB.77.075330} {\bibfield  {journal} {\bibinfo
  {journal} {Phys. Rev. B}\ }\textbf {\bibinfo {volume} {77}},\ \bibinfo
  {pages} {075330} (\bibinfo {year} {2008})}\BibitemShut {NoStop}%
\bibitem [{\citenamefont {Kemper}\ \emph {et~al.}(2013)\citenamefont {Kemper},
  \citenamefont {Moritz}, \citenamefont {Freericks},\ and\ \citenamefont
  {Devereaux}}]{Kemper2013NJP}%
  \BibitemOpen
  \bibfield  {author} {\bibinfo {author} {\bibfnamefont {A.~F.}\ \bibnamefont
  {Kemper}}, \bibinfo {author} {\bibfnamefont {B.}~\bibnamefont {Moritz}},
  \bibinfo {author} {\bibfnamefont {J.~K.}\ \bibnamefont {Freericks}}, \ and\
  \bibinfo {author} {\bibfnamefont {T.~P.}\ \bibnamefont {Devereaux}},\ }\href
  {http://stacks.iop.org/1367-2630/15/i=2/a=023003} {\bibfield  {journal}
  {\bibinfo  {journal} {New J. Phys.}\ }\textbf {\bibinfo {volume} {15}},\
  \bibinfo {pages} {023003} (\bibinfo {year} {2013})}\BibitemShut {NoStop}%
\bibitem [{\citenamefont {Higuchi}\ \emph {et~al.}(2014)\citenamefont
  {Higuchi}, \citenamefont {Stockman},\ and\ \citenamefont
  {Hommelhoff}}]{Higuchi2014}%
  \BibitemOpen
  \bibfield  {author} {\bibinfo {author} {\bibfnamefont {T.}~\bibnamefont
  {Higuchi}}, \bibinfo {author} {\bibfnamefont {M.~I.}\ \bibnamefont
  {Stockman}}, \ and\ \bibinfo {author} {\bibfnamefont {P.}~\bibnamefont
  {Hommelhoff}},\ }\href {\doibase 10.1103/PhysRevLett.113.213901} {\bibfield
  {journal} {\bibinfo  {journal} {Phys. Rev. Lett.}\ }\textbf {\bibinfo
  {volume} {113}},\ \bibinfo {pages} {213901} (\bibinfo {year}
  {2014})}\BibitemShut {NoStop}%
\bibitem [{\citenamefont {Vampa}\ \emph {et~al.}(2014)\citenamefont {Vampa},
  \citenamefont {McDonald}, \citenamefont {Orlando}, \citenamefont {Klug},
  \citenamefont {Corkum},\ and\ \citenamefont {Brabec}}]{Vampa2014PRL}%
  \BibitemOpen
  \bibfield  {author} {\bibinfo {author} {\bibfnamefont {G.}~\bibnamefont
  {Vampa}}, \bibinfo {author} {\bibfnamefont {C.~R.}\ \bibnamefont {McDonald}},
  \bibinfo {author} {\bibfnamefont {G.}~\bibnamefont {Orlando}}, \bibinfo
  {author} {\bibfnamefont {D.~D.}\ \bibnamefont {Klug}}, \bibinfo {author}
  {\bibfnamefont {P.~B.}\ \bibnamefont {Corkum}}, \ and\ \bibinfo {author}
  {\bibfnamefont {T.}~\bibnamefont {Brabec}},\ }\href {\doibase
  10.1103/PhysRevLett.113.073901} {\bibfield  {journal} {\bibinfo  {journal}
  {Phys. Rev. Lett.}\ }\textbf {\bibinfo {volume} {113}},\ \bibinfo {pages}
  {073901} (\bibinfo {year} {2014})}\BibitemShut {NoStop}%
\bibitem [{\citenamefont {Wu}\ \emph {et~al.}(2015)\citenamefont {Wu},
  \citenamefont {Ghimire}, \citenamefont {Reis}, \citenamefont {Schafer},\ and\
  \citenamefont {Gaarde}}]{Wu2015}%
  \BibitemOpen
  \bibfield  {author} {\bibinfo {author} {\bibfnamefont {M.}~\bibnamefont
  {Wu}}, \bibinfo {author} {\bibfnamefont {S.}~\bibnamefont {Ghimire}},
  \bibinfo {author} {\bibfnamefont {D.~A.}\ \bibnamefont {Reis}}, \bibinfo
  {author} {\bibfnamefont {K.~J.}\ \bibnamefont {Schafer}}, \ and\ \bibinfo
  {author} {\bibfnamefont {M.~B.}\ \bibnamefont {Gaarde}},\ }\href {\doibase
  10.1103/PhysRevA.91.043839} {\bibfield  {journal} {\bibinfo  {journal} {Phys.
  Rev. A}\ }\textbf {\bibinfo {volume} {91}},\ \bibinfo {pages} {043839}
  (\bibinfo {year} {2015})}\BibitemShut {NoStop}%
\bibitem [{\citenamefont {Tamaya}\ \emph {et~al.}(2016)\citenamefont {Tamaya},
  \citenamefont {Ishikawa}, \citenamefont {Ogawa},\ and\ \citenamefont
  {Tanaka}}]{Tamaya2016}%
  \BibitemOpen
  \bibfield  {author} {\bibinfo {author} {\bibfnamefont {T.}~\bibnamefont
  {Tamaya}}, \bibinfo {author} {\bibfnamefont {A.}~\bibnamefont {Ishikawa}},
  \bibinfo {author} {\bibfnamefont {T.}~\bibnamefont {Ogawa}}, \ and\ \bibinfo
  {author} {\bibfnamefont {K.}~\bibnamefont {Tanaka}},\ }\href {\doibase
  10.1103/PhysRevLett.116.016601} {\bibfield  {journal} {\bibinfo  {journal}
  {Phys. Rev. Lett.}\ }\textbf {\bibinfo {volume} {116}},\ \bibinfo {pages}
  {016601} (\bibinfo {year} {2016})}\BibitemShut {NoStop}%
\bibitem [{\citenamefont {Vampa}\ \emph
  {et~al.}(2015{\natexlab{b}})\citenamefont {Vampa}, \citenamefont {McDonald},
  \citenamefont {Orlando}, \citenamefont {Corkum},\ and\ \citenamefont
  {Brabec}}]{Vampa2015PRB}%
  \BibitemOpen
  \bibfield  {author} {\bibinfo {author} {\bibfnamefont {G.}~\bibnamefont
  {Vampa}}, \bibinfo {author} {\bibfnamefont {C.~R.}\ \bibnamefont {McDonald}},
  \bibinfo {author} {\bibfnamefont {G.}~\bibnamefont {Orlando}}, \bibinfo
  {author} {\bibfnamefont {P.~B.}\ \bibnamefont {Corkum}}, \ and\ \bibinfo
  {author} {\bibfnamefont {T.}~\bibnamefont {Brabec}},\ }\href {\doibase
  10.1103/PhysRevB.91.064302} {\bibfield  {journal} {\bibinfo  {journal} {Phys.
  Rev. B}\ }\textbf {\bibinfo {volume} {91}},\ \bibinfo {pages} {064302}
  (\bibinfo {year} {2015}{\natexlab{b}})}\BibitemShut {NoStop}%
\bibitem [{\citenamefont {Luu}\ and\ \citenamefont {W\"orner}(2016)}]{Luu2016}%
  \BibitemOpen
  \bibfield  {author} {\bibinfo {author} {\bibfnamefont {T.~T.}\ \bibnamefont
  {Luu}}\ and\ \bibinfo {author} {\bibfnamefont {H.~J.}\ \bibnamefont
  {W\"orner}},\ }\href {\doibase 10.1103/PhysRevB.94.115164} {\bibfield
  {journal} {\bibinfo  {journal} {Phys. Rev. B}\ }\textbf {\bibinfo {volume}
  {94}},\ \bibinfo {pages} {115164} (\bibinfo {year} {2016})}\BibitemShut
  {NoStop}%
\bibitem [{\citenamefont {Otobe}(2016)}]{Otobe2016}%
  \BibitemOpen
  \bibfield  {author} {\bibinfo {author} {\bibfnamefont {T.}~\bibnamefont
  {Otobe}},\ }\href {\doibase 10.1103/PhysRevB.94.235152} {\bibfield  {journal}
  {\bibinfo  {journal} {Phys. Rev. B}\ }\textbf {\bibinfo {volume} {94}},\
  \bibinfo {pages} {235152} (\bibinfo {year} {2016})}\BibitemShut {NoStop}%
\bibitem [{\citenamefont {Ikemachi}\ \emph {et~al.}(2017)\citenamefont
  {Ikemachi}, \citenamefont {Shinohara}, \citenamefont {Sato}, \citenamefont
  {Yumoto}, \citenamefont {Kuwata-Gonokami},\ and\ \citenamefont
  {Ishikawa}}]{Ikemachi2017}%
  \BibitemOpen
  \bibfield  {author} {\bibinfo {author} {\bibfnamefont {T.}~\bibnamefont
  {Ikemachi}}, \bibinfo {author} {\bibfnamefont {Y.}~\bibnamefont {Shinohara}},
  \bibinfo {author} {\bibfnamefont {T.}~\bibnamefont {Sato}}, \bibinfo {author}
  {\bibfnamefont {J.}~\bibnamefont {Yumoto}}, \bibinfo {author} {\bibfnamefont
  {M.}~\bibnamefont {Kuwata-Gonokami}}, \ and\ \bibinfo {author} {\bibfnamefont
  {K.~L.}\ \bibnamefont {Ishikawa}},\ }\href {\doibase
  10.1103/PhysRevA.95.043416} {\bibfield  {journal} {\bibinfo  {journal} {Phys.
  Rev. A}\ }\textbf {\bibinfo {volume} {95}},\ \bibinfo {pages} {043416}
  (\bibinfo {year} {2017})}\BibitemShut {NoStop}%
\bibitem [{\citenamefont {Osika}\ \emph {et~al.}(2017)\citenamefont {Osika},
  \citenamefont {Chac\'on}, \citenamefont {Ortmann}, \citenamefont {Su\'arez},
  \citenamefont {P\'erez-Hern\'andez}, \citenamefont {Szafran}, \citenamefont
  {Ciappina}, \citenamefont {Sols}, \citenamefont {Landsman},\ and\
  \citenamefont {Lewenstein}}]{Osika2017}%
  \BibitemOpen
  \bibfield  {author} {\bibinfo {author} {\bibfnamefont {E.~N.}\ \bibnamefont
  {Osika}}, \bibinfo {author} {\bibfnamefont {A.}~\bibnamefont {Chac\'on}},
  \bibinfo {author} {\bibfnamefont {L.}~\bibnamefont {Ortmann}}, \bibinfo
  {author} {\bibfnamefont {N.}~\bibnamefont {Su\'arez}}, \bibinfo {author}
  {\bibfnamefont {J.~A.}\ \bibnamefont {P\'erez-Hern\'andez}}, \bibinfo
  {author} {\bibfnamefont {B.}~\bibnamefont {Szafran}}, \bibinfo {author}
  {\bibfnamefont {M.~F.}\ \bibnamefont {Ciappina}}, \bibinfo {author}
  {\bibfnamefont {F.}~\bibnamefont {Sols}}, \bibinfo {author} {\bibfnamefont
  {A.~S.}\ \bibnamefont {Landsman}}, \ and\ \bibinfo {author} {\bibfnamefont
  {M.}~\bibnamefont {Lewenstein}},\ }\href {\doibase 10.1103/PhysRevX.7.021017}
  {\bibfield  {journal} {\bibinfo  {journal} {Phys. Rev. X}\ }\textbf {\bibinfo
  {volume} {7}},\ \bibinfo {pages} {021017} (\bibinfo {year}
  {2017})}\BibitemShut {NoStop}%
\bibitem [{\citenamefont {Hansen}\ \emph {et~al.}(2017)\citenamefont {Hansen},
  \citenamefont {Deffge},\ and\ \citenamefont {Bauer}}]{Hansen2017}%
  \BibitemOpen
  \bibfield  {author} {\bibinfo {author} {\bibfnamefont {K.~K.}\ \bibnamefont
  {Hansen}}, \bibinfo {author} {\bibfnamefont {T.}~\bibnamefont {Deffge}}, \
  and\ \bibinfo {author} {\bibfnamefont {D.}~\bibnamefont {Bauer}},\ }\href
  {\doibase 10.1103/PhysRevA.96.053418} {\bibfield  {journal} {\bibinfo
  {journal} {Phys. Rev. A}\ }\textbf {\bibinfo {volume} {96}},\ \bibinfo
  {pages} {053418} (\bibinfo {year} {2017})}\BibitemShut {NoStop}%
\bibitem [{\citenamefont {Tancogne-Dejean}\ \emph
  {et~al.}(2017{\natexlab{a}})\citenamefont {Tancogne-Dejean}, \citenamefont
  {M\"ucke}, \citenamefont {K\"artner},\ and\ \citenamefont
  {Rubio}}]{Tancogne-Dejean2017b}%
  \BibitemOpen
  \bibfield  {author} {\bibinfo {author} {\bibfnamefont {N.}~\bibnamefont
  {Tancogne-Dejean}}, \bibinfo {author} {\bibfnamefont {O.~D.}\ \bibnamefont
  {M\"ucke}}, \bibinfo {author} {\bibfnamefont {F.~X.}\ \bibnamefont
  {K\"artner}}, \ and\ \bibinfo {author} {\bibfnamefont {A.}~\bibnamefont
  {Rubio}},\ }\href {\doibase 10.1103/PhysRevLett.118.087403} {\bibfield
  {journal} {\bibinfo  {journal} {Phys. Rev. Lett.}\ }\textbf {\bibinfo
  {volume} {118}},\ \bibinfo {pages} {087403} (\bibinfo {year}
  {2017}{\natexlab{a}})}\BibitemShut {NoStop}%
\bibitem [{\citenamefont {Tancogne-Dejean}\ \emph
  {et~al.}(2017{\natexlab{b}})\citenamefont {Tancogne-Dejean}, \citenamefont
  {M{\"u}cke}, \citenamefont {K{\"a}rtner},\ and\ \citenamefont
  {Rubio}}]{Tancogne-Dejean2017}%
  \BibitemOpen
  \bibfield  {author} {\bibinfo {author} {\bibfnamefont {N.}~\bibnamefont
  {Tancogne-Dejean}}, \bibinfo {author} {\bibfnamefont {O.~D.}\ \bibnamefont
  {M{\"u}cke}}, \bibinfo {author} {\bibfnamefont {F.~X.}\ \bibnamefont
  {K{\"a}rtner}}, \ and\ \bibinfo {author} {\bibfnamefont {A.}~\bibnamefont
  {Rubio}},\ }\href {\doibase 10.1038/s41467-017-00764-5} {\bibfield  {journal}
  {\bibinfo  {journal} {Nat. Comm.}\ }\textbf {\bibinfo {volume} {8}},\
  \bibinfo {pages} {745} (\bibinfo {year} {2017}{\natexlab{b}})}\BibitemShut
  {NoStop}%
\bibitem [{\citenamefont {Ikemachi}\ \emph {et~al.}(2018)\citenamefont
  {Ikemachi}, \citenamefont {Shinohara}, \citenamefont {Sato}, \citenamefont
  {Yumoto}, \citenamefont {Kuwata-Gonokami},\ and\ \citenamefont
  {Ishikawa}}]{Ikemachi2018}%
  \BibitemOpen
  \bibfield  {author} {\bibinfo {author} {\bibfnamefont {T.}~\bibnamefont
  {Ikemachi}}, \bibinfo {author} {\bibfnamefont {Y.}~\bibnamefont {Shinohara}},
  \bibinfo {author} {\bibfnamefont {T.}~\bibnamefont {Sato}}, \bibinfo {author}
  {\bibfnamefont {J.}~\bibnamefont {Yumoto}}, \bibinfo {author} {\bibfnamefont
  {M.}~\bibnamefont {Kuwata-Gonokami}}, \ and\ \bibinfo {author} {\bibfnamefont
  {K.~L.}\ \bibnamefont {Ishikawa}},\ }\href {\doibase
  10.1103/PhysRevA.98.023415} {\bibfield  {journal} {\bibinfo  {journal} {Phys.
  Rev. A}\ }\textbf {\bibinfo {volume} {98}},\ \bibinfo {pages} {023415}
  (\bibinfo {year} {2018})}\BibitemShut {NoStop}%
\bibitem [{\citenamefont {Ikeda}\ \emph {et~al.}(2018)\citenamefont {Ikeda},
  \citenamefont {Chinzei},\ and\ \citenamefont {Tsunetsugu}}]{Ikeda2018PRA}%
  \BibitemOpen
  \bibfield  {author} {\bibinfo {author} {\bibfnamefont {T.~N.}\ \bibnamefont
  {Ikeda}}, \bibinfo {author} {\bibfnamefont {K.}~\bibnamefont {Chinzei}}, \
  and\ \bibinfo {author} {\bibfnamefont {H.}~\bibnamefont {Tsunetsugu}},\
  }\href {\doibase 10.1103/PhysRevA.98.063426} {\bibfield  {journal} {\bibinfo
  {journal} {Phys. Rev. A}\ }\textbf {\bibinfo {volume} {98}},\ \bibinfo
  {pages} {063426} (\bibinfo {year} {2018})}\BibitemShut {NoStop}%
\bibitem [{\citenamefont {Huttner}\ \emph {et~al.}(2017)\citenamefont
  {Huttner}, \citenamefont {Kira},\ and\ \citenamefont {Koch}}]{Huttner2017}%
  \BibitemOpen
  \bibfield  {author} {\bibinfo {author} {\bibfnamefont {U.}~\bibnamefont
  {Huttner}}, \bibinfo {author} {\bibfnamefont {M.}~\bibnamefont {Kira}}, \
  and\ \bibinfo {author} {\bibfnamefont {S.~W.}\ \bibnamefont {Koch}},\ }\href
  {\doibase 10.1002/lpor.201700049} {\bibfield  {journal} {\bibinfo  {journal}
  {Las. Photon. Rev.}\ }\textbf {\bibinfo {volume} {11}},\ \bibinfo {pages}
  {1700049} (\bibinfo {year} {2017})}\BibitemShut {NoStop}%
\bibitem [{\citenamefont {Kruchinin}\ \emph {et~al.}(2018)\citenamefont
  {Kruchinin}, \citenamefont {Krausz},\ and\ \citenamefont
  {Yakovlev}}]{Kruchinin2018}%
  \BibitemOpen
  \bibfield  {author} {\bibinfo {author} {\bibfnamefont {S.~Y.}\ \bibnamefont
  {Kruchinin}}, \bibinfo {author} {\bibfnamefont {F.}~\bibnamefont {Krausz}}, \
  and\ \bibinfo {author} {\bibfnamefont {V.~S.}\ \bibnamefont {Yakovlev}},\
  }\href {\doibase 10.1103/RevModPhys.90.021002} {\bibfield  {journal}
  {\bibinfo  {journal} {Rev. Mod. Phys.}\ }\textbf {\bibinfo {volume} {90}},\
  \bibinfo {pages} {021002} (\bibinfo {year} {2018})}\BibitemShut {NoStop}%
\bibitem [{\citenamefont {Ghimire}\ and\ \citenamefont
  {Reis}(2019)}]{Ghimire2019}%
  \BibitemOpen
  \bibfield  {author} {\bibinfo {author} {\bibfnamefont {S.}~\bibnamefont
  {Ghimire}}\ and\ \bibinfo {author} {\bibfnamefont {D.~A.}\ \bibnamefont
  {Reis}},\ }\href {\doibase 10.1038/s41567-018-0315-5} {\bibfield  {journal}
  {\bibinfo  {journal} {Nat. Phys.}\ }\textbf {\bibinfo {volume} {15}},\
  \bibinfo {pages} {10} (\bibinfo {year} {2019})}\BibitemShut {NoStop}%
\bibitem [{\citenamefont {Heissler}\ \emph {et~al.}(2014)\citenamefont
  {Heissler}, \citenamefont {Lugovoy}, \citenamefont {H\"orlein}, \citenamefont
  {Waldecker}, \citenamefont {Wenz}, \citenamefont {Heigoldt}, \citenamefont
  {Khrennikov}, \citenamefont {Karsch}, \citenamefont {Krausz}, \citenamefont
  {Abel},\ and\ \citenamefont {Tsakiris}}]{Heissler2014NJP}%
  \BibitemOpen
  \bibfield  {author} {\bibinfo {author} {\bibfnamefont {P.}~\bibnamefont
  {Heissler}}, \bibinfo {author} {\bibfnamefont {E.}~\bibnamefont {Lugovoy}},
  \bibinfo {author} {\bibfnamefont {R.}~\bibnamefont {H\"orlein}}, \bibinfo
  {author} {\bibfnamefont {L.}~\bibnamefont {Waldecker}}, \bibinfo {author}
  {\bibfnamefont {J.}~\bibnamefont {Wenz}}, \bibinfo {author} {\bibfnamefont
  {M.}~\bibnamefont {Heigoldt}}, \bibinfo {author} {\bibfnamefont
  {K.}~\bibnamefont {Khrennikov}}, \bibinfo {author} {\bibfnamefont
  {S.}~\bibnamefont {Karsch}}, \bibinfo {author} {\bibfnamefont
  {F.}~\bibnamefont {Krausz}}, \bibinfo {author} {\bibfnamefont
  {B.}~\bibnamefont {Abel}}, \ and\ \bibinfo {author} {\bibfnamefont {G.~D.}\
  \bibnamefont {Tsakiris}},\ }\href
  {http://stacks.iop.org/1367-2630/16/i=11/a=113045} {\bibfield  {journal}
  {\bibinfo  {journal} {New J. Phys.}\ }\textbf {\bibinfo {volume} {16}},\
  \bibinfo {pages} {113045} (\bibinfo {year} {2014})}\BibitemShut {NoStop}%
\bibitem [{\citenamefont {Yoshikawa}\ \emph {et~al.}(2017)\citenamefont
  {Yoshikawa}, \citenamefont {Tamaya},\ and\ \citenamefont
  {Tanaka}}]{Yoshikawa2017Science}%
  \BibitemOpen
  \bibfield  {author} {\bibinfo {author} {\bibfnamefont {N.}~\bibnamefont
  {Yoshikawa}}, \bibinfo {author} {\bibfnamefont {T.}~\bibnamefont {Tamaya}}, \
  and\ \bibinfo {author} {\bibfnamefont {K.}~\bibnamefont {Tanaka}},\ }\href
  {\doibase 10.1126/science.aam8861} {\bibfield  {journal} {\bibinfo  {journal}
  {Science}\ }\textbf {\bibinfo {volume} {356}},\ \bibinfo {pages} {736}
  (\bibinfo {year} {2017})}\BibitemShut {NoStop}%
\bibitem [{\citenamefont {Hafez}\ \emph {et~al.}(2018)\citenamefont {Hafez},
  \citenamefont {Kovalev}, \citenamefont {Deinert}, \citenamefont {Mics},
  \citenamefont {Green}, \citenamefont {Awari}, \citenamefont {Chen},
  \citenamefont {Germanskiy}, \citenamefont {Lehnert}, \citenamefont
  {Teichert}, \citenamefont {Wang}, \citenamefont {Tielrooij}, \citenamefont
  {Liu}, \citenamefont {Chen}, \citenamefont {Narita}, \citenamefont {Mullen},
  \citenamefont {Bonn}, \citenamefont {Gensch},\ and\ \citenamefont
  {Turchinovich}}]{Hafez2018}%
  \BibitemOpen
  \bibfield  {author} {\bibinfo {author} {\bibfnamefont {H.~A.}\ \bibnamefont
  {Hafez}}, \bibinfo {author} {\bibfnamefont {S.}~\bibnamefont {Kovalev}},
  \bibinfo {author} {\bibfnamefont {J.-C.}\ \bibnamefont {Deinert}}, \bibinfo
  {author} {\bibfnamefont {Z.}~\bibnamefont {Mics}}, \bibinfo {author}
  {\bibfnamefont {B.}~\bibnamefont {Green}}, \bibinfo {author} {\bibfnamefont
  {N.}~\bibnamefont {Awari}}, \bibinfo {author} {\bibfnamefont
  {M.}~\bibnamefont {Chen}}, \bibinfo {author} {\bibfnamefont {S.}~\bibnamefont
  {Germanskiy}}, \bibinfo {author} {\bibfnamefont {U.}~\bibnamefont {Lehnert}},
  \bibinfo {author} {\bibfnamefont {J.}~\bibnamefont {Teichert}}, \bibinfo
  {author} {\bibfnamefont {Z.}~\bibnamefont {Wang}}, \bibinfo {author}
  {\bibfnamefont {K.-J.}\ \bibnamefont {Tielrooij}}, \bibinfo {author}
  {\bibfnamefont {Z.}~\bibnamefont {Liu}}, \bibinfo {author} {\bibfnamefont
  {Z.}~\bibnamefont {Chen}}, \bibinfo {author} {\bibfnamefont {A.}~\bibnamefont
  {Narita}}, \bibinfo {author} {\bibfnamefont {K.}~\bibnamefont {Mullen}},
  \bibinfo {author} {\bibfnamefont {M.}~\bibnamefont {Bonn}}, \bibinfo {author}
  {\bibfnamefont {M.}~\bibnamefont {Gensch}}, \ and\ \bibinfo {author}
  {\bibfnamefont {D.}~\bibnamefont {Turchinovich}},\ }\href
  {http://dx.doi.org/10.1038/s41586-018-0508-1} {\bibfield  {journal} {\bibinfo
   {journal} {Nature (London)}\ }\textbf {\bibinfo {volume} {561}},\ \bibinfo
  {pages} {507} (\bibinfo {year} {2018})}\BibitemShut {NoStop}%
\bibitem [{\citenamefont {Chac{\'o}n}\ \emph {et~al.}(2018)\citenamefont
  {Chac{\'o}n}, \citenamefont {Zhu}, \citenamefont {Kelly}, \citenamefont
  {Dauphin}, \citenamefont {Pisanty}, \citenamefont {Pic{\'o}n}, \citenamefont
  {Ticknor}, \citenamefont {Ciappina}, \citenamefont {Saxena},\ and\
  \citenamefont {Lewenstein}}]{Chacon2018arXiv}%
  \BibitemOpen
  \bibfield  {author} {\bibinfo {author} {\bibfnamefont {A.}~\bibnamefont
  {Chac{\'o}n}}, \bibinfo {author} {\bibfnamefont {W.}~\bibnamefont {Zhu}},
  \bibinfo {author} {\bibfnamefont {S.~P.}\ \bibnamefont {Kelly}}, \bibinfo
  {author} {\bibfnamefont {A.}~\bibnamefont {Dauphin}}, \bibinfo {author}
  {\bibfnamefont {E.}~\bibnamefont {Pisanty}}, \bibinfo {author} {\bibfnamefont
  {A.}~\bibnamefont {Pic{\'o}n}}, \bibinfo {author} {\bibfnamefont
  {C.}~\bibnamefont {Ticknor}}, \bibinfo {author} {\bibfnamefont {M.~F.}\
  \bibnamefont {Ciappina}}, \bibinfo {author} {\bibfnamefont {A.}~\bibnamefont
  {Saxena}}, \ and\ \bibinfo {author} {\bibfnamefont {M.}~\bibnamefont
  {Lewenstein}},\ }\href {https://arxiv.org/abs/1807.01616} {\bibfield
  {journal} {\bibinfo  {journal} {arXiv:1807.01616}\ } (\bibinfo {year}
  {2018})}\BibitemShut {NoStop}%
\bibitem [{\citenamefont {Silva}\ \emph {et~al.}(2018)\citenamefont {Silva},
  \citenamefont {Blinov}, \citenamefont {Rubtsov}, \citenamefont {Smirnova},\
  and\ \citenamefont {Ivanov}}]{Silva2018NatPhoton}%
  \BibitemOpen
  \bibfield  {author} {\bibinfo {author} {\bibfnamefont {R.~E.~F.}\
  \bibnamefont {Silva}}, \bibinfo {author} {\bibfnamefont {I.~V.}\ \bibnamefont
  {Blinov}}, \bibinfo {author} {\bibfnamefont {A.~N.}\ \bibnamefont {Rubtsov}},
  \bibinfo {author} {\bibfnamefont {O.}~\bibnamefont {Smirnova}}, \ and\
  \bibinfo {author} {\bibfnamefont {M.}~\bibnamefont {Ivanov}},\ }\href
  {https://doi.org/10.1038/s41566-018-0129-0} {\bibfield  {journal} {\bibinfo
  {journal} {Nat. Photon.}\ }\textbf {\bibinfo {volume} {12}},\ \bibinfo
  {pages} {266} (\bibinfo {year} {2018})}\BibitemShut {NoStop}%
\bibitem [{\citenamefont {Murakami}\ \emph {et~al.}(2018)\citenamefont
  {Murakami}, \citenamefont {Eckstein},\ and\ \citenamefont
  {Werner}}]{Murakami2018PRL}%
  \BibitemOpen
  \bibfield  {author} {\bibinfo {author} {\bibfnamefont {Y.}~\bibnamefont
  {Murakami}}, \bibinfo {author} {\bibfnamefont {M.}~\bibnamefont {Eckstein}},
  \ and\ \bibinfo {author} {\bibfnamefont {P.}~\bibnamefont {Werner}},\ }\href
  {\doibase 10.1103/PhysRevLett.121.057405} {\bibfield  {journal} {\bibinfo
  {journal} {Phys. Rev. Lett.}\ }\textbf {\bibinfo {volume} {121}},\ \bibinfo
  {pages} {057405} (\bibinfo {year} {2018})}\BibitemShut {NoStop}%
\bibitem [{\citenamefont {Murakami}\ and\ \citenamefont
  {Werner}(2018)}]{Murakami2018PRB}%
  \BibitemOpen
  \bibfield  {author} {\bibinfo {author} {\bibfnamefont {Y.}~\bibnamefont
  {Murakami}}\ and\ \bibinfo {author} {\bibfnamefont {P.}~\bibnamefont
  {Werner}},\ }\href {\doibase 10.1103/PhysRevB.98.075102} {\bibfield
  {journal} {\bibinfo  {journal} {Phys. Rev. B}\ }\textbf {\bibinfo {volume}
  {98}},\ \bibinfo {pages} {075102} (\bibinfo {year} {2018})}\BibitemShut
  {NoStop}%
\bibitem [{\citenamefont {Tancogne-Dejean}\ \emph {et~al.}(2018)\citenamefont
  {Tancogne-Dejean}, \citenamefont {Sentef},\ and\ \citenamefont
  {Rubio}}]{Tancogne-Dejean2018}%
  \BibitemOpen
  \bibfield  {author} {\bibinfo {author} {\bibfnamefont {N.}~\bibnamefont
  {Tancogne-Dejean}}, \bibinfo {author} {\bibfnamefont {M.~A.}\ \bibnamefont
  {Sentef}}, \ and\ \bibinfo {author} {\bibfnamefont {A.}~\bibnamefont
  {Rubio}},\ }\href {\doibase 10.1103/PhysRevLett.121.097402} {\bibfield
  {journal} {\bibinfo  {journal} {Phys. Rev. Lett.}\ }\textbf {\bibinfo
  {volume} {121}},\ \bibinfo {pages} {097402} (\bibinfo {year}
  {2018})}\BibitemShut {NoStop}%
\bibitem [{\citenamefont {Zhu}\ \emph {et~al.}(2018)\citenamefont {Zhu},
  \citenamefont {Chacon},\ and\ \citenamefont {Zhu}}]{Zhu2018}%
  \BibitemOpen
  \bibfield  {author} {\bibinfo {author} {\bibfnamefont {W.}~\bibnamefont
  {Zhu}}, \bibinfo {author} {\bibfnamefont {A.}~\bibnamefont {Chacon}}, \ and\
  \bibinfo {author} {\bibfnamefont {J.-X.}\ \bibnamefont {Zhu}},\ }\href
  {https://arxiv.org/abs/1811.12334} {\bibfield  {journal} {\bibinfo  {journal}
  {arXiv:1811.12334}\ } (\bibinfo {year} {2018})}\BibitemShut {NoStop}%
\bibitem [{\citenamefont {Yu}\ \emph {et~al.}(2019)\citenamefont {Yu},
  \citenamefont {Hansen},\ and\ \citenamefont {Madsen}}]{Yu2019PRA}%
  \BibitemOpen
  \bibfield  {author} {\bibinfo {author} {\bibfnamefont {C.}~\bibnamefont
  {Yu}}, \bibinfo {author} {\bibfnamefont {K.~K.}\ \bibnamefont {Hansen}}, \
  and\ \bibinfo {author} {\bibfnamefont {L.~B.}\ \bibnamefont {Madsen}},\
  }\href {\doibase 10.1103/PhysRevA.99.013435} {\bibfield  {journal} {\bibinfo
  {journal} {Phys. Rev. A}\ }\textbf {\bibinfo {volume} {99}},\ \bibinfo
  {pages} {013435} (\bibinfo {year} {2019})}\BibitemShut {NoStop}%
\bibitem [{\citenamefont {Zhang}\ \emph {et~al.}(2018)\citenamefont {Zhang},
  \citenamefont {Si}, \citenamefont {Murakami}, \citenamefont {Bai},\ and\
  \citenamefont {George}}]{Zhang2018NatComm}%
  \BibitemOpen
  \bibfield  {author} {\bibinfo {author} {\bibfnamefont {G.~P.}\ \bibnamefont
  {Zhang}}, \bibinfo {author} {\bibfnamefont {M.~S.}\ \bibnamefont {Si}},
  \bibinfo {author} {\bibfnamefont {M.}~\bibnamefont {Murakami}}, \bibinfo
  {author} {\bibfnamefont {Y.~H.}\ \bibnamefont {Bai}}, \ and\ \bibinfo
  {author} {\bibfnamefont {T.~F.}\ \bibnamefont {George}},\ }\href
  {https://doi.org/10.1038/s41467-018-05535-4} {\bibfield  {journal} {\bibinfo
  {journal} {Nat. Comm.}\ }\textbf {\bibinfo {volume} {9}},\ \bibinfo {pages}
  {3031} (\bibinfo {year} {2018})}\BibitemShut {NoStop}%
\bibitem [{\citenamefont {Mukai}\ \emph {et~al.}(2014)\citenamefont {Mukai},
  \citenamefont {Hirori}, \citenamefont {Yamamoto}, \citenamefont {Kageyama},\
  and\ \citenamefont {Tanaka}}]{Mukai2014APL}%
  \BibitemOpen
  \bibfield  {author} {\bibinfo {author} {\bibfnamefont {Y.}~\bibnamefont
  {Mukai}}, \bibinfo {author} {\bibfnamefont {H.}~\bibnamefont {Hirori}},
  \bibinfo {author} {\bibfnamefont {T.}~\bibnamefont {Yamamoto}}, \bibinfo
  {author} {\bibfnamefont {H.}~\bibnamefont {Kageyama}}, \ and\ \bibinfo
  {author} {\bibfnamefont {K.}~\bibnamefont {Tanaka}},\ }\href {\doibase
  10.1063/1.4890475} {\bibfield  {journal} {\bibinfo  {journal} {Appl. Phys.
  Lett.}\ }\textbf {\bibinfo {volume} {105}},\ \bibinfo {pages} {022410}
  (\bibinfo {year} {2014})}\BibitemShut {NoStop}%
\bibitem [{\citenamefont {Ciappina}\ \emph {et~al.}(2017)\citenamefont
  {Ciappina}, \citenamefont {Perez-Hernandez}, \citenamefont {Landsman},
  \citenamefont {Okell}, \citenamefont {Zherebtsov}, \citenamefont {Forg},
  \citenamefont {Schotz}, \citenamefont {Seiffert}, \citenamefont {Fennel},
  \citenamefont {Shaaran}, \citenamefont {Zimmermann}, \citenamefont {Chacon},
  \citenamefont {Guichard}, \citenamefont {Zair}, \citenamefont {Tisch},
  \citenamefont {Marangos}, \citenamefont {Witting}, \citenamefont {Braun},
  \citenamefont {Maier}, \citenamefont {Roso}, \citenamefont {Kruger},
  \citenamefont {Hommelhoff}, \citenamefont {Kling}, \citenamefont {Krausz},\
  and\ \citenamefont {Lewenstein}}]{Ciappina2017RepProgPhys}%
  \BibitemOpen
  \bibfield  {author} {\bibinfo {author} {\bibfnamefont {M.~F.}\ \bibnamefont
  {Ciappina}}, \bibinfo {author} {\bibfnamefont {J.~A.}\ \bibnamefont
  {Perez-Hernandez}}, \bibinfo {author} {\bibfnamefont {A.~S.}\ \bibnamefont
  {Landsman}}, \bibinfo {author} {\bibfnamefont {W.~A.}\ \bibnamefont {Okell}},
  \bibinfo {author} {\bibfnamefont {S.}~\bibnamefont {Zherebtsov}}, \bibinfo
  {author} {\bibfnamefont {B.}~\bibnamefont {Forg}}, \bibinfo {author}
  {\bibfnamefont {J.}~\bibnamefont {Schotz}}, \bibinfo {author} {\bibfnamefont
  {L.}~\bibnamefont {Seiffert}}, \bibinfo {author} {\bibfnamefont
  {T.}~\bibnamefont {Fennel}}, \bibinfo {author} {\bibfnamefont
  {T.}~\bibnamefont {Shaaran}}, \bibinfo {author} {\bibfnamefont
  {T.}~\bibnamefont {Zimmermann}}, \bibinfo {author} {\bibfnamefont
  {A.}~\bibnamefont {Chacon}}, \bibinfo {author} {\bibfnamefont
  {R.}~\bibnamefont {Guichard}}, \bibinfo {author} {\bibfnamefont
  {A.}~\bibnamefont {Zair}}, \bibinfo {author} {\bibfnamefont {J.~W.~G.}\
  \bibnamefont {Tisch}}, \bibinfo {author} {\bibfnamefont {J.~P.}\ \bibnamefont
  {Marangos}}, \bibinfo {author} {\bibfnamefont {T.}~\bibnamefont {Witting}},
  \bibinfo {author} {\bibfnamefont {A.}~\bibnamefont {Braun}}, \bibinfo
  {author} {\bibfnamefont {S.~A.}\ \bibnamefont {Maier}}, \bibinfo {author}
  {\bibfnamefont {L.}~\bibnamefont {Roso}}, \bibinfo {author} {\bibfnamefont
  {M.}~\bibnamefont {Kruger}}, \bibinfo {author} {\bibfnamefont
  {P.}~\bibnamefont {Hommelhoff}}, \bibinfo {author} {\bibfnamefont {M.~F.}\
  \bibnamefont {Kling}}, \bibinfo {author} {\bibfnamefont {F.}~\bibnamefont
  {Krausz}}, \ and\ \bibinfo {author} {\bibfnamefont {M.}~\bibnamefont
  {Lewenstein}},\ }\href {http://stacks.iop.org/0034-4885/80/i=5/a=054401}
  {\bibfield  {journal} {\bibinfo  {journal} {Rep. Prog. Phys.}\ }\textbf
  {\bibinfo {volume} {80}},\ \bibinfo {pages} {054401} (\bibinfo {year}
  {2017})}\BibitemShut {NoStop}%
\bibitem [{\citenamefont {Hohlfeld}\ \emph {et~al.}(1997)\citenamefont
  {Hohlfeld}, \citenamefont {Matthias}, \citenamefont {Knorren},\ and\
  \citenamefont {Bennemann}}]{Hohlfeld1997PRL}%
  \BibitemOpen
  \bibfield  {author} {\bibinfo {author} {\bibfnamefont {J.}~\bibnamefont
  {Hohlfeld}}, \bibinfo {author} {\bibfnamefont {E.}~\bibnamefont {Matthias}},
  \bibinfo {author} {\bibfnamefont {R.}~\bibnamefont {Knorren}}, \ and\
  \bibinfo {author} {\bibfnamefont {K.~H.}\ \bibnamefont {Bennemann}},\ }\href
  {\doibase 10.1103/PhysRevLett.78.4861} {\bibfield  {journal} {\bibinfo
  {journal} {Phys. Rev. Lett.}\ }\textbf {\bibinfo {volume} {78}},\ \bibinfo
  {pages} {4861} (\bibinfo {year} {1997})}\BibitemShut {NoStop}%
\bibitem [{\citenamefont {Kimel}\ \emph {et~al.}(2005)\citenamefont {Kimel},
  \citenamefont {Kirilyuk}, \citenamefont {Usachev}, \citenamefont {Pisarev},
  \citenamefont {Balbashov},\ and\ \citenamefont {Rasing}}]{Kimel2005Nature}%
  \BibitemOpen
  \bibfield  {author} {\bibinfo {author} {\bibfnamefont {A.~V.}\ \bibnamefont
  {Kimel}}, \bibinfo {author} {\bibfnamefont {A.}~\bibnamefont {Kirilyuk}},
  \bibinfo {author} {\bibfnamefont {P.~A.}\ \bibnamefont {Usachev}}, \bibinfo
  {author} {\bibfnamefont {R.~V.}\ \bibnamefont {Pisarev}}, \bibinfo {author}
  {\bibfnamefont {A.~M.}\ \bibnamefont {Balbashov}}, \ and\ \bibinfo {author}
  {\bibfnamefont {T.}~\bibnamefont {Rasing}},\ }\href
  {https://doi.org/10.1038/nature03564} {\bibfield  {journal} {\bibinfo
  {journal} {Nature (London)}\ }\textbf {\bibinfo {volume} {435}},\ \bibinfo
  {pages} {655} (\bibinfo {year} {2005})}\BibitemShut {NoStop}%
\bibitem [{\citenamefont {Takayoshi}\ \emph
  {et~al.}(2014{\natexlab{a}})\citenamefont {Takayoshi}, \citenamefont {Aoki},\
  and\ \citenamefont {Oka}}]{Takayoshi2014PRBa}%
  \BibitemOpen
  \bibfield  {author} {\bibinfo {author} {\bibfnamefont {S.}~\bibnamefont
  {Takayoshi}}, \bibinfo {author} {\bibfnamefont {H.}~\bibnamefont {Aoki}}, \
  and\ \bibinfo {author} {\bibfnamefont {T.}~\bibnamefont {Oka}},\ }\href
  {\doibase 10.1103/PhysRevB.90.085150} {\bibfield  {journal} {\bibinfo
  {journal} {Phys. Rev. B}\ }\textbf {\bibinfo {volume} {90}},\ \bibinfo
  {pages} {085150} (\bibinfo {year} {2014}{\natexlab{a}})}\BibitemShut
  {NoStop}%
\bibitem [{\citenamefont {Takayoshi}\ \emph
  {et~al.}(2014{\natexlab{b}})\citenamefont {Takayoshi}, \citenamefont {Sato},\
  and\ \citenamefont {Oka}}]{Takayoshi2014PRBb}%
  \BibitemOpen
  \bibfield  {author} {\bibinfo {author} {\bibfnamefont {S.}~\bibnamefont
  {Takayoshi}}, \bibinfo {author} {\bibfnamefont {M.}~\bibnamefont {Sato}}, \
  and\ \bibinfo {author} {\bibfnamefont {T.}~\bibnamefont {Oka}},\ }\href
  {\doibase 10.1103/PhysRevB.90.214413} {\bibfield  {journal} {\bibinfo
  {journal} {Phys. Rev. B}\ }\textbf {\bibinfo {volume} {90}},\ \bibinfo
  {pages} {214413} (\bibinfo {year} {2014}{\natexlab{b}})}\BibitemShut
  {NoStop}%
\bibitem [{\citenamefont {Wolf}(2000)}]{Wolf2000BJP}%
  \BibitemOpen
  \bibfield  {author} {\bibinfo {author} {\bibfnamefont {W.~P.}\ \bibnamefont
  {Wolf}},\ }\href
  {http://www.scielo.br/scielo.php?script=sci_arttext&pid=S0103-97332000000400030&nrm=iso}
  {\bibfield  {journal} {\bibinfo  {journal} {Braz. J. Phys.}\ }\textbf
  {\bibinfo {volume} {30}},\ \bibinfo {pages} {794} (\bibinfo {year}
  {2000})}\BibitemShut {NoStop}%
\bibitem [{\citenamefont {Coldea}\ \emph {et~al.}(2010)\citenamefont {Coldea},
  \citenamefont {Tennant}, \citenamefont {Wheeler}, \citenamefont {Wawrzynska},
  \citenamefont {Prabhakaran}, \citenamefont {Telling}, \citenamefont
  {Habicht}, \citenamefont {Smeibidl},\ and\ \citenamefont
  {Kiefer}}]{Coldea2010Science}%
  \BibitemOpen
  \bibfield  {author} {\bibinfo {author} {\bibfnamefont {R.}~\bibnamefont
  {Coldea}}, \bibinfo {author} {\bibfnamefont {D.~A.}\ \bibnamefont {Tennant}},
  \bibinfo {author} {\bibfnamefont {E.~M.}\ \bibnamefont {Wheeler}}, \bibinfo
  {author} {\bibfnamefont {E.}~\bibnamefont {Wawrzynska}}, \bibinfo {author}
  {\bibfnamefont {D.}~\bibnamefont {Prabhakaran}}, \bibinfo {author}
  {\bibfnamefont {M.}~\bibnamefont {Telling}}, \bibinfo {author} {\bibfnamefont
  {K.}~\bibnamefont {Habicht}}, \bibinfo {author} {\bibfnamefont
  {P.}~\bibnamefont {Smeibidl}}, \ and\ \bibinfo {author} {\bibfnamefont
  {K.}~\bibnamefont {Kiefer}},\ }\href
  {http://science.sciencemag.org/content/327/5962/177} {\bibfield  {journal}
  {\bibinfo  {journal} {Science}\ }\textbf {\bibinfo {volume} {327}},\ \bibinfo
  {pages} {177} (\bibinfo {year} {2010})}\BibitemShut {NoStop}%
\bibitem [{\citenamefont {Vidal}(2007)}]{Vidal2007PRL}%
  \BibitemOpen
  \bibfield  {author} {\bibinfo {author} {\bibfnamefont {G.}~\bibnamefont
  {Vidal}},\ }\href {\doibase 10.1103/PhysRevLett.98.070201} {\bibfield
  {journal} {\bibinfo  {journal} {Phys. Rev. Lett.}\ }\textbf {\bibinfo
  {volume} {98}},\ \bibinfo {pages} {070201} (\bibinfo {year}
  {2007})}\BibitemShut {NoStop}%
\bibitem [{\citenamefont {White}(1992)}]{White1992PRL}%
  \BibitemOpen
  \bibfield  {author} {\bibinfo {author} {\bibfnamefont {S.~R.}\ \bibnamefont
  {White}},\ }\href {\doibase 10.1103/PhysRevLett.69.2863} {\bibfield
  {journal} {\bibinfo  {journal} {Phys. Rev. Lett.}\ }\textbf {\bibinfo
  {volume} {69}},\ \bibinfo {pages} {2863} (\bibinfo {year}
  {1992})}\BibitemShut {NoStop}%
\bibitem [{\citenamefont {Vidal}(2003)}]{Vidal2003PRL}%
  \BibitemOpen
  \bibfield  {author} {\bibinfo {author} {\bibfnamefont {G.}~\bibnamefont
  {Vidal}},\ }\href {\doibase 10.1103/PhysRevLett.91.147902} {\bibfield
  {journal} {\bibinfo  {journal} {Phys. Rev. Lett.}\ }\textbf {\bibinfo
  {volume} {91}},\ \bibinfo {pages} {147902} (\bibinfo {year}
  {2003})}\BibitemShut {NoStop}%
\bibitem [{\citenamefont {Jackson}(1998)}]{Jackson1998Book}%
  \BibitemOpen
  \bibfield  {author} {\bibinfo {author} {\bibfnamefont {J.~D.}\ \bibnamefont
  {Jackson}},\ }\href@noop {} {\emph {\bibinfo {title} {Classical
  Electrodynamics}}}\ (\bibinfo  {publisher} {Wiley},\ \bibinfo {address} {New
  York},\ \bibinfo {year} {1998})\BibitemShut {NoStop}%
\bibitem [{\citenamefont {Giamarchi}(2004)}]{Giamarchi2004Book}%
  \BibitemOpen
  \bibfield  {author} {\bibinfo {author} {\bibfnamefont {T.}~\bibnamefont
  {Giamarchi}},\ }\href@noop {} {\emph {\bibinfo {title} {Quantum physics in
  one dimension}}}\ (\bibinfo  {publisher} {Oxford university press},\ \bibinfo
  {address} {Oxford},\ \bibinfo {year} {2004})\BibitemShut {NoStop}%
\end{thebibliography}
\end{document}